\documentclass[twocolumn,aps,prl,groupeaddress,showpacs,longbibliography]{revtex4-2}
\usepackage{amsmath,amssymb,amsthm,amsfonts,float,graphics,epsfig,epstopdf,color,verbatim,tabularx,bm,multirow,appendix,url}
\let\oldproofname=\proofname
\renewcommand{\proofname}{\rm\bf{\oldproofname}}
\usepackage[utf8]{inputenc}
\usepackage[T1]{fontenc}
\usepackage{xcolor}
\usepackage{dsfont}
\usepackage{textcomp}
\usepackage{yfonts}
\usepackage{footnote}
\usepackage{bm}
\usepackage{mathrsfs}
\usepackage{graphicx}
\usepackage{verbatim}
\usepackage[colorlinks=true, citecolor=blue, linkcolor=blue, urlcolor=blue]{hyperref}
\usepackage{multirow}
\usepackage{enumitem,kantlipsum}
\usepackage{physics}
\usepackage{slashed}
\usepackage{braket}
\usepackage[normalem]{ulem}

\usepackage{tikz}
\usepackage[compat=1.1.0]{tikz-feynman}

\usetikzlibrary{calc}
\usetikzlibrary{shapes.multipart}

\graphicspath{ {images/} }

\newcommand{\be}{\begin{equation}}
\newcommand{\ee}{\end{equation}}
\newcommand{\bea}{\begin{eqnarray}}
\newcommand{\eea}{\end{eqnarray}}
\newcommand{\ii}{\mathrm{i}}

\renewcommand{\vec}[1]{{\mathbf #1}}

\newcommand{\comments}[1]{}

\newcommand{\stkout}[1]{\ifmmode\text{\sout{\ensuremath{#1}}}\else\sout{#1}\fi}

\DeclareMathOperator{\I}{i}
\DeclareMathOperator{\e}{e}


\begin{document}

\title{A Perturbative Approach to Symmetric Mass Generation}
\author{Simon Martin}
\email{simartin@ucsd.edu}
\affiliation{Department of Physics, University of California at San Diego, La Jolla, California 92093, USA}

\author{Tarun Grover}
\email{tagrover@ucsd.edu}
\affiliation{Department of Physics, University of California at San Diego, La Jolla, California 92093, USA}

\begin{abstract}
The  Landau paradigm has been a powerful framework for understanding phase transitions involving spontaneous symmetry breaking. In contrast, phase transitions between two symmetric phases, where neither phase breaks any symmetry, remain less explored. One intriguing class of such transitions involves ``symmetric mass generation'' (SMG), where interactions drive a transition from a gapless symmetric phase to a gapped symmetric phase. In this work, we develop a controlled perturbative approach to study a class of such transitions, based on an $\epsilon$-expansion around the critical dimension where the SMG-inducing-interaction becomes marginal. Applying this method to two distinct models, we identify a single-parameter-tuned transition in each case, which we conjecture captures the universal critical behavior of the SMG transition in these models. We compute universal quantities associated with these transitions.\end{abstract}

\maketitle

A large part of our understanding of quantum phase transitions builds on the intuition from the generalized Landau paradigm, where one condenses an order parameter charged under some global symmetry. Not all quantum phase transitions however are readily described in terms of order parameters. A prominent example is the phenomenon of ``symmetric mass generation'' (SMG) whose prototypical examples involve mass generation for gapless fermions without breaking any global symmetry spontaneously (see Ref.~\cite{wang2022symmetric} for a review). The presence of an SMG phase hinges on the absence of any anomalies that prohibit a symmetric, gapped phase without topological order, which may sometimes even prohibit any gapped phase at all. Such anomalies can be either non-perturbative, as in the context of interaction-reduced classification of symmetry protected topological phases~\cite{Fidkowski10a,Fidkowski2013non-abelian,wang2014interacting,metlitski2014interaction,Ryu2012interacting,Qi2013newclass,Ryu2013Interaction,Gu2014Effect}, or, they can be perturbative, which  relates to the question of regularizing chiral fermions on lattices~\cite{eichten1986chiral,Wen2013alattice,you2014interacting,BenTov2016Origin,Wang2023nonperturbative,tong2022comments,Shlomo2021gapped,Wang2019Solution,Meng2022symmetric,Catterall2021chiral,butt2025}. Although there has been remarkable progress in understanding the SMG phase itself, including concrete lattice models that host such a phase~\cite{slagle2015exotic,ayyar2015massive,ayyar2016origin,catterall2016fermion,he2016quantum,hou2023variational,liu2024disorder,Catterall2017novel}, analytical progress on the universal aspects of the SMG phase transition is rather limited. In this work, we pursue a controlled, perturbative renormalization group (RG) approach to the SMG critical point in a class of models. We primarily focus on a 2+1-D field theory motivated by the lattice models in Refs.~\cite{slagle2015exotic,ayyar2015massive,ayyar2016origin,catterall2016fermion,he2016quantum,hou2023variational} which involve four-fermion interactions favoring a gapped, symmetric phase. Additionally, we study a non-relativistic model of fermions, which is also expected to host SMG. Our main result is the identification of a new fixed point with a single relevant direction arising from a field-theory analog of the SMG-inducing interaction. This leads us to conjecture that this fixed point describes the SMG critical point in these models.

It is instructive to recall the simplest manifestation of SMG in the quantum mechanical problem of $N$ (an even number) Majorana modes $\chi_a$ with two symmetries: time reversal that acts as $\chi_a \to \chi_a, \ii \to -\ii$, and fermion parity that acts as $\chi_a \to -\chi_a, \ii \to \ii$~\cite{wang2022symmetric}. In the absence of interactions, the Hamiltonian is zero, since there exists no fermion bilinear that respects the aforementioned symmetries. Correspondingly, the ground state degeneracy is $2^{N/2}$. However, in the presence of interactions, and when the number of Majorana fermions is a multiple of eight, one can indeed find a Hamiltonian with a unique, symmetric ground state. For example, when $N = 8$, two such Hamiltonians are given by $H_{\pm} = \pm \left(c_1 c_2 c_3 c_4 + \textrm{h.c.}\right) $, where $c_i = \left(\chi_{2i-1} + \ii \chi_{2i}\right)/2$ are complex fermions. The unique, gapped ground state of $H_{\pm}$ is $|{\psi}_{\pm}\rangle = \frac{1}{\sqrt{2}}\left(|0 0 0 0 \rangle \mp |1 1 1 1 \rangle\right)$, where a basis state $|n_1 n_2 n_3 n_4\rangle$ is defined as the simultaneous eigenstate of the number operators $c^{\dagger}_i c^{\phantom{\dagger}}_i$. Relatedly, precisely when the number of modes is a multiple of eight, the 0+1-D system can be interpreted as the boundary of a trivial SPT phase with the aforementioned symmetries, and hence can be gapped out~\cite{Fidkowski10a}. This relation between gappable SPT boundary and SMG extends to higher dimensions as well. For example, 3+1-D topological superconductors in the class DIII have a $\mathbb{Z}_{16}$ classification~\cite{kapustin2015fermionic} and correspondingly, sixteen copies of Majorana fermions in 2+1-D with time-reversal and fermion parity can be gapped out without breaking any symmetry. Crucially, any four-fermion interaction at the free Majorana fixed point is perturbatively irrelevant and therefore, this raises the possibility of a phase transition between the non-interacting, gapless Majorana phase and the SMG phase~\cite{slagle2015exotic,ayyar2015massive,ayyar2016origin,catterall2016fermion,he2016quantum,hou2023variational}. 

A candidate theory for the SMG transition for the models in Refs.~\cite{slagle2015exotic,ayyar2015massive,ayyar2016origin,catterall2016fermion,he2016quantum,hou2023variational} was proposed in Refs.\cite{you2018symmetric,You2018frombosonic}. It involves fractionalizing the physical fermion in terms of a boson and a fermion which results in an $SU(4)$ QCD-Higgs gauge theory. The massless Dirac phase corresponds to the Higgs phase of this theory, where the bosonic partons condense, while the SMG phase was argued to be the confining phase, under the assumption that confinement does not break any global symmetries.   This  approach leads to a strongly-coupled field theory for the SMG transition, involving non-abelian gauge fields coupled to both bosonic and fermionic matter. Therefore, it is also desirable to find a weakly coupled description of this transition. Here we will pursue a different approach that leads to a weakly coupled candidate, and which is formulated directly in terms of the physical fermions.

Our basic observation is that in a field-theoretic description of the models for SMG in Refs.~\cite{slagle2015exotic,ayyar2015massive,ayyar2016origin,catterall2016fermion,he2016quantum,hou2023variational}, all four-fermion terms, including the ones that induce SMG, are simultaneously marginal at tree level in a certain spacetime dimension $D^{*}$. This allows a framework to search for the SMG critical point on the same footing as standard symmetry-breaking transitions. This is done by considering the theory in $D^{*} + \epsilon$ dimensions, and performing a controlled RG expansion in the parameter $\epsilon$, akin to the $2+\epsilon$ or $4-\epsilon$ expansion for the $O(N)$ model. In the absence of the SMG term, this approach recovers the standard Gross-Neveu-type transitions that correspond to spontaneous breaking of various global symmetries. Crucially, this approach does not rely on the presence of any order parameter. On that note, the $D = 2 + \epsilon$ expansion of the $O(N)$ non-linear sigma model has recently been argued to \textit{not} be smoothly connected to the Wilson-Fisher CFT in $D = 3$~\cite{jones2024explorations,de2025disturbing}. One reason why extending $O(N)$ models to fractional dimensions can be problematic is that these models aim to describe symmetry-breaking transitions in which topological defects may play a crucial role -- one that is not faithfully represented in non-integer dimensions. In contrast, the SMG critical point is  allowed precisely when topological defects associated even with any short-range ordering are featureless~\cite{wang2022symmetric,you2014interacting}. Therefore, an $\epsilon$-expansion-based approach appears more appropriate for an SMG transition.

To illustrate the idea, we will consider two different field theories: the first one describes a non-relativistic analog of the SMG phenomena, and corresponds to four flavors of fermions at unitarity supplemented with an SMG-type interaction of the form $\psi_1 \psi_2 \psi_3 \psi_4 + \textrm{h.c.}$. The second field theory is relativistic and corresponds to the continuum limit of the models in Refs.~\cite{slagle2015exotic,ayyar2015massive,ayyar2016origin,catterall2016fermion,he2016quantum,hou2023variational,you2018symmetric}. In both theories, we identify a putative SMG critical point with a single relevant direction and we calculate a few universal exponents. Let us begin our discussion with the non-relativistic theory, which turns out to be technically simpler.

\begin{figure}[h]
\begin{center}
\includegraphics[width=0.47\textwidth]{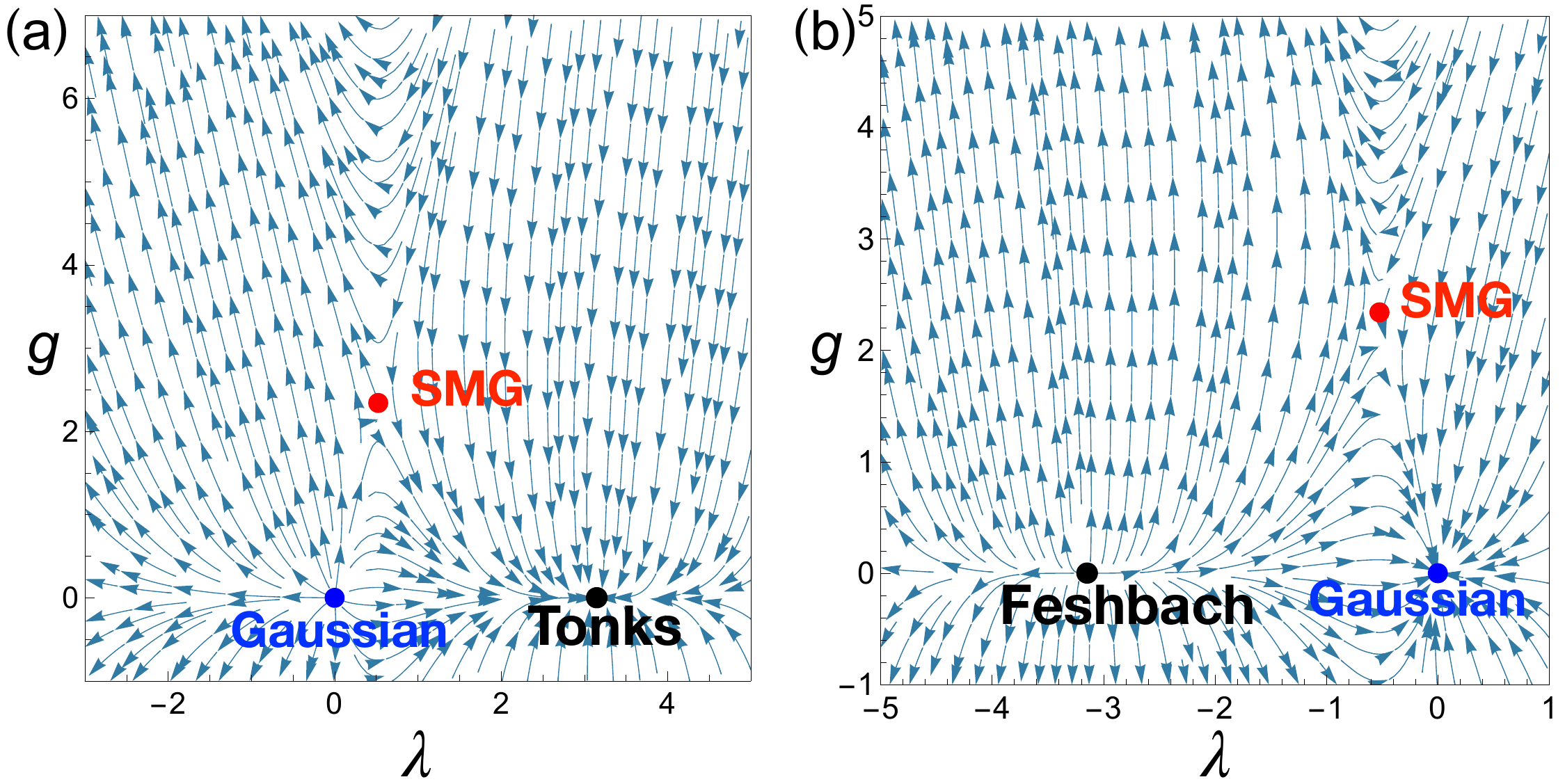}
\end{center}
\caption{RG flow in the $(\lambda,g)$ plane. (a) corresponds to $d = 1$ $(\epsilon=-1)$  and (b)  
corresponds to $d = 3$ $(\epsilon=1)$.}
\label{fig:RG_z2}
\end{figure}

\textit{A non-relativistic model with SMG-type transition:} Let us consider four flavors of non-relativistic complex fermions, denoted $\psi_{\alpha}$ with $\alpha=1,...,4$, which are governed by the following imaginary-time action $S = \int d\tau d^d\mathbf{x} \,\mathcal{L}$ in $d$ spatial dimensions, where

\begin{equation}
\mathcal{L} = \Bar{\psi}_{\alpha} \Big( \partial_{\tau} - \frac{1}{2m} \nabla^2 \Big) \psi_{\alpha} + \lambda (\Bar{\psi}_{\alpha} \psi_{\alpha})^2 + \mathcal{L}_{\text{SMG}} \label{Eq:Sz2}
\end{equation}

\noindent with $\mathcal{L}_{\text{SMG}} = g \Big( \psi_1 \psi_2 \psi_3 \psi_4 + \Bar{\psi}_4 \Bar{\psi}_3 \Bar{\psi}_2 \Bar{\psi}_1 \Big)$. The chemical potential $\mu$ is fine-tuned to zero (we will say more about this later).
The $U(1)$-symmetric theory, with $g=0$, has already been well studied in the literature \cite{nishida2007fermi,nikolic2007renormalization,nishida2007nonrelativistic}, at least for the case of $N_f=2$ flavors. In addition to the internal $SU(N_f)$ symmetry, the theory at $g = 0$ is invariant under Galilean transformations, protecting the value of the dynamical critical exponent to be $z=2$. The four-fermion interaction $\lambda$ is marginal in $d=2$ spatial dimensions and the theory can thus be studied in $d=2+\epsilon$. For $\epsilon>0$ ($d>2$), the Gaussian fixed point is stable and there is a quantum critical point at $\lambda^*<0$ describing a Feshbach resonance (i.e., fermions scattering in the unitarity limit). On the other hand, for $\epsilon<0$ ($d<2$), there is a stable IR fixed point at $\lambda^*>0$, so called `Tonks gas' (see $g = 0$ axis in Fig. \ref{fig:RG_z2}), which essentially again describes fermions at unitarity.

Let us now turn on the $U(1)$-breaking interaction $\mathcal{L}_{\text{SMG}}$. A nonzero $g$ breaks Galilean invariance and $z$ is no longer protected. We perform a 1-loop RG analysis of the full theory, Eq.~\ref{Eq:Sz2}, with a controlled $\epsilon$-expansion, using dimensional regularization in $d=2+\epsilon$ spatial dimensions and minimal subtraction (MS) scheme (see Appendix \ref{sec:app_nonrel} for details). The RG equations for $\lambda$ and $g$ are

\begin{align}
\begin{split}
\beta(\lambda) = \frac{d \lambda}{dl} &= -\epsilon \lambda - \frac{1}{4\pi} (4\lambda^2 + g^2) \\ \beta(g) = \frac{dg}{dl} &= -\epsilon g - \frac{6}{\pi} \lambda g \, .
\end{split} \label{eq:RGnonrel}
\end{align}

\noindent Moreover, the flow of the chemical potential is $\frac{d\mu}{dl}=2\mu$ and therefore, if one fine tunes $\mu$ to zero at the outset, it does not run. The RG equations are invariant under $g \rightarrow -g$, and we can therefore only focus on the case $g>0$. Fig.~\ref{fig:RG_z2} shows the RG flows in the $(\lambda,g)$ plane for: (a) $\epsilon=-1$ ($d=1$) and (b) $\epsilon=+1$ ($d=3$). Both scenarios lead to three fixed points (with $g\geq 0$). Crucially, in both cases, we find a new fixed point with a single relevant direction at  $(\lambda,g)=\big( -\frac{\pi}{6}\epsilon, \frac{\sqrt{5}}{3} \pi |\epsilon| \big)$. When $\epsilon = -1$ (i.e. $d = 1$), this fixed point describes the phase transition between Tonks gas and the phase corresponding to $g \gg 1 $, while when $\epsilon = +1$ (i.e. $d = 3$), it describes the phase transition between the Gaussian fixed-point and the phase at $g \gg 1$.

Let us now discuss the nature of the phase at $g \gg 1$. Since the theory is interacting and not exactly solvable, it is difficult to rigorously argue about the nature of the ground state in this limit. However, one can build intuition by finding the ground state of the Hamiltonian that is proportional to the relevant scaling operator at the fixed point labeled ``SMG'' in Fig.~\ref{fig:RG_z2}. We find that this ground state is unique and takes the form $\alpha |0 0 0 0 \rangle + \beta |1 1 1 1 \rangle$, where $\alpha, \beta$ are some non-zero coefficients (Appendix \ref{sec:app_nonrel}). This situation is not unlike turning on the relevant, Ising-symmetric perturbation at the Wilson-Fisher fixed point in $d < 4$ dimensions (with coefficient chosen to favor the symmetric phase), which is expected to result in a gapped, symmetric phase. Based on these observations and the topology of the RG flow shown in Fig. \ref{fig:RG_z2}, it is a reasonable expectation that the fixed point labeled ``SMG'' describes a single parameter tuned transition between a gapless, stable phase and the continuum analog of the SMG phase.

Let us next discuss universal characteristics of the putative SMG critical point. We find that the (inverse) correlation length exponent, which also characterizes the vanishing of the fermion gap as the critical point is approached, is $\nu^{-1} = \epsilon$ at 1-loop. We also obtain the dynamical critical exponent $z=2 + \frac{10}{81}\epsilon^2$ and the fermion anomalous dimension $\eta_{\psi} = \frac{5}{54}\epsilon^2$ (details in Appendix \ref{sec:app_nonrel}), both of which require a two-loop calculation. These deviate from the exact $U(1)$-symmetric values of $z=2$ and $\eta_{\psi}=0$ due to the breaking of Galilean invariance by the SMG interaction $\mathcal{L}_{\text{SMG}}$.

\textit{A relativistic model with SMG-type transition:}  The second theory we study is relativistic and is intended to capture the low-energy physics of the lattice model studied in Refs.~\cite{ayyar2015massive,ayyar2016origin,catterall2016fermion,he2016quantum,hou2023variational} which exhibit an SMG phase transition. This lattice Hamiltonian consists of four species (flavors) of spinless fermions at half-filling, hopping on a 2d honeycomb lattice: $H = H_0 + H_{\text{SMG}}$, with

\begin{align}
\begin{split}
H_0 &= - t \sum_{\langle r,r' \rangle} \sum_{\alpha=1}^4 \Big( c_{r,\alpha}^{\dag} c_{r',\alpha} + c^{\dag}_{r',\alpha} c_{r,\alpha} \Big) \\ H_{\text{SMG}} &= U \sum_{r} \Big( c_{r1} c_{r2} c_{r3} c_{r4} + c_{r4}^{\dag} c_{r3}^{\dag} c_{r2}^{\dag} c_{r1}^{\dag} \Big) \, ,
\end{split} \label{eq:H_SMG}
\end{align}

\noindent where $r,r'$ label the position on the honeycomb lattice, while the sum in the first term is performed over all nearest-neighbor sites. $\alpha=1,...,4$ labels the fermion flavor. The second term is the aforementioned $SU(4)$ flavor symmetric SMG term, so that when $|U|/t \to \infty$, the ground state is gapped and symmetric, and is explicitly given by $|\psi_{\pm}\rangle =\prod_{r} |\psi_{r,\pm}\rangle$ where $|{\psi}_{r,\pm}\rangle = \frac{1}{\sqrt{2}}\left(|0 0 0 0 \rangle_{r} \mp |1 1 1 1 \rangle_{r}\right)$. In addition to the $SU(4)$ flavor and the usual symmetries of the honeycomb lattice (rotation, reflection and translation), the Hamiltonian also possesses  an anti-unitary particle-hole symmetry, $c_{r,\alpha} \rightarrow (-1)^{r} c^{\dag}_{r,\alpha}$, $\ii \rightarrow -\I$, where $(-1)^{r}=\pm1$ for $r \in A/B$ sublattice. It has been argued in Ref. \cite{wang2022symmetric,you2018symmetric} that these symmetries imply that the low-energy theory of the system possesses a $\mathbb{Z}_{16}$ anomaly such that a gapped symmetric state is possible only if the low-energy theory has $8n$ number of complex Dirac fermions where $n \in \mathbb{Z}_{>0}$.

\begin{figure}[t]
\begin{center}
\includegraphics[width=0.47\textwidth]{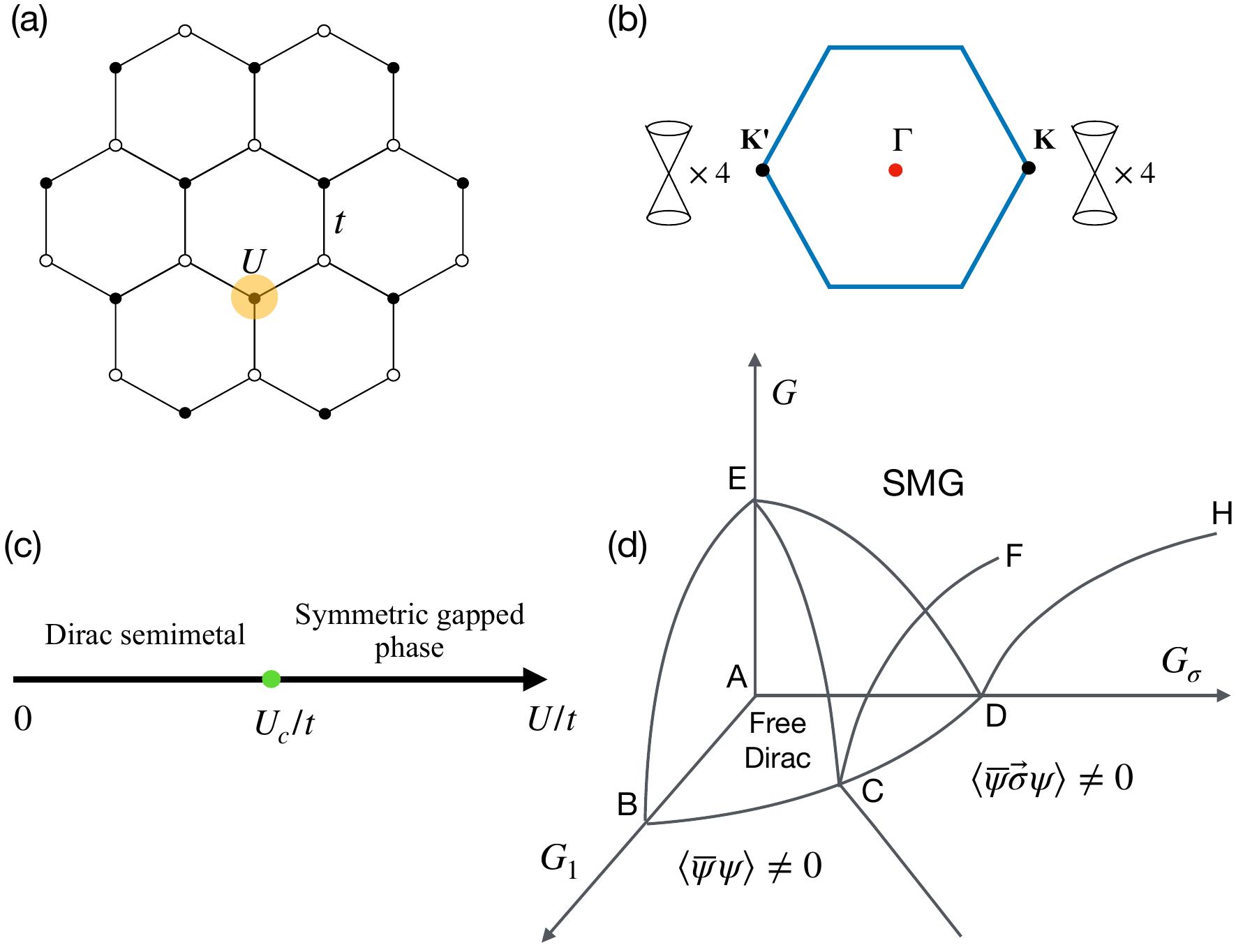}
\end{center}
\caption{(a) Honeycomb lattice, with intersite hopping $t$ and on-site interaction $U$ (Eq. \ref{eq:H_SMG}). (b) Brillouin zone, where each valley ($\mathbf{K, K'}$) has four Dirac cones. (c) Lattice phase diagram as a function of $U/t$, with the SMG QPT represented in green \cite{ayyar2016origin,catterall2016fermion,he2016quantum}. (d) Topology of the phase diagram based on the perturbative RG discussed in the main text. The surface EBC corresponds to the Gross-Neveu transition, the line CD corresponds to the Gross-Neveu-Heisenberg transition, and the surface ECD corresponds to the putative SMG critical point. Note that the GNH transition has co-dimension two because the SMG coupling is relevant at the GNH transition in the $U(1)$-symmetric plane $(G_1, G_\sigma)$.}
\label{fig:honeycomb}
\end{figure}

At $U=0$, $H$ describes the band structure of graphene with four species of spinless electrons. Each of them gives rise to two Dirac cones (two ``valleys'') and the low-energy theory therefore consists of eight species of $D=3$ free Dirac fermions (See Fig. \ref{fig:honeycomb} (b)). This Dirac semi-metal phase is (perturbatively) stable to four-fermion interactions. On the other hand, when $U/t \gg 1$, as already mentioned, one finds an SMG phase. Remarkably, Refs. \cite{ayyar2015massive,ayyar2016origin,catterall2016fermion,he2016quantum,hou2023variational} found evidence for a direct, second-order transition between the Dirac semi-metal phase and the SMG phase. Inspired by the aforementioned non-relativistic model, developing a perturbative approach to this transition will be our focus.

In the following, similar to Ref. \cite{you2018symmetric}, we will assume that the low-energy theory of $H$ is relativistic with an emergent $SU(2)$ valley symmetry. Under these assumptions, the imaginary-time action, up to quartic terms in fermion fields, is $S = S_0 + S_{\text{int},U(1)} + S_{\text{SMG}}$ where $S_0 = \int d^Dx \bar{\psi}_{i\alpha} \gamma_{\mu} \partial_{\mu} \psi_{i\alpha}$ is the free part ($x = (\tau,\vec{r})$ denotes spacetime coordinates), while $i=1,2$ labels the valley ($N_v=2$) and $\alpha=1,...,4$ labels the flavor ($N_f=4$), and $\psi_{i \alpha}$ is a two-component spinor. The second term contains the Fierz-complete four-fermion interactions respecting the global $SU(2)\times SU(4)$ symmetry that are in addition $U(1)$-symmetric ($\psi \to e^{i \theta} \psi$),

\begin{align}
\begin{split}
 &S_{\text{int},U(1)} \\ &= -\frac{G_1}{2} \int_x (\bar{\psi}_{i\alpha} \psi_{i\alpha})^2 - \frac{G_{\sigma}}{2} \int_x (\bar{\psi}_{i\alpha} \sigma_{ij}^A \psi_{j\alpha})^2 \\ &- \frac{G_{T}}{2} \int_x (\bar{\psi}_{i\alpha} T^A_{\alpha \beta} \psi_{i \beta})^2 - \frac{G_{\sigma T}}{2} \int_x (\bar{\psi}_{i\alpha} \sigma^A_{ij} T^B_{\alpha \beta} \psi_{j\beta})^2 \, ,
\end{split}
\end{align}

\noindent where $\sigma^A$ ($A=1,2,3$) are the three Pauli matrices acting in the $SU(2)$ valley space, while $T^A$ ($A=1,...,15$) are the fifteen $SU(4)$ generators acting in the flavor space. Finally, the third term, which is expected to lead to symmetric mass generation, is $S_{\text{SMG}} = \frac{G}{8} \epsilon_{ab} \epsilon_{cd} \epsilon_{ij} \epsilon_{kl} \epsilon_{\alpha \beta \gamma \delta} \int d^Dx \Big[ \psi_{ai\alpha} \psi_{bj\beta} \psi_{ck\gamma} \psi_{dl\delta} + \bar{\psi}_{ai\alpha} \bar{\psi}_{bj\beta} \bar{\psi}_{ck\gamma} \bar{\psi}_{dl\delta} \Big]$, where the first index $a=1,2$ labels the Dirac spinor component. This additional interaction is the low-energy version of the lattice on-site interaction $U$, as it is the only term one can write which is $SU(2)\times SU(4)$ and Lorentz-invariant \cite{you2018symmetric}, while breaking $U(1)$ down to $\mathbb{Z}_4$.

In $D=2$, the five distinct four-fermion interactions are marginal at the tree-level, and hence, this theory can be perturbatively studied in $D=2+\epsilon$. Our calculations will be performed using a ``hybrid'' scheme  (see \cite{grover2014emergent} for a $D=4-\epsilon$ analog): momentum integrals are computed in $D=2+\epsilon$ with $\epsilon$ small, but the spinor structure is taken to be the one of the target spacetime dimension, $D=3$ (see Appendix \ref{sec:app_rel} for details). A technical benefit of this method is that it minimizes the number of $U(1)$-symmetric operators. More importantly, it avoids introducing operators that have no physical meaning in $D=3$, where there is no notion of chirality/$\gamma^5$ Dirac matrix. Nevertheless, as discussed below, this scheme reproduces with a very good accuracy results of universal quantities in the $U(1)$-symmetric theory that have been computed with an alternative scheme in the literature~\cite{gracey2018large,ladovrechis2023gross}.

\begin{figure}[t]
	\centering
	\includegraphics[width=0.47\textwidth]{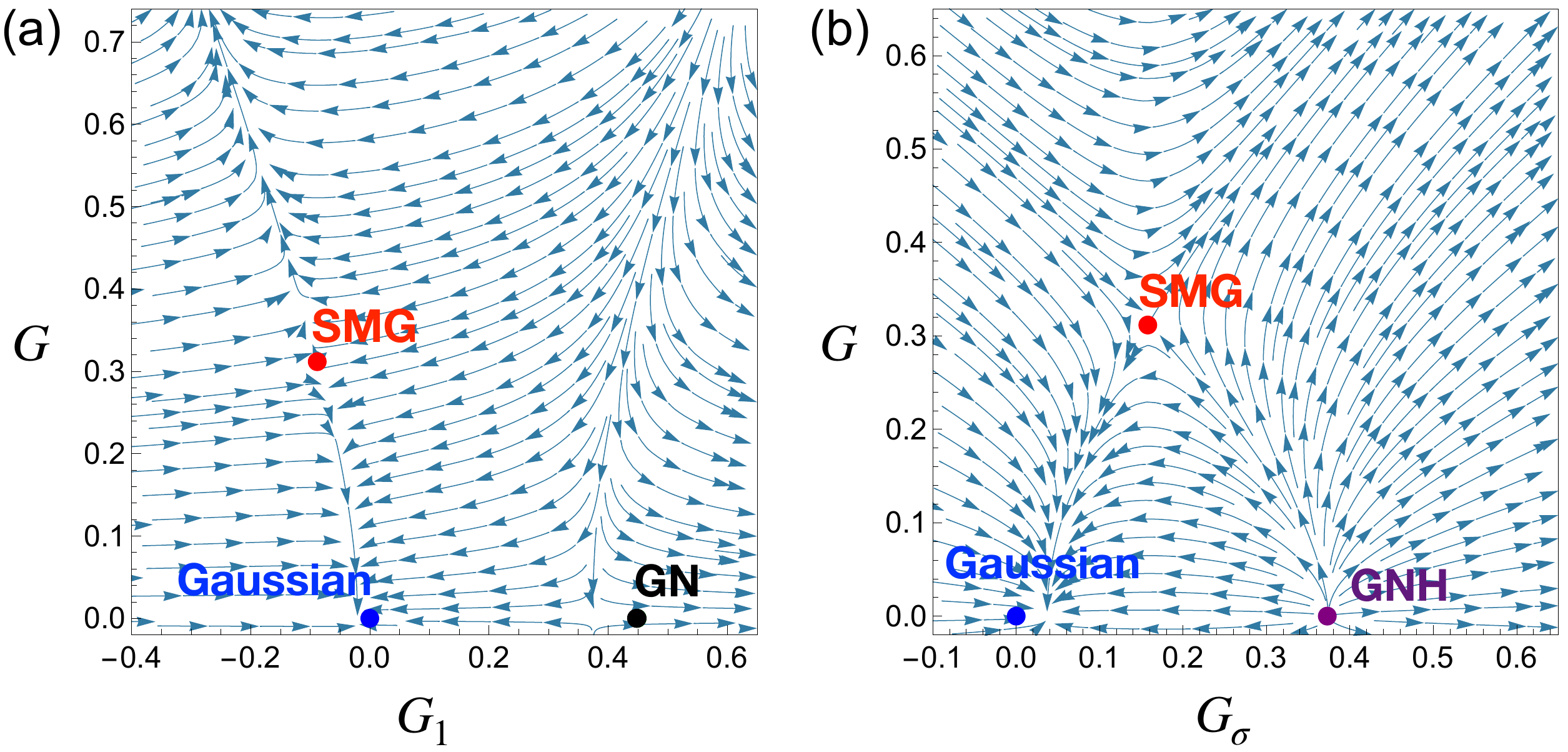}
	\caption{2d slices of the full 5d RG flow. In both cases, the vertical axis represents the SMG coupling $G$, while the horizontal axis corresponds to (a) the GN coupling $G_1$ or (b) the GNH coupling $G_{\sigma}$. In both flows, the three couplings not represented are set at their SMG fixed point value, ensuring that the SMG fixed point (red) lies in both slices. The projection of the Gaussian (blue), GN (black) and GNH (purple) fixed points are represented by dots.}
	\label{fig:RG2}
\end{figure}

We computed the $\beta$ functions for the five couplings entering the action $S$ at 1-loop using dimensional regularization with the MS scheme in $D=2+\epsilon$ spacetime dimensions. The resulting expressions have structural similarities with Eq. \ref{eq:RGnonrel} but are a bit cumbersome (see Appendix \ref{sec:app_rel}), and we turn directly to the results. As a confidence-building measure, let us first discuss the RG fixed points for the action $S_0 + S_{\text{int},U(1)}$. In this case, we find four real fixed points (and $2^4-4 = 12$ complex fixed points, which we ignore). The fixed point with no relevant direction is of course the free Dirac fermion CFT. There are then two fixed points with a single relevant direction. The first one of these has only $G_1$ non-zero and corresponds to the Gross-Neveu (GN) fixed point, which describes the phase transition between a Dirac semi-metal and an insulator, with parity/time-reversal breaking order-parameter $\langle \bar{\psi}_{i\alpha} \psi_{i\alpha} \rangle \neq 0$. The second fixed point with a single relevant direction, being mostly located along the $G_{\sigma}$ axis, is the Gross-Neveu-Heisenberg (GNH) fixed point. It describes a continuous phase transition from a Dirac semi-metal to an $SU(2)$ symmetry-broken phase, where $\langle \bar{\psi}_{i\alpha} \bm{\sigma}_{ij} \psi_{j\alpha} \rangle \neq 0$. Finally, the fixed point with two relevant directions is a multicritical point where the three stable phases meet.

Let us now turn on the SMG interaction. Similarly to the aforementioned non-relativistic model, the RG equations are invariant under $G \rightarrow -G$ and we can thus only focus on $G>0$. Remarkably, we obtain a unique fixed point with $G \neq 0$ with a single relevant direction. Based on the location of this fixed point (mostly along $G$) and the overall topology of the RG flow, we conjecture that this fixed point describes the quantum phase transition between the Dirac semi-metal and the symmetric-gapped phase of the honeycomb lattice. In addition, we obtain another fixed point with $G > 0$ with two relevant directions and which appears to be a multicritical point where the SMG phase, the GN phase, and the free Dirac phase meet. Fig. \ref{fig:RG2} shows two-dimensional slices of the full five-dimensional RG flow: (a) represents the $(G_1, G)$ plane, while (b) corresponds to the $(G_{\sigma},G)$ plane. For both flows, the three couplings not shown are set at their SMG fixed point value \footnote{Therefore, only the SMG fixed point (red) truly lies in the two planes, while the other dots correspond to the projection of other fixed points in these planes, resulting in a slight misalignment between the position of the markers and the apparent position of the fixed points from the flows.}.

An interesting outcome of our calculation is that the SMG interaction is \textit{relevant} at the $U(1)$-symmetric GNH fixed point, which therefore becomes a multicritical point with two relevant directions, as can be seen in Fig \ref{fig:RG2} (b). Similarly, the aforementioned $U(1)$-symmetric multicritical point  gets a third relevant direction. Based on these observations, and assuming that there are only four stable phases at low energies (gapless Dirac, SMG, $\langle \bar{\psi}_{i\alpha} \psi_{i\alpha} \rangle \neq 0$ and $\langle \bar{\psi}_{i\alpha} \bm{\sigma}_{ij} \psi_{j\alpha} \rangle \neq 0$), we sketch the schematic phase diagram in Fig. \ref{fig:honeycomb} (d). One limitation of our approach (and more generally, of any approach based on perturbing around the gapless Dirac fermion) is that it can only describe phase transitions where the fermions are gapless. Therefore, it cannot access the transition between the GNH phase ($\langle \bar{\psi}_{i\alpha} \bm{\sigma}_{ij} \psi_{j\alpha} \rangle \neq 0$) and the SMG phase which is expected to be described by the standard $O(3)$ Wilson-Fisher fixed point.

To characterize the identified critical points (GN, GNH and putative SMG), we compute the correlation length exponent $\nu$ and the fermion anomalous dimension $\eta_{\psi}$ (Appendix \ref{sec:app_rel}). Although 2+$\epsilon$ expansion results for the GN and GNH fixed points have been obtained in previous work (Refs.  \cite{gracey2018large,ladovrechis2023gross}), our RG procedure is different since we maintain the spinor structure of the $D=3$ theory. Therefore, it is interesting to compare our results with Refs. \cite{gracey2018large,ladovrechis2023gross} for the GN and GNH fixed points. We find that at all three fixed points, $\nu^{-1} = \epsilon$, which, for the GN and GNH fixed points, agrees with Refs. \cite{gracey2018large,ladovrechis2023gross} \footnote{For the GNH fixed point, this exponent $\nu$ is calculated for the theory in the $U(1)$-preserving subspace}. Similar to the non-relativistic theory, the anomalous dimension vanishes at 1-loop, and we must therefore go to 2-loop to obtain a non-zero result. For the GN fixed point, we get $\eta_{\psi}^{\text{GN}} = \frac{15}{392}\epsilon^2 \approx 0.03827 \epsilon^2$ which again agrees perfectly with Refs. \cite{gracey2018large,ladovrechis2023gross}. For the GNH fixed point, we find $\eta_{\psi}^{\text{GNH}}\approx 0.07366\epsilon^2$, which is  close to the value obtained in \cite{ladovrechis2023gross}, $0.07489 \epsilon^2$, despite our different RG scheme. We have also computed $\eta_{\psi}^{\text{GNH}}$ for a large-$N_f$ version of the $U(1)$ symmetric theory, and we find that  $\eta_{\psi}^{\text{GNH}}$ matches exactly with \cite{ladovrechis2023gross,gracey2018large} at least up to $\mathcal{O}(1/N^3_f)$  (See Appendix \ref{sec:app_rel}). This validates the identification of the Gross-Neveu-Heisenberg fixed point made before. Finally, at the SMG fixed point, the fermion anomalous dimension is found to be $\eta_{\psi}^{\text{SMG}} \approx 0.04650\epsilon^2$. This is not a very large number at $\epsilon = 1$, in contrast to the numerical results on the SMG phase transition, and perhaps indicates that an accurate estimation of $\eta_{\psi}$ requires one to go to higher orders in $\epsilon$. The exponent $\nu$ after setting $\epsilon = 1$ is close to what is seen numerically (approximately unity) \cite{ayyar2016origin,catterall2016fermion,he2016quantum}. 

Our work motivates several questions/directions. Our approach seems oblivious to the $\mathbb{Z}_{16}$ anomaly of the relativistic theory, and it would be valuable to understand the nature of the phase that emerges at large values of the SMG-type interaction in settings where the SMG phase is prohibited by anomalies. In fact, in a theory analogous to our relativistic action but with just two species two-component Dirac fermions (see Appendix \ref{sec:boundary_SPT}), one again finds a fixed point similar to the putative SMG critical point. However, this theory cannot have an SMG phase or more generally, even a gapped ground state~\cite{wang2014interacting}. Therefore, it is not clear what phase one enters as the system leaves the gapless Dirac phase. It should be noted, however, that unlike the action studied in the main text, this problem describes the boundary of a non-trivial SPT phase, and an $\epsilon$-expansion based approach may be less reliable due to non-trivial topological defects~\cite{wang2022symmetric,you2014interacting}. It will also be worth exploring numerical simulation of a lattice regularized version of the non-relativistic theory we first discussed (e.g. using DMRG in 1+1-D). Another question is the topology of the phase diagram for the relativistic model (Fig. \ref{fig:honeycomb}(d)) -- within the $\epsilon$-expansion, the SMG interaction is relevant at the GNH fixed point in the $U(1)$ symmetric subspace, but it is not obvious whether this conclusion holds at $\epsilon = 1$. This question can be settled using sign-problem free Quantum Monte Carlo (QMC) simulations similar to Refs.~\cite{ayyar2016origin,catterall2016fermion,he2016quantum}. It will also be interesting to test emergent Lorentz invariance at low energies in lattice models of SMG, such as Eq. \ref{eq:H_SMG}.

\begin{acknowledgments}
\textit{Acknowledgments} - We thank Cenke Xu, John McGreevy, Max Metlitski and Yi-Zhuang You for helpful discussions and comments. T.G. is supported by
the National Science Foundation under Grant No. DMR-2521369.
\end{acknowledgments}

\newpage 
\onecolumngrid
\appendix

\section{RG Analysis of Model 1: Fermions at Unitarity}\label{sec:app_nonrel}

\subsection{Setting up the RG calculation}

This appendix presents details of the RG analysis for the theory of fermions at unitarity, with the imaginary-time action

\begin{align}
\begin{split}
S &= \int d\tau d^d\mathbf{x} \, \bar{\psi}_{\alpha} \Big( \partial_{\tau} - \frac{1}{2m} \nabla^2 \Big) \psi_{\alpha} + \lambda \int d\tau d^d\mathbf{x} \, (\bar{\psi}_{\alpha} \psi_{\alpha})^2 + \frac{g}{4!} \epsilon_{\alpha \beta \gamma \delta} \int d\tau d^d\mathbf{x} \, \big( \psi_{\alpha} \psi_{\beta} \psi_{\gamma} \psi_{\delta} + \bar{\psi}_{\alpha} \bar{\psi}_{\beta} \bar{\psi}_{\gamma} \bar{\psi}_{\delta} \big) \\ &= S_0[\psi] + S_{\lambda}[\psi] + S_{g}[\psi] \, ,
\end{split}
\end{align}

\noindent where $\alpha=1,...,4$ labels the four flavors of non-relativistic fermions. The SMG interaction has been written in a manifestly $SU(4)$-invariant form. The difference in scaling between space and time is taken into account by introducing  different mass dimensions to position and imaginary-time: $[\mathbf{x}]=-1$ and $[\tau]=-z$. For $g=0$, the theory has full Galilean invariance, which protects $z=2$. However, the presence of the SMG term breaks this symmetry and $z$ is therefore allowed to deviate from two. Note that taking into account a varying $z$ allows one to absorb $m$ by appropriate rescalings of the coordinates, field and couplings \cite{yerzhakov2018disordered,yerzhakov2021random,dey2022quantum} and we can thus set $m=1$ in the above action. 

Since we will need to go to 2-loop, it is convenient to do the RG calculation using dimensional regularization with minimal subtraction scheme in $d$ spatial dimensions. The action written in terms of bare quantities takes the following form

\begin{align}
\begin{split}
S &= \int d\tau_B d^d\mathbf{x}_B \, \bar{\psi}_{B,\alpha} \Big( \partial_{\tau_B} - \frac{1}{2} \nabla^2_B \Big) \psi_{B,\alpha} + \lambda_B \int d\tau_B d^d\mathbf{x}_B \, (\bar{\psi}_{B,\alpha} \psi_{B,\alpha})^2 \\ &\hspace{2cm} + \frac{g_B}{4!} \epsilon_{\alpha \beta \gamma \delta} \int d\tau_B d^d\mathbf{x}_B \, \big( \psi_{B,\alpha} \psi_{B,\beta} \psi_{B,\gamma} \psi_{B,\delta} + \bar{\psi}_{B,\alpha} \bar{\psi}_{B,\beta} \bar{\psi}_{B,\gamma} \bar{\psi}_{B,\delta} \big) \, .
\end{split}
\end{align}

\noindent We are here working with a  scheme similar to \cite{thomson2017quantum,yerzhakov2018disordered,yerzhakov2021random}, where the calculation of $z$ requires the introduction of bare imaginary-time and space coordinates. From power-counting, the various quantities in the action scale as

\begin{equation}
[\mathbf{x}_B]=-1 \, , \quad [\tau_B]=-z \, , \quad [\psi_{B,\alpha}]=\frac{d}{2} \, , \quad [\lambda_B] = [g_B] = 2-d = -\epsilon \, ,
\end{equation}

\noindent where we have introduced $\epsilon = d-2$ and used the fact that $z=2$ at tree level. The renormalized action takes the following form

\begin{align}
\begin{split}
S &= \int d\tau d^d\mathbf{x} \, \bar{\psi}_{\alpha} \Big( Z_1 \partial_{\tau} - \frac{Z_2}{2} \nabla^2 \Big) \psi_{\alpha} + Z_3 \mu^{-\epsilon} \lambda \int d\tau d^d\mathbf{x} \, (\bar{\psi}_{\alpha} \psi_{\alpha})^2 \\ &\hspace{2cm} + Z_4 \mu^{-\epsilon} \frac{g}{4!} \epsilon_{\alpha \beta \gamma \delta} \int d\tau d^d\mathbf{x} \, \big( \psi_{\alpha} \psi_{\beta} \psi_{\gamma} \psi_{\delta} + \bar{\psi}_{\alpha} \bar{\psi}_{\beta} \bar{\psi}_{\gamma} \bar{\psi}_{\delta} \big) \, ,
\end{split}
\end{align}

\noindent where $\mu$ is the RG scale ($[\mu]=1$) and the two renormalized couplings are dimensionless. The renormalization constants are written as $Z_i = 1 + \delta_i$, with $i=1,...,4$. Let us choose $\mathbf{x}_B=\mathbf{x}$ and thus define $\tau_B = \theta \tau$. By comparing the bare and renormalized action, we find

\begin{align}
\begin{split}
\psi_{B,\alpha}=Z_1^{1/2} \psi_{\alpha} \, , \quad \theta = \frac{Z_2}{Z_1} \, \quad \lambda_B = \mu^{-\epsilon} Z_1^{-1} Z_2^{-1} Z_3 \lambda \, , \quad g_B = \mu^{-\epsilon} Z_1^{-1} Z_2^{-1} Z_4 g \, ,
\end{split}
\end{align}

\noindent from which we identify the wavefunction renormalization $Z_{\psi} = Z_1$. The anomalous dimensions are defined as $\gamma_i = \mu \frac{d \ln Z_i}{d\mu}$ with $i=1,...,4$. Knowing this, the dynamical critical exponent is defined as $z = 2 + \mu \frac{d \ln \tau}{d\mu}$. Using the fact that $\tau = \theta^{-1} \tau_B$ and that $\tau_B$ is $\mu$-independent, it follows that

\begin{equation}
z = 2 + \gamma_1 - \gamma_2 \, .
\end{equation}

\noindent Moreover, using the fact that bare couplings are $\mu$-independent, one gets the RG equations for $\lambda$ and $g$

\begin{align} \label{Eq:beta_functions}
\begin{split}
-\beta(\lambda) = \mu \frac{d\lambda}{d\mu} &= (\epsilon + \gamma_1 + \gamma_2 - \gamma_3) \lambda \\-\beta(g) =  \mu \frac{dg}{d\mu} &= (\epsilon + \gamma_1 + \gamma_2 - \gamma_4) g \, . 
\end{split}
\end{align}

\noindent Note the sign convention we are using here, where the $\beta$ functions are defined in terms of the IR RG scale $l$, so that $\beta(\lambda) = \frac{d\lambda}{dl}$, $\beta(g) = \frac{dg}{dl}$. Using this, one can simplify the expression for the anomalous dimensions using the chain rule

\begin{equation} \label{Eq:gamma_i}
\gamma_i = \frac{1}{Z_i} \mu \frac{d Z_i}{d\mu} = \frac{1}{Z_i} \sum_{j} \frac{dZ_i}{dg_j} \mu \frac{dg_j}{d\mu} = -\frac{1}{Z_i} \sum_j \frac{dZ_i}{dg_j} \beta(g_j) \approx \epsilon \sum_j \frac{dZ_i}{dg_j} g_j + ... \, ,
\end{equation}

\noindent where $g_j = (\lambda,g)$.

The goal is now to compute the four renormalization constants $Z_1$, $Z_2$, $Z_3$ and $Z_4$. This requires the calculation of the effective action $\Gamma[\psi_{c}]$ by integrating out fluctuations $\psi_f$ about the classical configuration $\psi_c$ respecting the equations of motion. The counterterms are then fixed by requiring $\Gamma[\psi_c]$ to be finite at each order in $\lambda$ and $g$. The first step is to expand the fermion field as $\psi_{\alpha} = \psi_{c,\alpha} + \psi_{f,\alpha}$ and to replace it in the renormalized action. For simplicity, we denote $\psi_{c,\alpha} \equiv c_{\alpha}$, $\psi_{f,\alpha}\equiv f_{\alpha}$. The free part of the action becomes

\begin{align}
\begin{split}
S_0[c+f] = S_0[c] + S_0^{(2)}[f] + ... = \int_x \bar{c}_{\alpha} \Big( Z_1 \partial_{\tau} - \frac{Z_2}{2} \nabla^2 \Big) c_{\alpha} + \int_x \bar{f}_{\alpha} \Big( \partial_{\tau} - \frac{1}{2} \nabla^2 \Big) f_{\alpha} + ... \, ,
\end{split}
\end{align}

\noindent where $x=(\tau,\mathbf{x})$ and $\int_x = \int d\tau d^d\mathbf{x}$. Terms linear in $f_{\alpha}$ vanish from the equation of motion, while terms combining fluctuations and counterterms do not contribute at the order we are working at. Denoting $p\cdot x = -p_0 x_0+\mathbf{p}\cdot \mathbf{x}$ and Fourier transforming $S_0^{(2)}[f]$ with the convention $f_{\alpha}(x) = \int_p \e^{\ii p \cdot x} f_{\alpha}(p)$, with $p=(p_0,\mathbf{p})$ and $\int_p = \int \frac{dp_0}{2\pi} \frac{d^d\mathbf{p}}{(2\pi)^d}$, allows us to identify the free propagator for the fluctuations

\begin{equation}
S_0^{(2)}[f] = \int \frac{dp_0}{2\pi} \int \frac{d^d\mathbf{p}}{(2\pi)^d} \bar{f}_{\alpha}(p) \Big( -\ii p_0 + \frac{\mathbf{p}^2}{2} \Big) f_{\alpha}(p) \implies \Tilde{G}(\ii p_0,\mathbf{p}) = \frac{1}{-\ii p_0 + \frac{\mathbf{p}^2}{2}} \, .
\end{equation}

Expanding the fermion field in the two interaction terms yields

\begin{align} \label{Eq:S_lambda_expanded}
\begin{split}
S_{\lambda}[c+f] &= S_{\lambda}[c] + S_{\lambda}^{(2)}[c,f] + S_{\lambda}^{(3)}[c,f] + S_{\lambda}^{(4)}[f] + ... \\ &= Z_3 \mu^{-\epsilon} \lambda \int_x (\bar{c}_{\alpha} c_{\alpha})^2 + \mu^{-\epsilon} \lambda \int_x \Big( 2 \bar{c}_{\alpha} c_{\alpha} \bar{f}_{\beta} f_{\beta} + 2 \bar{c}_{\alpha} f_{\alpha} \bar{f}_{\beta} c_{\beta} + \bar{c}_{\alpha} f_{\alpha} \bar{c}_{\beta} f_{\beta} + \bar{f}_{\alpha} c_{\alpha} \bar{f}_{\beta} c_{\beta} \Big) \\ & \hspace{1cm} + 2 \mu^{-\epsilon} \lambda
 \int_x \Big( \bar{c}_{\alpha} f_{\alpha} \bar{f}_{\beta} f_{\beta} + \bar{f}_{\alpha} f_{\alpha} \bar{f}_{\beta} c_{\beta} \Big) + \mu^{-\epsilon} \lambda \int_x (\bar{f}_{\alpha} f_{\alpha})^2 + ... \, ,
\end{split}
\end{align}

\begin{align}\label{Eq:S_g_expanded}
\begin{split}
S_g[c+f] &= S_g[c] + S_g^{(2)}[c,f] + S_g^{(3)}[c,f] + S_g^{(4)}[f] + ... \\ &= Z_4 \mu^{-\epsilon} \frac{g}{4!} \epsilon_{\alpha \beta \gamma \delta} \int_x \Big( c_{\alpha} c_{\beta} c_{\gamma} c_{\delta} + \bar{c}_{\alpha} \bar{c}_{\beta} \bar{c}_{\gamma} \bar{c}_{\delta} \Big) + \mu^{-\epsilon} \frac{g}{4} \epsilon_{\alpha \beta \gamma \delta} \int_x \Big( c_{\alpha} c_{\beta} f_{\gamma} f_{\delta} + \bar{c}_{\alpha} \bar{c}_{\beta} \bar{f}_{\gamma} \bar{f}_{\delta} \Big) \\ &\hspace{1cm} + \mu^{-\epsilon} \frac{g}{6} \epsilon_{\alpha \beta \gamma \delta} \int_x \Big( c_{\alpha} f_{\beta} f_{\gamma} f_{\delta} + \bar{c}_{\alpha} \bar{f}_{\beta} \bar{f}_{\gamma} \bar{f}_{\delta} \Big) + \mu^{-\epsilon} \frac{g}{4!} \epsilon_{\alpha \beta \gamma \delta} \int_x \Big( f_{\alpha} f_{\beta} f_{\gamma} f_{\delta} + \bar{f}_{\alpha} \bar{f}_{\beta} \bar{f}_{\gamma} \bar{f}_{\delta} \Big) + ... \, .
\end{split}
\end{align}

\noindent Let us regroup terms with the same number of powers of $f_{\alpha}$ by defining $S[c] = S_0[c]+S_{\lambda}[c]+S_{g}[c]$, $S_{\text{int}}^{(2)}[c,f] = S_{\lambda}^{(2)}[c,f] + S_{g}^{(2)}[c,f]$, $S_{\text{int}}^{(3)}[c,f] = S_{\lambda}^{(3)}[c,f] + S_{g}^{(3)}[c,f]$ and $S_{\text{int}}^{(4)}[f] = S_{\lambda}^{(4)}[f] + S_{g}^{(4)}[f]$. The fully expanded action is thus

\begin{equation}
S[c+f] = S[c] + S_0^{(2)}[f] + S_{\text{int}}^{(2)}[c,f] + S_{\text{int}}^{(3)}[c,f] + S_{\text{int}}^{(4)}[f] \, .
\end{equation}

The effective action is obtained by integrating out the fluctuations

\begin{equation}
\e^{-\Gamma[c]} = \int \mathcal{D}[\bar{f},f] \e^{-S_0^{(2)}[f]} \e^{-S_{\text{int}}^{(2)}[c,f] - S_{\text{int}}^{(3)}[c,f] - S_{\text{int}}^{(4)}[f]} \, .
\end{equation}

\noindent At second order in the cumulant expansion, the effective action is

\begin{align}
\begin{split}
\Gamma[c] = S[c] + \ev{S_{\text{int}}^{(2)}[c,f] + S_{\text{int}}^{(3)}[c,f] + S_{\text{int}}^{(4)}[f]}_f - \frac{1}{2} \ev{\Big(S_{\text{int}}^{(2)}[c,f] + S_{\text{int}}^{(3)}[c,f] + S_{\text{int}}^{(4)}[f]\Big)^2}_f^c + ... \, ,
\end{split}
\end{align}

\noindent where the expectation values are over the fluctuations, and only connected correlators are considered. The expectation value at linear order in the cumulant expansion is easily treated. Clearly, $\ev{S_{\text{int}}^{(3)}[c,f]}_f = 0$ while $\ev{S_{\text{int}}^{(4)}[f]}_f$ is just a constant. The remaining term is

\begin{equation}
\ev{S_{\text{int}}^{(2)}[c,f]}_f^c = \ev{S_{\lambda}^{(2)}[c,f]}_f^c = -2 (N_f-1) \mu^{-\epsilon} \lambda \int_p \Tilde{G}(p) \int_x \bar{c}_{\alpha} c_{\alpha} \, ,
\end{equation}

\noindent where

\begin{equation}
\int_p \Tilde{G}(p) = \int_{-\infty}^{\infty} \frac{dp_0}{2\pi} \int \frac{d^d\mathbf{p}}{(2\pi)^d} \frac{1}{-\ii p_0 + \frac{\mathbf{p}^2}{2}} \, .
\end{equation}

\noindent This clearly does not yield a $1/\epsilon$ pole and can thus be discarded. Hence, the linear term in the cumulant expansion is trivial.

\subsection{Quadratic order in cumulant expansion}

We now move to the second order term in the cumulant expansion

\begin{align} \label{Eq:Order_2_cumulant}
\begin{split}
\ev{\Big(S_{\text{int}}^{(2)}[c,f] + S_{\text{int}}^{(3)}[c,f] + S_{\text{int}}^{(4)}[f]\Big)^2}_f^c = \ev{\Big(S_{\text{int}}^{(2)}[c,f]\Big)^2}_f^c + \ev{\Big(S_{\text{int}}^{(3)}[c,f]\Big)^2}_f^c + 2 \ev{S_{\text{int}}^{(2)}[c,f] S_{\text{int}}^{(4)}[f]}_f^c + ... \, ,
\end{split}
\end{align}

\noindent where ellipsis denote vanishing or constant contributions. The first term has four classical (external) fields and thus renormalizes the interactions, while the last two terms renormalize the propagator.

\subsubsection{Renormalization of the interactions}

Expanding the first term of Eq. \ref{Eq:Order_2_cumulant} yields

\begin{equation}
\ev{\Big(S_{\text{int}}^{(2)}[c,f]\Big)^2}_f^c = \ev{\Big( S_{\lambda}^{(2)}[c,f] \Big)^2}_f^c + \ev{\Big( S_{g}^{(2)}[c,f] \Big)^2}_f^c + 2 \ev{S_{\lambda}^{(2)}[c,f] S_{g}^{(2)}[c,f]}_f^c \equiv D_1^{(2)} + D_2^{(2)} + 2 D_3^{(2)} \, .
\end{equation}

For $D_1^{(2)}$, using the expression for $S_{\lambda}^{(2)}[c,f]$ from Eq. \ref{Eq:S_lambda_expanded}, one gets

\begin{align} \label{Eq:D_1^2}
\begin{split}
D_1^{(2)} &= \mu^{-2\epsilon} \lambda^2 \int_{x,y} \Bigg[ 4 \ev{(\bar{c}_{\alpha} c_{\alpha} \bar{f}_{\beta} f_{\beta})(x) (\bar{c}_{\gamma} c_{\gamma} \bar{f}_{\delta} f_{\delta})(y)}_f^c + 8 \ev{(\bar{c}_{\alpha} c_{\alpha} \bar{f}_{\beta} f_{\beta})(x) (\bar{c}_{\gamma} f_{\gamma} \bar{f}_{\delta} c_{\delta})(y)}_f^c \\ &\hspace{2cm} + 4 \ev{(\bar{c}_{\alpha} f_{\alpha} \bar{f}_{\beta} c_{\beta})(x) (\bar{c}_{\gamma} f_{\gamma} \bar{f}_{\delta} c_{\delta})(y)}_f^c + 2 \ev{(\bar{c}_{\alpha} f_{\alpha} \bar{c}_{\beta} f_{\beta})(x) (\bar{f}_{\gamma} c_{\gamma} \bar{f}_{\delta} c_{\delta})(y)}_f^c \Bigg] \, .
\end{split}
\end{align}

\begin{figure}[H]
\centering
\includegraphics[width=0.7\hsize]{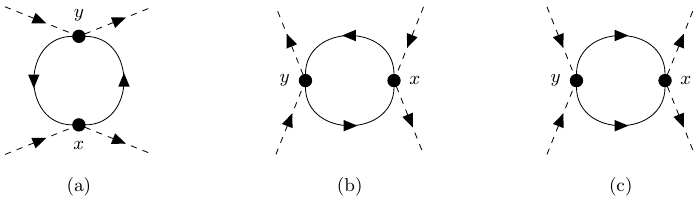}
\caption{Three Feynman diagrams contributing to the renormalization of the interactions at 1-loop. Dashed lines represent external classical fields while full lines represent fluctuations. A black dot indicates an insertion of the $\lambda$-vertex. Diagrams (a) and (b) correspond to the first three expectation values of $D_1^{(2)}$, which all contain a factor of $G(x-y) G(y-x)$. Diagram (c) corresponds to the fourth expectation value of $D_1^{(2)}$ and contains $G^2(x-y)$.}
\label{Fig:Feynman_diagram_1}
\end{figure}

\noindent The first three expectation values, corresponding to diagrams (a) and (b) of Fig. \ref{Fig:Feynman_diagram_1}, contain a factor of $G(x-y) G(y-x)$, which after Fourier transforming leads to

\begin{equation} \label{Eq:Int_G2p}
\int_{-\infty}^{\infty} \frac{dp_0}{2\pi} \int \frac{d^d\mathbf{p}}{(2\pi)^d} G^2(\ii p_0,\mathbf{p}) = \int_{-\infty}^{\infty} \frac{dp_0}{2\pi} \int \frac{d^d\mathbf{p}}{(2\pi)^d} \frac{1}{(-\ii p_0 + \frac{\mathbf{p}^2}{2})^2} = 0 \, 
\end{equation}

\noindent where the integral over $p_0$ vanishes since there is no pole in the upper-half of the complex plane. This is an important feature which will be used again later. Hence, $D_1^{(2)}$ only gets a nonzero contribution from the last expectation value (Fig. \ref{Fig:Feynman_diagram_1} (c))

\begin{align}\label{Eq:D1}
\begin{split}
D_1^{(2)} = -4 \mu^{-2\epsilon} \lambda^2 \int_{x,y} (\bar{c}_{\alpha} \bar{c}_{\beta})(x) (c_{\alpha} c_{\beta})(y) G^2(x-y) \approx 4 \mu^{-2\epsilon} \lambda^2 \int_x (\bar{c}_{\alpha} c_{\alpha})^2 \int_p \Tilde{G}(p) \Tilde{G}(-p) \, ,
\end{split}
\end{align}

\noindent where the second equality is obtained after expanding at leading order in the separation $l=x-y$ and Fourier transforming. The evaluation of the integral over $p$ yields

\begin{align} \label{Eq:Int_GpGmp}
\begin{split}
\mu^{-\epsilon} \int_p \Tilde{G}(p) \Tilde{G}(-p) &= \mu^{-\epsilon} \int_{-\infty}^{\infty} \frac{dp_0}{2\pi} \int \frac{d^d\mathbf{p}}{(2\pi)^d} \frac{1}{(-\ii p_0+\frac{\mathbf{p}^2}{2}) (\ii p_0+\frac{\mathbf{p}^2}{2})} \\ &= \mu^{-\epsilon} \int \frac{d^d\mathbf{p}}{(2\pi)^d} \frac{1}{\mathbf{p}^2+\Delta} \\ &= \frac{\mu^{-\epsilon}}{(4\pi)^{d/2}} \Gamma(1-d/2) \Delta^{\frac{d}{2}-1} \\ &= -\frac{1}{2\pi \epsilon} + \text{finite} \, ,
\end{split}
\end{align}

\noindent where $\Delta$ is an arbitrary scale introduced to extract the UV divergence (one does not need to introduce it if the exact Fourier transform is kept in Eq. \ref{Eq:D1}, where the external momentum plays the role of the IR regulator). Therefore, one finds that

\begin{equation}
D_1^{(2)} = 4 \mu^{-\epsilon} \lambda^2 \Bigg( -\frac{1}{2\pi \epsilon} + \text{finite} \Bigg) \int_x (\bar{c}_{\alpha} c_{\alpha})^2= -\frac{2}{\pi} \mu^{-\epsilon} \lambda^2 \frac{1}{\epsilon} \int_x (\bar{c}_{\alpha} c_{\alpha})^2 \, ,
\end{equation}

\noindent where finite terms will be omitted for now on. 

\begin{figure}[H]
\centering
\includegraphics[width=0.7\hsize]{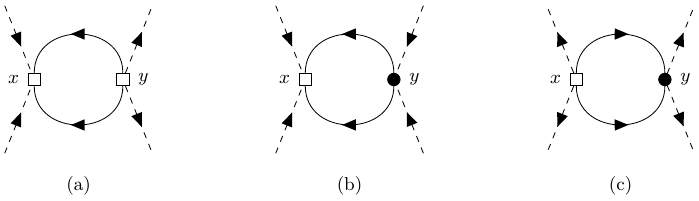}
\caption{Three Feynman diagrams contributing to the renormalization of the interactions at 1-loop. Dashed lines represent external classical fields while full lines represent fluctuations. A black dot indicates an insertions of a $\lambda$-vertex, while a square denotes a $g$-vertex. Diagram (a) corresponds to $D_2^{(2)}$, while diagrams (b) and (c) represent the two expectation values of $D_3^{(2)}$.}
\label{Fig:Feynman_diagram_2}
\end{figure}

Moving on to $D_2^{(2)}$ (Fig. \ref{Fig:Feynman_diagram_2} (a)), we get

\begin{align}
\begin{split}
D_2^{(2)} &= \mu^{-2\epsilon} \frac{g^2}{8} \epsilon_{\alpha \beta \gamma \delta} \epsilon_{\alpha' \beta' \gamma' \delta'} \int_{x,y} \ev{(c_{\alpha} c_{\beta} f_{\gamma} f_{\delta})(x) (\bar{c}_{\alpha'} \bar{c}_{\beta'} \bar{f}_{\gamma'} \bar{f}_{\delta'})(y)}_f^c \\ &\approx - \mu^{-2\epsilon} \frac{g^2}{4} \epsilon_{\alpha \beta \gamma \delta} \epsilon_{\alpha' \beta' \gamma \delta} \int_x c_{\alpha} c_{\beta} \bar{c}_{\alpha'} \bar{c}_{\beta'} \int_p \Tilde{G}(p) \Tilde{G}(-p) \\ &= -\frac{1}{2\pi} \mu^{-\epsilon} g^2 \frac{1}{\epsilon} \int_x (\bar{c}_{\alpha} c_{\alpha})^2 \, ,
\end{split}
\end{align}

\noindent where Eq. \ref{Eq:Int_GpGmp} has been used along with the identity $\epsilon_{\alpha \beta \gamma \delta} \epsilon_{\alpha' \beta' \gamma \delta} = 2 (\delta_{\alpha \alpha'} \delta_{\beta \beta'} - \delta_{\alpha \beta'} \delta_{\beta \alpha'})$ and the fermions have been brought back to the original order. Finally, we have for $D_3^{(2)}$ (Fig. \ref{Fig:Feynman_diagram_2} (b) and (c))

\begin{align}
\begin{split}
D_3^{(2)} &= \mu^{-2\epsilon} \frac{\lambda g}{4} \epsilon_{\alpha \beta \gamma \delta} \int_{x,y} \Bigg[ \ev{(c_{\alpha} c_{\beta} f_{\gamma} f_{\delta})(x) (\bar{f}_{\alpha'} c_{\alpha'} \bar{f}_{\beta'} c_{\beta'})(y)}_f^c + \ev{(\bar{c}_{\alpha} \bar{c}_{\beta} \bar{f}_{\gamma} \bar{f}_{\delta})(x) (\bar{c}_{\alpha'} f_{\alpha'} \bar{c}_{\beta'} f_{\beta'})(y)}_f^c \Bigg] \\ &\approx \mu^{-2\epsilon} \frac{\lambda g}{2} \int_p \Tilde{G}(p) \Tilde{G}(-p) \epsilon_{\alpha \beta \gamma \delta} \int_x \Big( c_{\alpha} c_{\beta} c_{\gamma} c_{\delta} + \bar{c}_{\alpha} \bar{c}_{\beta} \bar{c}_{\gamma} \bar{c}_{\delta} \Big) \\ &= -\frac{1}{4\pi} \mu^{-\epsilon} \lambda g \frac{1}{\epsilon} \epsilon_{\alpha \beta \gamma \delta} \int_x \Big( c_{\alpha} c_{\beta} c_{\gamma} c_{\delta} + \bar{c}_{\alpha} \bar{c}_{\beta} \bar{c}_{\gamma} \bar{c}_{\delta} \Big) \, .
\end{split}
\end{align}

\noindent Note that all the expectation values in $D_2^{(2)}$ and $D_3^{(2)}$ are nonzero because none of the diagrams contain $G(x-y) G(y-x)$. Combining the results for $D_1^{(2)}$, $D_2^{(2)}$ and $D_3^{(2)}$, the 1-loop contribution to the effective action renormalizing the interactions is

\begin{equation}
\begin{split}
\ev{\Big(S_{\text{int}}^{(2)}[c,f]\Big)^2}_f^c = -\frac{\mu^{-\epsilon}}{2\pi} (4\lambda^2+g^2) \frac{1}{\epsilon} \int_x (\bar{c}_{\alpha} c_{\alpha})^2 - \frac{\mu^{-\epsilon}}{2\pi} \lambda g \frac{1}{\epsilon} \epsilon_{\alpha \beta \gamma \delta} \int_x \Big( c_{\alpha} c_{\beta} c_{\gamma} c_{\delta} + \bar{c}_{\alpha} \bar{c}_{\beta} \bar{c}_{\gamma} \bar{c}_{\delta} \Big) \, .
 \end{split}
\end{equation}

\subsubsection{Renormalization of the propagator}

\begin{figure}[H]
\centering
\includegraphics[width=0.7\hsize]{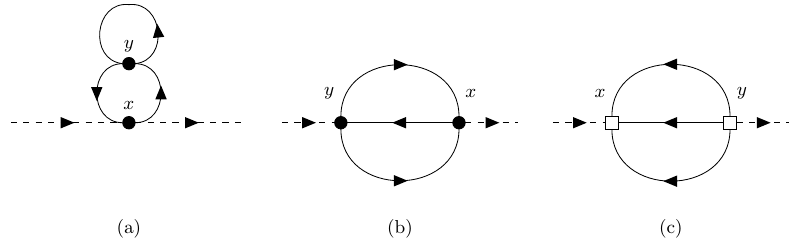}
\caption{Three Feynman diagrams contributing to the renormalization fo the propagator. Dashed lines represent external classical fields while full lines represent fluctuations. A black dot indicates an insertions of a $\lambda$-vertex, while a square denotes a $g$-vertex. Diagrams (a) and (b) vanish due to the presence of $G(x-y)G(y-x)$. Diagram (c) is nonzero and contributes to $D_2^{(3)}$.}
\label{Fig:Feynman_diagram_3}
\end{figure}

Let us now turn our attention to the two terms of Eq. \ref{Eq:Order_2_cumulant} contributing to the renormalization of the propagator (the second and the third expectation values). The second one of these, $\ev{S_{\text{int}}^{(2)}[c,f] S_{\text{int}}^{(4)}[f]}_f^c $, has a vertex with four fluctuation fields $f_{\alpha}$ and thus corresponds to diagram (a) of Fig. \ref{Fig:Feynman_diagram_3}.

\noindent This diagram contains $G(x-y)G(y-x)$ and therefore vanishes according to Eq. \ref{Eq:Int_G2p}. We then turn our attention to the other contribution

\begin{align}
\begin{split}
\ev{\Big(S_{\text{int}}^{(3)}[c,f]\Big)^2}_f^c = \ev{\Big( S_{\lambda}^{(3)}[c,f] \Big)^2}_f^c + \ev{\Big( S_{g}^{(3)}[c,f] \Big)^2}_f^c \equiv D_1^{(3)} + D_2^{(3)} \, ,
\end{split}
\end{align}

\noindent where there is no diagram associated with the cross-term with $S_{\lambda}^{(3)}[c,f]$ and $S_{g}^{(3)}[c,f]$. $D_1^{(3)}$ and $D_2^{(3)}$ respectively correspond to Fig. \ref{Fig:Feynman_diagram_3} (b) and (c). One can see that $D_1^{(3)}$ contains $G(x-y) G(y-x)$ and is thus zero. This is consistent with the fact that the self-energy for the $U(1)$-symmetric theory vanishes at every order \cite{nikolic2007renormalization}. We are thus only left with $D_2^{(3)}$ (Fig. \ref{Fig:Feynman_diagram_3} (c))

\begin{align}
\begin{split}
D_2^{(3)} &= 2 \mu^{-2\epsilon} \frac{g^2}{6^2} \epsilon_{\alpha \beta \gamma \delta} \epsilon_{\alpha' \beta' \gamma' \delta'} \int_{x,y} \ev{(c_{\alpha} f_{\beta} f_{\gamma} f_{\delta})(x) (\bar{c}_{\alpha'} \bar{f}_{\beta'} \bar{f}_{\gamma'} \bar{f}_{\delta'})(y)}_f^c \\ &= - \mu^{-2\epsilon} \frac{g^2}{3} \epsilon_{\alpha \beta \gamma \delta} \epsilon_{\alpha' \beta \gamma \delta} \int_{x,y} \bar{c}_{\alpha'}(y) c_{\alpha}(x) G^3(x-y) \\ &= - 2 g^2 \mu^{-2\epsilon} \int_{x,y} \bar{c}_{\alpha}(y) c_{\alpha}(x) G^3(x-y) \\ &= -2 g^2 \mu^{-2\epsilon} \int_q \bar{c}_{\alpha}(q) c_{\alpha}(q) \int_k \Tilde{G}(k) \int_p \Tilde{G}(-p-k-q) \Tilde{G}(p) \, ,
\end{split}
\end{align}

\noindent where the identity $\epsilon_{\alpha \beta \gamma \delta} \epsilon_{\alpha' \beta \gamma \delta} = 6 \delta_{\alpha \alpha'}$ has been used. Let us focus on the two loop integrals. Integrating over $p_0$ and $k_0$ yields

\begin{align}
\begin{split}
&\int_k \Tilde{G}(k) \int_p \Tilde{G}(-p-k-q) \Tilde{G}(p) \\ &= \int_{-\infty}^{\infty} \frac{dk_0}{2\pi} \int \frac{d^d\mathbf{k}}{(2\pi)^d} \frac{1}{-\ii k_0 + \frac{\mathbf{k}^2}{2}}\int_{-\infty}^{\infty} \frac{dp_0}{2\pi} \int \frac{d^d\mathbf{p}}{(2\pi)^d} \frac{1}{\big[\I(p_0+k_0+q_0) + \frac{(\mathbf{p}+\mathbf{k}+\mathbf{q})^2}{2}\big] \big[ -\ii p_0 + \frac{\mathbf{p}^2}{2} \big]} \\ &= \int \frac{d^d\mathbf{k}}{(2\pi)^d} \int \frac{d^d\mathbf{p}}{(2\pi)^d} \int_{-\infty}^{\infty} \frac{dk_0}{2\pi} \frac{1}{-\ii k_0 + \frac{\mathbf{k}^2}{2}} \frac{1}{\I(k_0+q_0) + \frac{1}{2}(\mathbf{p}^2 + (\mathbf{p}+\mathbf{k}+\mathbf{q})^2)} \\ &= \int \frac{d^d\mathbf{k}}{(2\pi)^d} \int \frac{d^d\mathbf{p}}{(2\pi)^d} \frac{1}{\ii q_0 + \frac{1}{2} \big( \mathbf{k}^2 + \mathbf{p}^2 + (\mathbf{p}+\mathbf{k}+\mathbf{q})^2 \big)} \, .
\end{split}
\end{align}

\noindent To compute the remaining integrals over the spatial momenta, we perform two changes of variables. First, shift $\mathbf{p} \rightarrow \mathbf{p} - \mathbf{q}/3$, $\mathbf{k} \rightarrow \mathbf{k} - \mathbf{q}/3$ and secondly, send $\mathbf{p} \rightarrow \mathbf{p}-\mathbf{k}/2$. The remaining integrals now take a simple form and can both be evaluated in $d=2+\epsilon$

\begin{align}
\begin{split}
\int \frac{d^d\mathbf{k}}{(2\pi)^d} \int \frac{d^d\mathbf{p}}{(2\pi)^d} \frac{1}{\ii q_0 + \frac{1}{2} \big( \mathbf{k}^2 + \mathbf{p}^2 + (\mathbf{p}+\mathbf{k}+\mathbf{q})^2 \big)} &= \int \frac{d^d\mathbf{k}}{(2\pi)^d} \int \frac{d^d\mathbf{p}}{(2\pi)^d} \frac{1}{\ii q_0 + \frac{1}{6} \mathbf{q}^2 + \frac{3}{4} \mathbf{k}^2 + \mathbf{p}^2} \\ &= \frac{\Gamma(1-\frac{d}{2})}{(4\pi)^{d/2}} \int \frac{d^d\mathbf{k}}{(2\pi)^d} \Big( \ii q_0 + \frac{1}{6} \mathbf{q}^2 + \frac{3}{4} \mathbf{k}^2 \Big)^{\frac{d}{2}-1} \\ &= \frac{\Gamma(1-\frac{d}{2})}{(4\pi)^{d/2}} \Bigg(\frac{3}{4} \Bigg)^{\frac{d}{2}-1} \int \frac{d^d\mathbf{k}}{(2\pi)^d} \frac{1}{\big[ \mathbf{k}^2 + \frac{4}{3} (\ii q_0 + \frac{1}{6} \mathbf{q}^2) \big]^{1-\frac{d}{2}}} \\ &= \frac{\Gamma(1-\frac{d}{2})}{(4\pi)^{d/2}} \Bigg(\frac{3}{4} \Bigg)^{\frac{d}{2}-1} \frac{1}{(4\pi)^{d/2}} \frac{\Gamma(1-d)}{\Gamma(1-\frac{d}{2})} \Big[ \frac{4}{3} \big( \ii q_0 + \frac{1}{6} \mathbf{q}^2 \big) \Big]^{d-1} \, .
\end{split}
\end{align}

\noindent Expanding in $d=2+\epsilon$ therefore yields

\begin{equation}
D_2^{(3)} = \ev{\Big(S_{\text{int}}^{(3)}[c,f]\Big)^2}_f^c = -\frac{g^2}{6\pi^2} \frac{1}{\epsilon} \int_q \bar{c}_{\alpha}(q) \Big( \ii q_0 + \frac{1}{6} \mathbf{q}^2 \Big) c_{\alpha}(q) = \frac{g^2}{6\pi^2} \frac{1}{\epsilon} \int_c \bar{c}_{\alpha} \Big( \partial_{\tau} + \frac{1}{6} \nabla^2 \Big) c_{\alpha} \, .
\end{equation}

\noindent Note the factor of $+1/6$ instead of $-1/2$ in front of the spatial gradient, which highlights the fact that the dynamical exponent will deviate from the tree-level value $z=2$.

\subsection{Calculation of RG functions}

The effective action at quadratic order is thus

\begin{align}
\begin{split}
\Gamma[c] &= \int_x \bar{c}_{\alpha} \Big( Z_1 \partial_{\tau} - \frac{Z_2}{2} \nabla^2 \Big) c_{\alpha} + Z_3 \mu^{-\epsilon} \lambda \int_x (\bar{c}_{\alpha} c_{\alpha})^2 + Z_4 \mu^{-\epsilon} \frac{g}{4!} \epsilon_{\alpha \beta \gamma \delta} \int_x \Big( c_{\alpha} c_{\beta} c_{\gamma} c_{\delta} + \bar{c}_{\alpha} \bar{c}_{\beta} \bar{c}_{\gamma} \bar{c}_{\delta} \Big) \\ &\hspace{0.5cm} + \frac{1}{2} \frac{\mu^{-\epsilon}}{2\pi} (4\lambda^2 + g^2) \frac{1}{\epsilon} \int_x (\bar{c}_{\alpha} c_{\alpha})^2 + \frac{1}{2} \frac{\mu^{-\epsilon}}{2\pi} \lambda g \frac{1}{\epsilon} \epsilon_{\alpha \beta \gamma \delta} \int_x \Big( c_{\alpha} c_{\beta} c_{\gamma} c_{\delta} + \bar{c}_{\alpha} \bar{c}_{\beta} \bar{c}_{\gamma} \bar{c}_{\delta} \Big) - \frac{1}{2} \frac{g^2}{6\pi^2} \frac{1}{\epsilon} \int_x \bar{c}_{\alpha} \Big( \partial_{\tau} + \frac{1}{6} \nabla^2 \Big) c_{\alpha} \, ,
\end{split}
\end{align}

\noindent from which we obtain the renormalization constants by canceling the UV poles with the counterterms

\begin{align}
\begin{split}
Z_1 &= 1+\delta_1 = 1 + \frac{g^2}{12\pi^2} \frac{1}{\epsilon} \\ Z_2 &= 1+\delta_2 = 1 - \frac{g^2}{36\pi^2} \frac{1}{\epsilon} \\ Z_3 &= 1+\delta_3 = 1 - \frac{1}{4\pi}\Big( 4\lambda + \frac{g^2}{\lambda} \Big) \frac{1}{\epsilon} \\ Z_4 &= 1 + \delta_4 = 1 - \frac{6}{\pi} \lambda \frac{1}{\epsilon} \, .
\end{split}
\end{align}

\noindent Using Eq. \ref{Eq:gamma_i}, the anomalous dimensions are thus

\begin{equation}
\gamma_1 = \frac{g^2}{6\pi^2} \, , \quad \gamma_2 = -\frac{g^2}{18\pi^2} \, , \quad \gamma_3 = -\frac{\lambda}{\pi} - \frac{g^2}{4\pi \lambda} \, , \quad \gamma_4 = -\frac{6\lambda}{\pi} \, ,
\end{equation}

\noindent and from Eq. \ref{Eq:beta_functions}, the $\beta$ functions at quadratic order (neglecting $\gamma_1$ and $\gamma_2$ yielding cubic contributions) are therefore

\begin{equation}
\beta(\lambda) = -\epsilon \lambda - \frac{1}{4\pi} (4\lambda^2+g^2) \, , \qquad \beta(g) = -\epsilon g - \frac{6}{\pi} \lambda g \, .
\end{equation}

\noindent Note also that from a tree-level power-counting, the RG equation for the chemical potential (which is fine tuned to 0 in our model) is simply $\beta(\mu) = 2\mu$, which does not receive any loop corrections, at least to the order we are working at. The fermion anomalous dimension and the dynamical critical exponent are also

\begin{equation}
\gamma_{\psi} = \gamma_1 = \frac{g^2}{6\pi^2} \, , \qquad z = 2 + \gamma_1-\gamma_2 = 2 + \frac{2g^2}{9\pi^2} \, .
\end{equation}

\subsection{Fixed point analysis}

Solving the two RG equations for fixed points gives four solutions. Two of them correspond to the $U(1)$-symmetric fixed points, that is the Gaussian fixed point $(\lambda,g)=(0,0)$ and the non-trivial fixed point $(\lambda,g) = (-\pi \epsilon,0)$, which is either the Feshbach resonance ($d>2$) or the Tonks gas ($d<2$) fixed point, depending on the sign of $\epsilon$. Note that for both of these, $\eta_{\psi}=0$, which is an exact result \cite{nishida2007nonrelativistic,nikolic2007renormalization}, while $z=2$, since the theory with $g=0$ is Galilean-invariant. 

The two other solutions have a nonzero $g$ and are located in $(\lambda,g) = (-\frac{\pi}{6}\epsilon,\pm \frac{\sqrt{5}}{3} \pi \epsilon)$. These two fixed points map to each other under $g \rightarrow -g$, due to the fact that the RG equations are invariant under this transformation. This means that we can only focus on one of them (we take $g>0$). By diagonalizing the $2\times 2$ stability matrix, one finds that this new fixed point has a single relevant direction, with eigenvalue $e_+ = \epsilon$, leading to a correlation length exponent $\nu^{-1} = \epsilon$. Moreover, this fixed point has a non-trivial fermion anomalous dimension $\eta_{\psi} = \frac{5}{54}\epsilon^2$ and a dynamical critical exponent of $z=2 + \frac{10}{81}\epsilon^2$, which deviates from $z=2$ since the theory is no longer Galilean invariant.

\subsection{Existence of the SMG phase}

One way to build intuition for the phase diagram in the vicinity of the aforementioned fixed point with non-zero $g$ is to consider a ``single-site'' Hamiltonian that is proportional to the relevant scaling operator at this fixed-point. This Hamiltonian takes the form

\begin{equation}
H = U \mathcal{O}_+ = U \Big[ c_{\lambda} (\psi_{\alpha}^{\dag} \psi_{\alpha})^2 + c_g (\psi_1 \psi_2 \psi_3 \psi_4 + \psi_4^{\dag} \psi_3^{\dag} \psi_2^{\dag} \psi_1^{\dag}) \Big] \, ,
\end{equation}

\noindent where $\mathcal{O}_+ = c_{\lambda} \mathcal{O}_{\lambda} + c_g \mathcal{O}_g$ is the relevant scaling operator at the SMG fixed point and $c_{\lambda}$ and $c_g$ are the two components of the left eigenvector of the stability matrix. To obtain the ground state of $H$, we work in the number basis, defined as

\begin{equation}
|n_1 n_2 n_3 n_4 \rangle = (\psi_1^{\dag})^{n_1} (\psi_2^{\dag})^{n_2} (\psi_3^{\dag})^{n_3} (\psi_4^{\dag})^{n_4} |0000\rangle \, .
\end{equation}

\noindent Evaluating the matrix elements of $H$ in this basis and diagonalizing the resulting $16 \times 16$ matrix results in the following spectrum

\begin{align}
\begin{split}
e_1 &= 0 \, , \quad d_1 = 4 \\ e_2 &= 2 U c_{\lambda} \, , \quad d_2 = 6 \\ e_3 &= 6 U c_{\lambda} , \quad d_3 = 4 \\ e_4 &= U \Big[ 6c_{\lambda} - \sqrt{c_g^2 + 36c_{\lambda}^2} \Big] \, , \quad d_4 = 1 \\ e_5 &= U \Big[ 6c_{\lambda} + \sqrt{c_g^2 + 36c_{\lambda}^2} \Big] \, , \quad d_5 = 1 \, ,
\end{split}
\end{align}

\noindent where $e_i$ it the $i$-th eigenvalue and $d_i$ is the associated degeneracy.  It is easy to verify that for any real value of $c_{\lambda}$ and $U$ and a nonzero $c_g$, the lowest energy is always either $e_4$ or $e_5$, confirming that the ground state is indeed non-degenerate and gapped. The independence of the values of the parameters highlights the fact that the SMG interaction is dominant and will always govern the low-energy physics. The eigenvectors associated with $e_4$ and $e_5$ are respectively

\begin{equation}
v_4 = \frac{1}{\sqrt{1+\Big( \frac{6c_{\lambda}+\sqrt{c_g^2+36c_{\lambda}^2}}{c_g} \Big)^2}} \Bigg[ -\frac{6c_{\lambda} + \sqrt{c_g^2+36c_{\lambda}^2}}{c_g} \ket{0000} + \ket{1111} \Bigg] \, ,
\end{equation}

\begin{equation}
v_5 = \frac{1}{\sqrt{1+\Big( \frac{6c_{\lambda}-\sqrt{c_g^2+36c_{\lambda}^2}}{c_g} \Big)^2}} \Bigg[ -\frac{6c_{\lambda} - \sqrt{c_g^2+36c_{\lambda}^2}}{c_g} \ket{0000} + \ket{1111} \Bigg] \, ,
\end{equation}

\noindent which have a form similar to $|{\psi}_{\pm}\rangle = \frac{1}{\sqrt{2}}\left(|0 0 0 0 \rangle \mp |1 1 1 1 \rangle\right)$, but with different relative coefficients (note that $v_4$ and $v_5$ reduce to $|\psi_{\pm} \rangle$ when $c_{\lambda}=0$).

\section{RG Analysis of Model 2: Dirac Fermions with $SU(2)\times SU(4)$ Symmetry} \label{sec:app_rel}

\subsection{Setting up the RG calculation}

This appendix presents details of the RG calculation for the theory of Dirac fermions with $SU(2)\times SU(4)$ global symmetry. The RG equations for the couplings are obtained at 1-loop, while the renormalization of the propagator is done at 2-loop (so as to obtain the anomalous exponent for the fermion). The action presented in the main text was written in terms of the five four-fermion operators

\begin{align}\label{Eq:Order_parameter_operators}
\begin{split}
\mathcal{O}_1 &= (\bar{\psi}_{i\alpha} \psi_{i\alpha})^2 \\ \mathcal{O}_{\sigma} &= (\bar{\psi}_{i\alpha} \sigma_{ij}^A \psi_{j\alpha}) (\bar{\psi}_{k\beta} \sigma_{kl}^A \psi_{l\beta}) \\ \mathcal{O}_{T} &= (\bar{\psi}_{i\alpha} T_{\alpha\beta}^A \psi_{i\beta}) (\bar{\psi}_{j\gamma} T_{\gamma \delta}^A \psi_{j\delta}) \\ \mathcal{O}_{\sigma T} &= (\bar{\psi}_{i\alpha} \sigma_{ij}^A T_{\alpha\beta}^B \psi_{j\beta}) (\bar{\psi}_{k\gamma} \sigma_{kl}^A T_{\gamma \delta}^B \psi_{l\delta}) \\ \mathcal{O}_{\text{SMG}} &= \epsilon_{ab} \epsilon_{cd} \epsilon_{ij} \epsilon_{kl} \epsilon_{\alpha \beta \gamma \delta} \Big( \psi_{ai\alpha} \psi_{bj\beta} \psi_{ck\gamma} \psi_{dl\delta} + \bar{\psi}_{ai\alpha} \bar{\psi}_{bj\beta} \bar{\psi}_{ck\gamma} \bar{\psi}_{dl\delta} \Big) \, ,
\end{split}
\end{align}

\noindent where $a=1,2$ label the component of the Dirac spinor, $i=1,2$ labels the valley ($N_v$) and $\alpha=1,...,4$ labels the flavor ($N_f$). $\sigma^A$ and $T^A$ are respectively the generators of $SU(2)$ and $SU(4)$ in the fundamental representation. As will be demonstrated, imposing symmetry under $SU(2)\times SU(4)$ and Lorentz invariance implies that the RG equations at 1-loop are closed for these five operators. However, it turns out that the RG calculation is technically easier to perform with a different basis of operators. Instead of $\mathcal{O}_{\sigma}$, $\mathcal{O}_{T}$ and $\mathcal{O}_{\sigma T}$, we work with

\begin{align}
\begin{split}
\mathcal{O}_2 = (\bar{\psi}_{i\alpha} \psi_{j\alpha}) (\bar{\psi}_{j\beta} \psi_{i\beta}) \, , \quad \mathcal{O}_3 = (\bar{\psi}_{i\alpha} \psi_{i\beta}) (\bar{\psi}_{j\beta} \psi_{j\alpha}) \, , \quad \mathcal{O}_4 = (\bar{\psi}_{i\alpha} \psi_{j\beta}) (\bar{\psi}_{j\beta} \psi_{i\alpha}) \, .
\end{split}
\end{align}

\noindent These can be expressed in terms of the original operators using the following two Fierz identities

\begin{equation} \label{Eq:Fierz_identities}
\sigma_{ij}^A \sigma_{kl}^A = 2\Big( \delta_{il} \delta_{jk} - \frac{1}{N_v} \delta_{ij} \delta_{kl} \Big) \, , \quad T_{\alpha\beta}^A T_{\gamma \delta}^A = 2 \Big( \delta_{\alpha \delta} \delta_{\beta \gamma} - \frac{1}{N_f} \delta_{\alpha \beta} \delta_{\gamma \delta} \Big) \, ,
\end{equation}

\noindent which will be done at the end of the calculation. Therefore, we are considering the following action with five interaction terms

\begin{align}
\begin{split}
S &= \int d^Dx \bar{\psi}_{i\alpha} \slashed{\partial} \psi_{i\alpha} -\frac{g_1}{2} \int d^Dx \mathcal{O}_1 -\frac{g_2}{2} \int d^Dx \mathcal{O}_2 -\frac{g_3}{2} \int d^Dx \mathcal{O}_3 -\frac{g_4}{2} \int d^Dx \mathcal{O}_4 +\frac{g}{8} \int d^Dx \mathcal{O}_{\text{SMG}} \, .
\end{split}
\end{align}

\noindent The RG is performed using a hybrid scheme, where as in the previous theory, momentum integrals are performed in dimensional regularization with $D=2+\epsilon$ spacetime dimensions and we work with the MS scheme. However, we are using the spinor structure of the target dimension $D=3$, which means that there are three Dirac gamma matrices, which are simply taken to be the Pauli matrices $\gamma_{\mu}=(\sigma^3,\sigma^1,\sigma^2)$, respecting the Clifford algebra $\{ \gamma_{\mu},\gamma_{\nu} \} = 2\delta_{\mu\nu}$. One justification for this method is that in our target dimension $D = 3$, there is no chirality and thus no $\gamma^5$ matrix. From a technical perspective, this approach also leads to a fewer number of allowed operators compared to an alternative approach employed in Ref. \cite{ladovrechis2023gross} where one also allow terms with $\gamma^5$ matrix.

The imaginary-time action in terms of bare quantities is

\begin{align}
\begin{split}
S &= \int_x \bar{\psi}_{B,i\alpha} \slashed{\partial} \psi_{B,i\alpha} -\frac{g_{B,1}}{2} \int_x \mathcal{O}_{B,1} -\frac{g_{B,2}}{2} \int_x \mathcal{O}_{B,2} -\frac{g_{B,3}}{2} \int_x \mathcal{O}_{B,3} -\frac{g_{B,4}}{2} \int_x \mathcal{O}_{B,4} +\frac{g_B}{8} \int_x \mathcal{O}_{B,\text{SMG}} \, ,
\end{split}
\end{align}

\noindent where $\mathcal{O}_{B,i}$ ($i=1,...,4$) and $\mathcal{O}_{B,\text{SMG}}$ are the operators in terms of the bare field $\psi_{B,i\alpha}$. In terms of renormalized quantities, the action becomes

\begin{align}
\begin{split}
S &= Z_{\psi} \int_x \bar{\psi}_{i\alpha} \slashed{\partial} \psi_{i\alpha} - Z_1 \mu^{-\epsilon} \frac{g_1}{2} \int_x \mathcal{O}_1 - Z_2 \mu^{-\epsilon} \frac{g_2}{2} \int_x \mathcal{O}_2 - Z_3 \mu^{-\epsilon} \frac{g_3}{2} \int_x \mathcal{O}_3 - Z_4 \mu^{-\epsilon} \frac{g_4}{2} \int_x \mathcal{O}_4 + Z_g \mu^{-\epsilon} \frac{g}{8} \int_x \mathcal{O}_{\text{SMG}} \\ &\equiv S_0 + S_{\text{int},1} + S_{\text{int},2} + S_{\text{int},3} + S_{\text{int},4} + S_{\text{int,SMG}} \, ,
\end{split}
\end{align}

\noindent where the renormalized couplings are dimensionless and $\mu$ is the RG mass scale. The wavefunction renormalization $Z_{\psi}$ is defined through $\psi_{B,i\alpha} = Z_{\psi}^{1/2} \psi_{i\alpha}$ and the renormalization constants are written as $Z_i = 1+\delta_i$ for $i=1,...,4$, $Z_{\psi} = 1+\delta_{\psi}$ and $Z_g = 1+\delta_g$. Comparing the two versions of the action yields

\begin{equation}
g_{B,i} = \mu^{-\epsilon} Z_{\psi}^{-2} Z_i g_i \, , \qquad g_B = \mu^{-\epsilon} Z_{\psi}^{-2} Z_g g \, ,
\end{equation}

\noindent and since the bare couplings are $\mu$-independent, it follows that the RG equations are

\begin{align} \label{Eq:Beta_functions}
\begin{split}
-\beta(g_i) = \mu \frac{dg_i}{d\mu} = (\epsilon + 2 \gamma_{\psi} - \gamma_i)g_i \, , \qquad - \beta(g) = \mu \frac{dg}{d\mu} = (\epsilon + 2 \gamma_{\psi} - \gamma_g)g \, ,
\end{split}
\end{align}

\noindent where as before, the anomalous dimensions are defined as $\gamma_X = \mu \frac{d\ln Z_X}{d\mu}$, with $Z_X = \{Z_i,Z_g,Z_{\psi}\}$. Using the leading order term in the RG equations, the expression for the anomalous dimensions simplifies at the first non-trivial order to become

\begin{equation} \label{Eq:Anomalous_dim}
\gamma_X = \epsilon \Bigg( \sum_{j=1}^4 \frac{dZ_X}{dg_j}g_j + \frac{dZ_X}{dg} g \Bigg) \, .
\end{equation}

We now compute the renormalization constants $Z_i$, $Z_g$ and $Z_{\psi}$ by requiring the effective action $\Gamma[\psi_c]$ to be finite order-by-order. To do so, the fermion field is expanded around its classical configuration, $\psi_{i\alpha}=\psi_{c,i\alpha} + \psi_{f,i\alpha}$ and the effective action is obtained by averaging over the fluctuations. For simplicity, we denote $\psi_{c,i\alpha} \equiv c_{i\alpha}$, $\psi_{f,i\alpha}\equiv f_{i\alpha}$. We now expand the action, starting with the free part $S_0$

\begin{equation}
S_0[s+f] = Z_{\psi} \int_x \, \Bar{c}_{i\alpha} \slashed{\partial} c_{i\alpha} + \int_x \Bar{f}_{i\alpha} \slashed{\partial} f_{i\alpha} + ... \equiv S_0[s] + S_0^{(2)}[f] + ... \, ,
\end{equation}

\noindent where additional terms vanish or are unimportant at the order we work at. Fourier transforming $S_0^{(2)}[f]$ with our convention $f_{i\alpha}(x) = \int_p \e^{\ii x\cdot p} f_{i\alpha}(p)$ allows to identify the free propagator for the fluctuations

\begin{equation}
S_0^{(2)}[f] = \int \frac{d^Dp}{(2\pi)^D} \Bar{f}_i \ii \gamma_{\mu} p_{\mu} f_i \implies \Tilde{G}(p) = \frac{1}{\ii \gamma_{\mu} p_{\mu}} = - \frac{\ii \slashed{p}}{p^2} \, .
\end{equation}

Expanding the interaction terms yields

\begin{align}
\begin{split}
S_{\text{int},1}[c+f] &= S_{\text{int},1}[c] + S_{\text{int},1}^{(2)}[c,f] + S_{\text{int},1}^{(3)}[c,f] + ... \\ S_{\text{int},2}[c+f] &= S_{\text{int},2}[c] + S_{\text{int},2}^{(2)}[c,f] + S_{\text{int},2}^{(3)}[c,f] + ... \\ S_{\text{int},3}[c+f] &= S_{\text{int},3}[c] + S_{\text{int},3}^{(2)}[c,f] + S_{\text{int},3}^{(3)}[c,f] + ... \\ S_{\text{int},4}[c+f] &= S_{\text{int},4}[c] + S_{\text{int},4}^{(2)}[c,f] + S_{\text{int},4}^{(3)}[c,f] + ... \\ S_{\text{int,SMG}}[c+f] &= S_{\text{int,SMG}}[c] + S_{\text{int,SMG}}^{(2)}[c,f] + S_{\text{int,SMG}}^{(3)}[c,f] + ... \, ,
\end{split}
\end{align}

\noindent with the purely classical terms being

\begin{align}
\begin{split}
S_{\text{int},1}[c] &= - Z_1 \mu^{-\epsilon} \frac{g_1}{2} \int_x \mathcal{O}_1(c) \, , \quad S_{\text{int},2}[c] = - Z_2 \mu^{-\epsilon} \frac{g_2}{2} \int_x \mathcal{O}_2(c) \, , \quad S_{\text{int},3}[c] = - Z_3 \mu^{-\epsilon} \frac{g_3}{2} \int_x \mathcal{O}_3(c) \, , \\ &\hspace{1cm} S_{\text{int},4}[c] = - Z_4 \mu^{-\epsilon} \frac{g_4}{2} \int_x \mathcal{O}_4(c) \, , \quad S_{\text{int,SMG}}[c] =  Z_g \mu^{-\epsilon} \frac{g}{8} \int_x \mathcal{O}_{\text{SMG}}(c) \, ,
\end{split}
\end{align}

\noindent where $\mathcal{O}_i(c)$ denotes the operator $\mathcal{O}_i$ evaluated in terms of the classical field. The interaction terms quadratic in fluctuations are

\begin{align} \label{Eq:Sint2cf}
\begin{split}
S_{\text{int},1}^{(2)}[c,f] &= - \mu^{-\epsilon}\frac{g_1}{2} \int_x \Big( 2 \Bar{c}_{ai\alpha} c_{ai\alpha} \Bar{f}_{bj\beta} f_{bj\beta} + 2 \Bar{c}_{ai\alpha} f_{ai\alpha} \Bar{f}_{bj\beta} c_{bj\beta} + \bar{c}_{ai\alpha} f_{ai\alpha} \bar{c}_{bj\beta} f_{bj\beta} + \Bar{f}_{ai\alpha} c_{ai\alpha} \Bar{f}_{bj\beta} c_{bj\beta} \Big) \\ S_{\text{int},2}^{(2)}[c,f] &= -  \mu^{-\epsilon}\frac{g_2}{2} \int_x \Big( 2 \bar{c}_{ai\alpha} c_{aj\alpha} \Bar{f}_{bj\beta} f_{bi\beta} + 2 \bar{c}_{ai\alpha} f_{aj\alpha} \Bar{f}_{bj\beta} c_{bi\beta} + \bar{c}_{ai\alpha} f_{aj\alpha} \bar{c}_{bj\beta} f_{bi\beta} + \Bar{f}_{ai\alpha} c_{aj\alpha} \Bar{f}_{bj\beta} c_{bi\beta} \Big) \\ S_{\text{int},3}^{(2)}[c,f] &= -  \mu^{-\epsilon}\frac{g_3}{2} \int_x \Big( 2 \bar{c}_{ai\alpha} c_{ai\beta} \Bar{f}_{bj\beta} f_{bj\alpha} + 2 \bar{c}_{ai\alpha} f_{ai\beta} \Bar{f}_{bj\beta} c_{bj\alpha} + \bar{c}_{ai\alpha} f_{ai\beta} \bar{c}_{bj\beta} f_{bj\alpha} + \Bar{f}_{ai\alpha} c_{ai\beta} \Bar{f}_{bj\beta} c_{bj\alpha} \Big) \\ S_{\text{int},4}^{(2)}[c,f] &= -  \mu^{-\epsilon}\frac{g_4}{2} \int_x \Big( 2 \bar{c}_{ai\alpha} c_{aj\beta} \Bar{f}_{bj\beta} f_{bi\alpha} + 2 \bar{c}_{ai\alpha} f_{aj\beta} \Bar{f}_{bj\beta} c_{bi\alpha} + \bar{c}_{ai\alpha} f_{aj\beta} \bar{c}_{bj\beta} f_{bi\alpha} + \Bar{f}_{ai\alpha} c_{aj\beta} \Bar{f}_{bj\beta} c_{bi\alpha} \Big) \\ S_{\text{int,SMG}}^{(2)}[c,f] &=  \mu^{-\epsilon}\frac{g}{8} \epsilon_{ab} \epsilon_{cd} \epsilon_{ij} \epsilon_{cd} \epsilon_{\alpha \beta \gamma \delta} \int_x \Big( 2 c_{ai\alpha} c_{bj\beta} f_{ck\gamma} f_{dl\delta} + 4 c_{ai\alpha} f_{bj\beta} c_{ck\gamma} f_{dl\delta} \\ &\hspace{4cm} + 2 \bar{c}_{ai\alpha} \bar{c}_{bj\beta} \bar{f}_{ck\gamma} \bar{f}_{dl\delta} + 4 \bar{c}_{ai\alpha} \bar{f}_{bj\beta} \bar{c}_{ck\gamma} \bar{f}_{dl\delta} \big) \, ,
\end{split}
\end{align}

\noindent while the cubic ones are

\begin{align} \label{Eq:Sint3cf}
\begin{split}
S_{\text{int},1}^{(3)}[c,f] &= -\mu^{-\epsilon} g_1 \int_x \Big( \bar{c}_{ai\alpha} f_{ai\alpha} \bar{f}_{bj\beta} f_{bj\beta} + \bar{f}_{ai\alpha} f_{ai\alpha} \bar{f}_{bj\beta} c_{bj\beta} \Big) \\ S_{\text{int},2}^{(3)}[c,f] &= - \mu^{-\epsilon} g_2 \int_x \Big( \bar{c}_{ai\alpha} f_{aj\alpha} \bar{f}_{bj\beta} f_{bi\beta} + \bar{f}_{ai\alpha} f_{aj\alpha} \bar{f}_{bj\beta} c_{bi\beta} \Big) \\ S_{\text{int},3}^{(3)}[c,f] &= - \mu^{-\epsilon} g_3 \int_x \Big( \bar{c}_{ai\alpha} f_{ai\beta} \bar{f}_{bj\beta} f_{bj\alpha} + \bar{f}_{ai\alpha} f_{ai\beta} \bar{f}_{bj\beta} c_{bj\alpha} \Big) \\ S_{\text{int},4}^{(3)}[c,f] &= - \mu^{-\epsilon} g_4 \int_x \Big( \bar{c}_{ai\alpha} f_{aj\beta} \bar{f}_{bj\beta} f_{bi\alpha} + \bar{f}_{ai\alpha} f_{aj\beta} \bar{f}_{bj\beta} c_{bi\alpha} \Big) \\ S_{\text{int,SMG}}^{(3)}[c,f] &= \mu^{-\epsilon} \frac{g}{2} \epsilon_{ab} \epsilon_{cd} \epsilon_{ij} \epsilon_{kl} \epsilon_{\alpha \beta \gamma \delta} \int_x \Big( c_{ai\alpha} f_{bj\beta} f_{ck\gamma} f_{dl\delta} + \bar{c}_{ai\alpha} \bar{f}_{bj\beta} \bar{f}_{ck\gamma} \bar{f}_{dl\delta} \Big) \, .
\end{split}
\end{align}

\noindent Terms with four powers of fluctuations $f_{i\alpha}$ can contribute at 2-loop to diagrams of the form of Fig. \ref{Fig:Feynman_diagram_3} (a). However, these kind of diagrams vanish with massless Dirac fermions since a mass term cannot be generated due to parity/time-reversal symmetry. Hence, quartic terms in $f_{i\alpha}$ can be ignored from the expanded action. Before continuing, let us finally define

\begin{align}
\begin{split}
S[c] &\equiv S_0 + S_{\text{int},1}[c] + S_{\text{int},2}[c] + S_{\text{int},3}[c] + S_{\text{int},4}[c] + S_{\text{int,SMG}}[c] \\ S_{\text{int}}^{(2)}[c,f] &\equiv S_{\text{int},1}^{(2)}[c,f] + S_{\text{int},2}^{(2)}[c,f] + S_{\text{int},3}^{(2)}[c,f] + S_{\text{int},4}^{(2)}[c,f] + S_{\text{int,SMG}}^{(2)}[c,f] \\ S_{\text{int}}^{(3)}[c,f] &\equiv S_{\text{int},1}^{(3)}[c,f] + S_{\text{int},2}^{(3)}[c,f] + S_{\text{int},3}^{(3)}[c,f] + S_{\text{int},4}^{(3)}[c,f] + S_{\text{int,SMG}}^{(3)}[c,f] \, .
\end{split}
\end{align}

We are now in a position to obtain the effective action $\Gamma[c]$, which takes the following form at the second order in the cumulant expansion

\begin{equation} \label{Cumulant_Exp}
\Gamma[c] = S[c] + \ev{S_{\text{int}}^{(2)}[c,f] + S_{\text{int}}^{(3)}[c,f]}_f - \frac{1}{2} \ev{\Big( S_{\text{int}}^{(2)}[c,f]+ S_{\text{int}}^{(3)}[c,f]\Big)^2}_f^c + ... \, ,
\end{equation}

\noindent where correlation functions in the last term are connected. The expectation value at the leading order in the cumulant expansion contributes to the renormalization of the propagator. However, as in any theory with four-fermion interactions, the resulting `seagull' Feynman diagram does not contain external momentum in the loop, which means that it can only renormalize the mass, which itself cannot be generated due to parity/time-reversal symmetry. This means that $\ev{S_{\text{int}}^{(2)}[c,f] + S_{\text{int}}^{(3)}[c,f]}_f=0$ and the effective action thus becomes

\begin{equation}\label{Eq:SEff}
\Gamma[c] = S[c] - \frac{1}{2} \ev{\Big( S_{\text{int}}^{(2)}[c,f] \Big)^2}_f^c - \frac{1}{2} \ev{\Big( S_{\text{int}}^{(3)}[c,f] \Big)^2}_f^c + ... \, ,
\end{equation}

\noindent where the cross-term at second order obviously vanishes. The first expectation value renormalizes the interaction vertices at 1-loop, while the last term renormalizes the propagator at 2-loop. Let us evaluate these two terms separately.

\subsection{1-loop corrections to the interaction vertices}

Expanding the first expectation value of Eq. \ref{Eq:SEff} yields the following 15 terms

\begin{align} \label{Eq:EV_Sint2}
\begin{split}
\ev{\Big( S_{\text{int}}^{(2)}[c,f] \Big)^2}_f^c &= \ev{\Big( S_{\text{int},1}^{(2)} + S_{\text{int},2}^{(2)} + S_{\text{int},3}^{(2)} + S_{\text{int},4}^{(2)} + S_{\text{int,SMG}}^{(2)} \Big)^2}_f^c \\ &= \ev{\Big( S_{\text{int},1}^{(2)} \Big)^2}_f^c + \ev{\Big( S_{\text{int},2}^{(2)} \Big)^2}_f^c + \ev{\Big( S_{\text{int},3}^{(2)} \Big)^2}_f^c + \ev{\Big( S_{\text{int},4}^{(2)} \Big)^2}_f^c \\ &\hspace{0.5cm} + 2 \ev{S_{\text{int},1}^{(2)}S_{\text{int},2}^{(2)}}_f^c + 2 \ev{S_{\text{int},1}^{(2)}S_{\text{int},3}^{(2)}}_f^c + 2 \ev{S_{\text{int},1}^{(2)}S_{\text{int},4}^{(2)}}_f^c \\ &\hspace{0.5cm} + 2 \ev{S_{\text{int},2}^{(2)}S_{\text{int},3}^{(2)}}_f^c + 2 \ev{S_{\text{int},2}^{(2)}S_{\text{int},4}^{(2)}}_f^c + 2 \ev{S_{\text{int},3}^{(2)}S_{\text{int},4}^{(2)}}_f^c \\ &\hspace{0.5cm} + \ev{\Big( S_{\text{int,SMG}}^{(2)} \Big)^2}_f^c + 2 \ev{S_{\text{int,SMG}}^{(2)}S_{\text{int},1}^{(2)}}_f^c + 2 \ev{S_{\text{int,SMG}}^{(2)}S_{\text{int},2}^{(2)}}_f^c \\ &\hspace{0.5cm} + 2 \ev{S_{\text{int,SMG}}^{(2)}S_{\text{int},3}^{(2)}}_f^c + 2 \ev{S_{\text{int,SMG}}^{(2)}S_{\text{int},4}^{(2)}}_f^c \\ &\equiv T_1^{(2)} + T_2^{(2)} + T_3^{(2)} + T_4^{(2)} + 2 T_5^{(2)} + 2 T_6^{(2)} + 2 T_7^{(2)} + 2 T_8^{(2)} \\ &\hspace{0.5cm}+ 2T_9^{(2)} + 2 T_{10}^{(2)} + T_{11}^{(2)} + 2 T_{12}^{(2)} + 2 T_{13}^{(2)} + 2 T_{14}^{(2)} + 2 T_{15}^{(2)} \, .
\end{split}
\end{align}

\noindent Each of these must be evaluated using the expressions from Eq. \ref{Eq:Sint2cf}. The calculation is quite tedious and we only highlight the most important steps. Note also that for the first ten terms, which are those corresponding to the $U(1)$-symmetric subspace of the theory with $g=0$, we work with an arbitrary number of valleys $N_v$ and flavors $N_f$. This will prove to be useful for an important check later.

Let us start with the first term, which evaluates to

\begin{align}
\begin{split}
T_1^{(2)} &= \ev{\Big( S_{\text{int},1}^{(2)}[c,f] \Big)^2}_f^c \\ &= \mu^{-2\epsilon} \frac{g_1^2}{4} \int_{x,y} \Big\langle \Big[ 4(\bar{c}_{ai\alpha} c_{ai\alpha} \bar{f}_{bj\beta} f_{bj\beta})(x) (\bar{c}_{a'i'\alpha'} c_{a'i'\alpha'} \bar{f}_{b'j'\beta'} f_{b'j'\beta'})(y) \\ &\hspace{2cm} + 8(\bar{c}_{ai\alpha} c_{ai\alpha} \bar{f}_{bj\beta} f_{bj\beta})(x) (\bar{c}_{a'i'\alpha'} f_{a'i'\alpha'} \bar{f}_{b'j'\beta'} c_{b'j'\beta'})(y) \\ &\hspace{2cm} + 4(\bar{c}_{ai\alpha} f_{ai\alpha} \bar{f}_{bj\beta} c_{bj\beta})(x) (\bar{c}_{a'i'\alpha'} f_{a'i'\alpha'} \bar{f}_{b'j'\beta'} c_{b'j'\beta'})(y) \\ &\hspace{2cm} + 2(\bar{c}_{ai\alpha} f_{ai\alpha} \bar{c}_{bj\beta} f_{bj\beta})(x) (\bar{f}_{a'i'\alpha'} c_{a'i'\alpha'} \bar{f}_{b'j'\beta'} c_{b'j'\beta'})(y) \Big] \Big\rangle_f^c \\ &\equiv \mu^{-\epsilon} g_1^2 \Big( I_{1,1} + 2 I_{1,2} + I_{1,3} + \frac{1}{2} I_{1,4} \Big) \, .
\end{split}
\end{align}

\noindent These four terms are similar in structure to those in Eq. \ref{Eq:D_1^2} and are thus represented by the same diagrams as in Fig. \ref{Fig:Feynman_diagram_1}, with the additional spin structure of Dirac fermions. The first integral yields

\begin{align} \label{Eq:I1,1}
\begin{split}
I_{1,1} &= -\delta_{jj'} \delta_{jj'} \delta_{\beta \beta'} \delta_{\beta \beta'} \mu^{-\epsilon} \int_{x,y} (\bar{c}_{i\alpha} c_{i\alpha})(x) (\bar{c}_{i'\alpha'} c_{i'\alpha'})(y) G_{bb'}(x-y) G_{b'b}(y-x) \\ &\approx -N_v N_f \mu^{-\epsilon} \int_p \Tr[\tilde{G}(p) \tilde{G}(p)] \int_x (\bar{c}_{i\alpha} c_{i\alpha})^2 \\ &= 2 N_v N_f \mu^{-\epsilon} \int_p \frac{1}{p^2} \int_x (\bar{c}_{i\alpha} c_{i\alpha})^2 \, ,
\end{split}
\end{align}

\noindent where the second equality follows by expanding the previous expression at the leading order in the separation $l=x-y$ and Fourier transforming from $l$ to $p$, while the third equality follows from using $\slashed{p}\slashed{p}=p^2$ and the remaining trace over Dirac spinors is $\Tr[\mathds{1}]=D$. At 1-loop with the MS scheme, one can simply set $D=2$. The remaining momentum integral, which is the same appearing in all the diagrams renormalizing the interactions, is

\begin{align} \label{Eq:Int_Gp}
\begin{split}
\mu^{-\epsilon} \int \frac{d^Dp}{(2\pi)^D} \frac{1}{p^2+\Delta} = \frac{\mu^{-\epsilon}}{(4\pi)^{D/2}} \Gamma(1-D/2) \Delta^{\frac{D}{2}-1} = -\frac{1}{2\pi \epsilon} + \text{finite} \, ,
\end{split}
\end{align}

\noindent where $\Delta$ is an arbitrary (squared) mass scale that has been introduced to regulate the IR and thus to extract the UV divergence. As in the nonrelativistic theory, one does not need to introduce this arbitrary scale if the exact Fourier transform is performed in Eq. \ref{Eq:I1,1}, where the external momentum plays the role of the IR regulator. For now on, finite parts will be implicit. Therefore

\begin{equation}
I_{1,1} = - \frac{N_v N_f}{\pi} \frac{1}{\epsilon} \int_x (\bar{c}_{i\alpha} c_{i\alpha})^2 \, .
\end{equation}

\noindent Proceeding similarly, one gets for $I_{1,2}$

\begin{equation}
I_{1,2} = \frac{1}{2\pi} \frac{1}{\epsilon} \int_x (\bar{c}_{i\alpha} c_{i\alpha})^2 \, ,
\end{equation}

\noindent while for the last two integrals, one can show that

\begin{align}
\begin{split}
I_{1,3} + \frac{1}{2} I_{1,4} = \int_p \int_x \bar{c}_{i\alpha} \tilde{G}(p) c_{i\alpha} \bar{c}_{j\beta} \tilde{G}(p) c_{j\beta} + \int_p \int_x \bar{c}_{i\alpha} \tilde{G}(p) c_{i\alpha} \bar{c}_{j\beta} \tilde{G}(-p) c_{j\beta} = 0 \, ,
\end{split}
\end{align}

\noindent since $\tilde{G}(-p) = - \tilde{G}(p)$ and where $I_{1,4}$ gets a factor of 2 from two identical Wick contractions. Combining everything thus yields

\begin{equation}
T_1^{(2)} = -\mu^{-\epsilon} \frac{(N_v N_f-1)}{\pi} g_1^2 \frac{1}{\epsilon} \int_x \mathcal{O}_1(c) \, .
\end{equation}

We now move on to the evaluation of $T_2^{(2)}$

\begin{align}
\begin{split}
T_2^{(2)} &= \ev{\Big( S_{\text{int},2}^{(2)}[c,f] \Big)^2}_f^c \\ &= \mu^{-2\epsilon} \frac{g_2^2}{4} \int_{x,y} \Big\langle \Big[ 4(\bar{c}_{ai\alpha} c_{aj\alpha} \bar{f}_{bj\beta} f_{bi\beta})(x) (\bar{c}_{a'i'\alpha'} c_{a'j'\alpha'} \bar{f}_{b'j'\beta'} f_{b'i'\beta'})(y) \\ &\hspace{2cm} + 8(\bar{c}_{ai\alpha} c_{aj\alpha} \bar{f}_{bj\beta} f_{bi\beta})(x) (\bar{c}_{a'i'\alpha'} f_{a'j'\alpha'} \bar{f}_{b'j'\beta'} c_{b'i'\beta'})(y) \\ &\hspace{2cm} + 4(\bar{c}_{ai\alpha} f_{aj\alpha} \bar{f}_{bj\beta} c_{bi\beta})(x) (\bar{c}_{a'i'\alpha'} f_{a'j'\alpha'} \bar{f}_{b'j'\beta'} c_{b'i'\beta'})(y) \\ &\hspace{2cm} + 2(\bar{c}_{ai\alpha} f_{aj\alpha} \bar{c}_{bj\beta} f_{bi\beta})(x) (\bar{f}_{a'i'\alpha'} c_{a'j'\alpha'} \bar{f}_{b'j'\beta'} c_{b'i'\beta'})(y) \Big] \Big\rangle_f^c \\ &\equiv \mu^{-\epsilon} g_2^2 \Big( I_{2,1} + 2 I_{2,2} + I_{2,3} + \frac{1}{2} I_{2,4} \Big) \, .
\end{split}
\end{align}

\noindent This expression can also be represented by the diagrams of Fig. \ref{Fig:Feynman_diagram_1}. The calculation of the first two integrals is very similar as before and yields

\begin{equation}
I_{2,1} = -\frac{N_f}{\pi}\frac{1}{\epsilon} \int_x (\bar{c}_{i\alpha} c_{j\alpha})(\bar{c}_{j\beta} c_{i\beta}) \, , \qquad I_{2,2} = \frac{1}{2\pi} \frac{1}{\epsilon} \int_x (\bar{c}_{i\alpha} c_{i\alpha})^2 \, .
\end{equation}

\noindent However, this time $I_{2,3}$ and $I_{2,4}$ do not cancel and their calculation actually requires a crucial step, which we now highlight. After a few manipulations, $I_{2,3}$ becomes

\begin{align}
\begin{split}
I_{2,3} \approx -N_v \mu^{-\epsilon} \int_p \Tilde{G}_{ab'}(p) \tilde{G}_{a'b}(p) \int_x (\bar{c}_{ai\alpha} c_{bi\beta}) (\bar{c}_{a'j\beta} c_{b'j\alpha}) = N_v \mu^{-\epsilon} \int_p \frac{p_ {\mu} p_{\nu}}{p^4} \gamma_{ab'}^{\mu} \gamma_{a'b}^{\nu} \int_x (\bar{c}_{ai\alpha} c_{bi\beta}) (\bar{c}_{a'j\beta} c_{b'j\alpha}) \, .
\end{split}
\end{align}

\noindent Because of rotational invariance, we can make the replacement $p_{\mu} p_{\nu} \rightarrow \frac{1}{D} \delta_{\mu \nu}p^2$ in the integrand. The integral over $p$ now takes the form of Eq. \ref{Eq:Int_Gp}, leading to

\begin{equation}
I_{2,3} = -\frac{N_v}{4\pi} \frac{1}{\epsilon} \gamma_{ab'}^{\mu} \gamma_{a'b}^{\mu} \int_x (\bar{c}_{ai\alpha} c_{bi\beta}) (\bar{c}_{a'j\beta} c_{b'j\alpha}) \, .
\end{equation}

\noindent Note that the resulting operator in terms of the classical fields does not have the structure of any of the initial operators, because of the gamma matrices. This is where we use the crucial fact that our gamma matrices live in $D=3$. Since these are simply the three Pauli matrices, they respect the $SU(2)$ completeness relation

\begin{equation} \label{Eq:Completeness_gamma_matrix}
\gamma_{ab}^{\mu} \gamma_{cd}^{\mu} = \sigma_{ab}^{A} \sigma_{cd}^{A} = 2 \delta_{ad} \delta_{bc} - \delta_{ab} \delta_{cd} \, .
\end{equation}

\noindent Applying this to the above equation and after a few manipulations, we get

\begin{equation}
I_{2,3}=-\frac{N_v}{4\pi} \frac{1}{\epsilon} \Bigg( 2\int_x (\bar{c}_{i\alpha} c_{i\beta}) (\bar{c}_{j\beta} c_{j\alpha}) + \int_x (\bar{c}_{i\alpha} c_{j\alpha}) (\bar{c}_{j\beta} c_{i\beta}) \Bigg) \, ,
\end{equation}

\noindent which now only contains operators we started with. Let us then recapitulate our procedure: Integrals over momentum are first performed in $D=2+\epsilon$ spacetime dimensions without worrying about the spinor structure. Once this is done, the dimension of spacetime over which spinors are defined is promoted from $2+\epsilon$ to $3$, which allows one to treat the gamma matrices as the three Pauli matrices. This method will be used extensively for the rest of the calculation and as will be highlighted later, leads to values of universal quantities that agree very well with the literature. The evaluation of $I_{2,4}$ also requires the use of Eq. \ref{Eq:Completeness_gamma_matrix}, from which one can show that

\begin{align}
\begin{split}
I_{2,4} = \frac{1}{2\pi} \frac{1}{\epsilon} \Bigg( 2\int_x (\bar{c}_{i\alpha} c_{j\beta})(\bar{c}_{j\beta} c_{i\alpha}) + \int_x (\bar{c}_{i\alpha} c_{i\alpha})^2 \Bigg) \, .
\end{split}
\end{align}

\noindent Therefore, combining the results of the four integrals yields

\begin{align}
\begin{split}
T_2^{(2)} = -\mu^{-\epsilon} \frac{g_2^2}{\pi} \frac{1}{\epsilon} \Bigg[ -\frac{5}{4} \int_x \mathcal{O}_1(c) + \frac{(N_v+4 N_f)}{4} \int_x \mathcal{O}_2(c) + \frac{N_v}{2} \int_x \mathcal{O}_3(c) - \frac{1}{2} \int_x \mathcal{O}_4(c) \Bigg] \, .
\end{split}
\end{align}

The calculation of the next eight terms/diagrams ($T_3^{(2)}$ to $T_{10}^{(2)}$) is performed using the same tools required for the evaluation of the first two terms, which results in

\begin{align}
\begin{split}
T_3^{(2)} &= -\mu^{-\epsilon}\frac{g_3^2}{\pi} \frac{1}{\epsilon} \Bigg[ -\frac{5}{4} \int_x \mathcal{O}_1(c) + \frac{N_f}{2} \int_x \mathcal{O}_2(c) + \frac{(N_f+4 N_v)}{4} \int_x \mathcal{O}_3(c) - \frac{1}{2} \int_x \mathcal{O}_4(c) \Bigg] \, , \\ 
T_4^{(2)} &= -\mu^{-\epsilon}\frac{g_4^2}{\pi} \frac{1}{\epsilon} \Bigg[ -\frac{(5-2N_v N_f)}{4} \int_x \mathcal{O}_1(c) + \frac{(2+N_v N_f)}{4} \int_x \mathcal{O}_4(c) \Bigg] \, , \\
T_5^{(2)} &= -\mu^{-\epsilon} \frac{g_1 g_2}{\pi} \frac{1}{\epsilon} \Bigg[ \frac{(2N_f-N_v)}{2} \int_x \mathcal{O}_1(c) - \frac{1}{2} \int_x \mathcal{O}_2(c) \Bigg] \, , \\
T_6^{(2)} &= -\mu^{-\epsilon}\frac{g_1 g_3}{\pi} \frac{1}{\epsilon} \Bigg[ \frac{(2N_v-N_f)}{2} \int_x \mathcal{O}_1(c) - \frac{1}{2} \int_x \mathcal{O}_3(c) \Bigg] \, , \\
T_7^{(2)} &= -\mu^{-\epsilon} \frac{g_1 g_4}{\pi} \frac{1}{\epsilon} \Bigg[ \frac{(2-N_v N_f)}{2} \int_x \mathcal{O}_1(c) - \frac{1}{2} \int_x \mathcal{O}_4(c) \Bigg] \, , \\
T_8^{(2)} &= -\mu^{-\epsilon}\frac{g_2 g_3}{\pi} \frac{1}{\epsilon} \Bigg[ 
\int_x \mathcal{O}_1(c) - \frac{N_f}{2} \int_x \mathcal{O}_2(c) - \frac{N_v}{2} \int_x \mathcal{O}_3(c) \Bigg] \, , \\
T_9^{(2)} &= -\mu^{-\epsilon}\frac{g_2 g_4}{\pi} \frac{1}{\epsilon} \Bigg[ \frac{(N_v-N_f)}{2} \int_x \mathcal{O}_1(c) + \frac{1}{2} \int_x \mathcal{O}_2(c) - \frac{3}{4} \int_x \mathcal{O}_3(c) + \frac{N_v}{4} \int_x \mathcal{O}_4(c) \Bigg] \, , \\
T_{10}^{(2)} &= -\mu^{-\epsilon}\frac{g_3 g_4}{\pi} \frac{1}{\epsilon} \Bigg[ \frac{(N_f-N_v)}{2} \int_x \mathcal{O}_1(c) - \frac{3}{4} \int_x \mathcal{O}_2(c) + \frac{1}{2} \int_x \mathcal{O}_3(c) + \frac{N_f}{4} \int_x \mathcal{O}_4(c) \Bigg] \, .
\end{split}
\end{align}

\noindent Note that the sum of the first ten terms of Eq. \ref{Eq:EV_Sint2} is symmetric under the exchanges $g_2 \leftrightarrow g_3$, $\mathcal{O}_2 \leftrightarrow \mathcal{O}_3$ and $N_v \leftrightarrow N_f$, which constitutes a non-trivial check of our results.

We now compute $T_{11}^{(2)}$ which involves the square of the SMG vertex

\begin{align}
\begin{split}
T_{11}^{(2)} &= \ev{\Big( S_{\text{int,SMG}}^{(2)}[c,f] \Big)^2}_f^c \\ &= 2 \mu^{-2\epsilon} \frac{g^2}{8^2} \epsilon_{ab} \epsilon_{cd} \epsilon_{ij} \epsilon_{kl} \epsilon_{\alpha \beta \gamma \delta} \epsilon_{a'b'} \epsilon_{c'd'} \epsilon_{i'j'} \epsilon_{k'l'} \epsilon_{\alpha' \beta' \gamma' \delta'} \\ &\hspace{0.5cm} \int_{x,y} \Big\langle \Big[ 4(c_{ai\alpha} c_{bj\beta} f_{ck\gamma} f_{dl\delta})(x) (\bar{c}_{a'i'\alpha'} \bar{c}_{b'j'\beta'} \bar{f}_{c'k'\gamma'} \bar{f}_{d'l'\delta'})(y) \\ &\hspace{1.5cm}+ 8(c_{ai\alpha} c_{bj\beta} f_{ck\gamma} f_{dl\delta})(x) (\bar{c}_{a'i'\alpha'} \bar{f}_{b'j'\beta'} \bar{c}_{c'k'\gamma'} \bar{f}_{d'l'\delta'})(y) \\ &\hspace{1.5cm}+ 8 (c_{ai\alpha} f_{bj\beta} c_{ck\gamma} f_{dl\delta})(x) (\bar{c}_{a'i'\alpha'} \bar{c}_{b'j'\beta'} \bar{f}_{c'k'\gamma'} \bar{f}_{d'l'\delta'})(y) \\ &\hspace{1.5cm}+ 16 (c_{ai\alpha} f_{bj\beta} c_{ck\gamma} f_{dl\delta})(x) (\bar{c}_{a'i'\alpha'} \bar{f}_{b'j'\beta'} \bar{c}_{c'k'\gamma'} \bar{f}_{d'l'\delta'})(y) \Big] \Big\rangle_f^c \\ &\equiv \mu^{-\epsilon} \frac{g^2}{8} \Big[ I_{11,1} + 2 I_{11,2} + 2 I_{11,3} + 4 I_{11,4} \Big] \, .
\end{split}
\end{align}

\noindent This expression can be schematically represented by diagrams of the form of Fig. \ref{Fig:Feynman_diagram_2} (a). Let us start with the first integral

\begin{align}
\begin{split}
I_{11,1} &= -2 \mu^{-\epsilon} \epsilon_{ab} \epsilon_{cd} \epsilon_{ij} \epsilon_{kl} \epsilon_{\alpha \beta \gamma \delta} \epsilon_{a'b'} \epsilon_{c'd'} \epsilon_{i'j'} \epsilon_{k'l'} \epsilon_{\alpha' \beta' \gamma' \delta'} \delta_{kk'} \delta_{ll'} \delta_{\gamma \gamma'} \delta_{\delta \delta'} \\ &\hspace{1cm} \int_{x,y} (c_{ai\alpha} c_{bj\beta})(x) (\bar{c}_{a'i'\alpha'} \bar{c}_{b'j'\beta'})(y) G_{cc'}(x-y) G_{dd'}(x-y) \\ &\approx -8 \mu^{-\epsilon} \epsilon_{ab} \epsilon_{cd} \epsilon_{a'b'} \epsilon_{c'd'} \epsilon_{ij} \epsilon_{kl} \big( \delta_{\alpha \alpha'} \delta_{\beta \beta'} - \delta_{\alpha \beta'} \delta_{\alpha' \beta} \big) \\ &\hspace{1cm} \int_p \Tilde{G}_{cc'}(p) \Tilde{G}_{dd'}(-p) \int_x (c_{ai\alpha} c_{bj\beta}) (\bar{c}_{a'k\alpha'} \bar{c}_{b'l\beta'}) \, ,
\end{split}
\end{align}

\noindent where we have used the Levi-Civita identities $\epsilon_{ij} \epsilon_{ij} = 2$ and $\epsilon_{\alpha \beta \gamma \delta} \epsilon_{\alpha'\beta' \gamma \delta} = 2 \big( \delta_{\alpha \alpha'} \delta_{\beta \beta'} - \delta_{\alpha \beta'} \delta_{\alpha' \beta} \big)$. The evaluation of the momentum integral yields

\begin{align} \label{Eq:Int_Gp_Gmp}
\begin{split}
\mu^{-\epsilon} \int_p \Tilde{G}_{cc'}(p) \Tilde{G}_{dd'}(-p) = \mu^{-\epsilon} \gamma^{\mu}_{cc'} \gamma_{dd'}^{\nu} \int_p \frac{p_{\mu} p_{\nu}}{p^4} = \gamma^{\mu}_{cc'} \gamma^{\mu}_{cc'} \frac{\mu^{-\epsilon}}{2}\int_p \frac{1}{p^2} = -\frac{1}{4\pi} \frac{1}{\epsilon} \Big( 2\delta_{cd'} \delta_{dc'}-\delta_{cc'}\delta_{dd'} \Big) \, .
\end{split}
\end{align}

\noindent Hence, replacing in the above expression and simplifying leads to

\begin{align}
\begin{split}
I_{11,1} = -\frac{24}{\pi} \frac{1}{\epsilon} \epsilon_{ab} \epsilon_{cd} \epsilon_{ij} \epsilon_{kl} \Bigg[ \int_x (\bar{c}_{ai\alpha} \bar{c}_{bj\beta}) (c_{ck\alpha} c_{dl\beta}) \Bigg] = -\frac{24}{\pi} \frac{1}{\epsilon} \Bigg[ -\int_x \mathcal{O}_1(c) + \int_x \mathcal{O}_2(c) + \int_x \mathcal{O}_3(c) - \int_x \mathcal{O}_4(c) \Bigg] \, ,
\end{split}
\end{align}

\noindent where the second equality follows from using $\epsilon_{ab} \epsilon_{cd} = \delta_{ac} \delta_{bd} - \delta_{ad} \delta_{bc}$ and similarly for $\epsilon_{ij}\epsilon_{kl}$. The calculation of $I_{11,2}$ and $I_{11,3}$ is very similar and yields

\begin{equation}
I_{11,2} = I_{11,3} = -\frac{6}{\pi} \frac{1}{\epsilon} \Bigg[ -\int_x \mathcal{O}_1(c) + \int_x \mathcal{O}_2(c) + \int_x \mathcal{O}_3(c) - \int_x \mathcal{O}_4(c) \Bigg] \, .
\end{equation}

\noindent Finally, the calculation of $I_{11,4}$ is actually the simplest, as all the Levi-Civita symbols disappear (there is no need to use the anti-symmetrization in terms of delta functions), resulting in

\begin{align}
\begin{split}
I_{11,4} = -\frac{1}{\pi}\frac{1}{\epsilon} \Big[ 2\int_x \mathcal{O}_2(c) + \int_x \mathcal{O}_3(c) - \int_x \mathcal{O}_1(c) - 2 \int_x \mathcal{O}_4(c) \Big] \, .
\end{split}
\end{align}

\noindent Therefore, combining the four integrals and simplifying yields

\begin{equation}
T_{11}^{(2)} = -\mu^{-\epsilon}\frac{g^2}{2\pi} \frac{1}{\epsilon} \Bigg[ -13\int_x \mathcal{O}_1(c) + 14 \int_x \mathcal{O}_2(c) + 13 \int_x \mathcal{O}_3(c) -14 \int_x \mathcal{O}_4(c) \Bigg] \, .
\end{equation}

\noindent This shows that the SMG interaction by itself generates the four $U(1)$-symmetric interactions at 1-loop.

Let us finally move on to the evaluation of the last four terms, $T_{12}^{(2)}$ to $T_{15}^{(2)}$, which all have a very similar structure and can be represented by diagrams (b) and (c) of Fig. \ref{Fig:Feynman_diagram_2}. For the first of these, we have

\begin{align}
\begin{split}
T_{12}^{(2)} &= \ev{S_{\text{int,SMG}}^{(2)}[c,f] S_{\text{int},1}^{(2)}[c,f]}_f^c \\ &= -\mu^{-2\epsilon} \frac{g g_1}{16} \epsilon_{ab} \epsilon_{cd} \epsilon_{ij} \epsilon_{kl} \epsilon_{\alpha \beta \gamma \delta} \int_{x,y} \Big\langle \Big[ 2(c_{ai\alpha} c_{bj\beta} f_{ck\gamma} f_{dl\delta})(x) (\bar{f}_{a'i'\alpha'} c_{a'i'\alpha'} \bar{f}_{b'j'\beta'} c_{b'j'\beta'})(y) \\ &\hspace{5cm} + 4 (c_{ai\alpha} f_{bj\beta} c_{ck\gamma} f_{dl\delta})(x) (\bar{f}_{a'i'\alpha'} c_{a'i'\alpha'} \bar{f}_{b'j'\beta'} c_{b'j'\beta'})(y) \\ &\hspace{5cm} + 2 (\bar{c}_{ai\alpha} \bar{c}_{bj\beta} \bar{f}_{ck\gamma} \bar{f}_{dl\delta})(x) (\bar{c}_{a'i'\alpha'} f_{a'i'\alpha'} \bar{c}_{b'j'\beta'} f_{b'j'\beta'})(y) \\ &\hspace{5cm} + 4 (\bar{c}_{ai\alpha} \bar{f}_{bj\beta} \bar{c}_{ck\gamma} \bar{f}_{dl\delta})(x) (\bar{c}_{a'i'\alpha'} f_{a'i'\alpha'} \bar{c}_{b'j'\beta'} f_{b'j'\beta'})(y) \Big] \Big\rangle_f^c \\ &\equiv - \mu^{-\epsilon}\frac{g g_1}{8} \Big[ I_{12,1} + 2I_{12,2} + I_{12,3} + 2I_{12,4} \Big] \, .
\end{split}
\end{align}

\noindent The first integral evaluates to

\begin{align}
\begin{split}
I_{12,1} &= 2 \mu^{-\epsilon} \epsilon_{ab} \epsilon_{cd} \epsilon_{ij} \epsilon_{kl} \epsilon_{\alpha \beta \gamma \delta} \int_{x,y} \delta_{ki'} \delta_{lj'} \delta_{\gamma \alpha'} \delta_{\delta \beta'} (c_{ai \alpha} c_{bj\beta})(x) (c_{a' i' \alpha'} c_{b'j' \beta'})(y) G_{ca'}(x-y) G_{db'}(x-y) \\ &\approx 2 \epsilon_{ab} \epsilon_{cd} \epsilon_{ij} \epsilon_{kl} \epsilon_{\alpha \beta \gamma \delta} \mu^{-\epsilon} \int_p \tilde{G}_{ca'}(p) \tilde{G}_{db'}(-p) \int_x c_{ai \alpha} c_{bj\beta} c_{a' k \gamma} c_{b'l \delta} \\ &= \frac{3}{2\pi} \frac{1}{\epsilon} \epsilon_{ab} \epsilon_{cd} \epsilon_{ij} \epsilon_{kl} \epsilon_{\alpha \beta \gamma \delta} \int_x c_{ai\alpha} c_{bj\beta} c_{ck\gamma} c_{dl\delta} \, ,
\end{split}
\end{align}

\noindent where the last equality follows from using Eq. \ref{Eq:Int_Gp_Gmp}. For $I_{12,2}$, the calculation is very similar and one arrives at

\begin{align} \label{Eq:I_12,2}
\begin{split}
I_{12,2}= \frac{1}{2\pi} \frac{1}{\epsilon} \epsilon_{ab} \epsilon_{cd} \epsilon_{ij} \epsilon_{kl} \epsilon_{\alpha \beta \gamma \delta} \Bigg[ \int_x (c_{ai\alpha} c_{bj\beta}) (c_{ck\gamma} c_{dl\delta}) + 2 \int_x (c_{ai\alpha} c_{bl\delta}) (c_{ck\gamma} c_{dj\beta}) \Bigg] \, .
\end{split}
\end{align}

\noindent The second term can be brought in the desired form by noticing that

\begin{equation} \label{OSMGTilde}
\epsilon_{ab} \epsilon_{cd} \epsilon_{ij} \epsilon_{kl} \epsilon_{\alpha \beta \gamma \delta} (c_{ai\alpha} c_{bl\delta} c_{ck\gamma} c_{dj\beta}) = -\frac{1}{2} \epsilon_{ab} \epsilon_{cd} \epsilon_{ij} \epsilon_{kl} \epsilon_{\alpha \beta \gamma \delta} ( c_{ai\alpha} c_{bj\beta} c_{ck\gamma} c_{dl\delta}) \, .
\end{equation}

\noindent By replacing in Eq. \ref{Eq:I_12,2}, we see that $I_{12,2}$ vanishes. For the two remaining integrals, one can show that $I_{12,3}$ and $I_{12,4}$ are respectively the same as $I_{12,1}$ and $I_{12,2}$, but with the exchange $c_{ai\alpha} \rightarrow \bar{c}_{ai\alpha}$. It thus follows that

\begin{equation}
T_{12}^{(2)} = -\mu^{-\epsilon} \frac{3}{16\pi} g g_1 \frac{1}{\epsilon} \int_x \mathcal{O}_{\text{SMG}}(c) \, .
\end{equation}

The calculation of the last three terms is very analogous as the one for $T_{12}^{(2)}$ and the results are

\begin{equation}
T_{13}^{(2)} = \mu^{-\epsilon}\frac{6}{16\pi} g g_2 \frac{1}{\epsilon} \int_x \mathcal{O}_{\text{SMG}}(c) \, , \quad T_{14}^{(2)} = \mu^{-\epsilon}\frac{3}{16\pi} g g_3 \frac{1}{\epsilon} \int_x \mathcal{O}_{\text{SMG}}(c) \, , \quad T_{15}^{(2)} = -\mu^{-\epsilon}\frac{6}{16\pi} g g_4 \frac{1}{\epsilon} \int_x \mathcal{O}_{\text{SMG}}(c) \, .
\end{equation}

Therefore, by combining the 15 terms of Eq. \ref{Eq:EV_Sint2} and regrouping terms with the same operator, we get

\begin{align}\label{Eq:EV_Sint2_2}
\begin{split}
&\ev{\Big( S_{\text{int}}^{(2)}[s,f] \Big)^2}_f^c \\ &= -\frac{\mu^{-\epsilon}}{\pi} \Bigg[ (N_v N_f-1) g_1^2 - \frac{5}{4} g_2^2 - \frac{5}{4} g_3^2 + \frac{(-5+2N_v N_f)}{4}  g_4^2 + (2N_f - N_v)g_1 g_2 + (2N_v-N_f) g_1 g_3 \\ &\hspace{1cm} + (2-N_v N_f) g_1 g_4 + 2 g_2 g_3 + (N_v-N_f) g_2 g_4 + (N_f-N_v) g_3 g_4 - \frac{13}{2} g^2 \Bigg] \frac{1}{\epsilon} \int_x \mathcal{O}_1(c)
\\ &\hspace{0.3cm}- \frac{\mu^{-\epsilon}}{\pi} \Bigg[ \frac{(N_v+4N_f)}{4} g_2^2 + \frac{N_f}{2} g_3^2 - g_1 g_2 - N_f g_2 g_3 + g_2 g_4 - \frac{3}{2} g_3 g_4 + 7 g^2 \Bigg] \frac{1}{\epsilon} \int_x \mathcal{O}_2(c)
\\ &\hspace{0.3cm}- \frac{\mu^{-\epsilon}}{\pi} \Bigg[ \frac{N_v}{2} g_2^2 + \frac{(4N_v+N_f)}{4} g_3^2 - g_1 g_3 - N_v g_2 g_3 - \frac{3}{2} g_2 g_4 + g_3 g_4 + \frac{13}{2} g^2 \Bigg] \frac{1}{\epsilon} \int_x \mathcal{O}_3(c)
\\ &\hspace{0.3cm}- \frac{\mu^{-\epsilon}}{\pi} \Bigg[ -\frac{1}{2} g_2^2 - \frac{1}{2} g_3^2 + \frac{(2+N_v N_f)}{4} g_4^2 - g_1 g_4 + \frac{N_v}{2} g_2 g_4 + \frac{N_f}{2} g_3 g_4 - 7 g^2 \Bigg] \frac{1}{\epsilon} \int_x \mathcal{O}_4(c)
\\ &\hspace{0.3cm} - \frac{\mu^{-\epsilon}}{\pi} \Bigg[ \frac{3}{8} g g_1 - \frac{6}{8} g g_2 - \frac{3}{8} g g_3 + \frac{6}{8} g g_4 \Bigg] \frac{1}{\epsilon} \int_x \mathcal{O}_{\text{SMG}}
\\ &\equiv -\frac{\mu^{-\epsilon}}{\pi} C_1^{(2)} \frac{1}{\epsilon} \int_x \mathcal{O}_1(c) - \frac{\mu^{-\epsilon}}{\pi} C_2^{(2)}(c) \frac{1}{\epsilon} \int_x \mathcal{O}_2(c) - \frac{\mu^{-\epsilon}}{\pi} C_3^{(2)} \frac{1}{\epsilon} \int_x \mathcal{O}_3(c) \\ &\hspace{0.5cm} - \frac{\mu^{-\epsilon}}{\pi} C_4^{(2)} \frac{1}{\epsilon} \int_x \mathcal{O}_4(c) - \frac{\mu^{-\epsilon}}{\pi} C_{\text{SMG}}^{(2)} \frac{1}{\epsilon} \int_x \mathcal{O}_{\text{SMG}}(c) \, ,
\end{split}
\end{align}

\noindent where the coefficients $C^{(2)}_i$ (with $i=1,...,4$) and $C_{\text{SMG}}^{(2)}$ correspond to the brackets in the previous equality.

\subsection{2-loop correction to the propagator}

We now move on to the evaluation of the second expectation value in Eq. \ref{Eq:SEff}, which renormalizes the propagator. Expanding yields the following 11 terms

\begin{align} \label{Eq:EV_Sint3}
\begin{split}
\ev{\Big( S_{\text{int}}^{(3)}[c,f] \Big)^2}_f^c &= \ev{\Big( S_{\text{int},1}^{(3)} + S_{\text{int},2}^{(3)} + S_{\text{int},3}^{(3)} + S_{\text{int},4}^{(3)} + S_{\text{int,SMG}}^{(3)} \Big)^2}_f^c \\ &= \ev{\Big( S_{\text{int},1}^{(3)} \Big)^2}_f^c + \ev{\Big( S_{\text{int},2}^{(3)} \Big)^2}_f^c + \ev{\Big( S_{\text{int},3}^{(3)} \Big)^2}_f^c + \ev{\Big( S_{\text{int},4}^{(3)} \Big)^2}_f^c \\ &\hspace{0.5cm} + 2 \ev{S_{\text{int},1}^{(3)}S_{\text{int},2}^{(3)}}_f^c + 2 \ev{S_{\text{int},1}^{(3)}S_{\text{int},3}^{(3)}}_f^c + 2 \ev{S_{\text{int},1}^{(3)}S_{\text{int},4}^{(3)}}_f^c \\ &\hspace{0.5cm} + 2 \ev{S_{\text{int},2}^{(3)}S_{\text{int},3}^{(3)}}_f^c + 2 \ev{S_{\text{int},2}^{(3)}S_{\text{int},4}^{(3)}}_f^c + 2 \ev{S_{\text{int},3}^{(3)}S_{\text{int},4}^{(3)}}_f^c + \ev{\Big( S_{\text{int,SMG}}^{(3)} \Big)^2}_f^c \\ &\equiv T_1^{(3)} + T_2^{(3)} + T_3^{(3)} + T_4^{(3)} + 2 T_5^{(3)} + 2 T_6^{(3)} + 2 T_7^{(3)} + 2 T_8^{(3)}+ 2T_9^{(3)} + 2 T_{10}^{(3)} + T_{11}^{(3)} \, ,
\end{split}
\end{align}

\noindent where the four cross-terms with $S^{(3)}_{\text{int,SMG}}$ vanish, which can be seen from Eq. \ref{Eq:Sint3cf}. Note that for the first ten terms, we keep working with general $N_v$ and $N_f$.

Let us start with the first term

\begin{align}
\begin{split}
T_1^{(3)} &= \ev{\Big( S_{\text{int},1}^{(3)}[c,f] \Big)^2}_f^c \\ &= 2 \mu^{-2\epsilon} g_1^2 \int_{x,y} \ev{(\bar{c}_{ai\alpha} f_{ai\alpha} \bar{f}_{bj\beta} f_{bj\beta})(x) (\bar{f}_{a'i'\alpha'} f_{a' i' \alpha'} \bar{f}_{b'j'\beta'} c_{b'j'\beta'})(y)}_f^c \\ &= 2\mu^{-2\epsilon} g_1^2 \int_{x,y} \Bigg( -N_v N_f \Tr[G(x-y) G(y-x)] \bar{c}_{i\alpha}(x) G(x-y) c_{i\alpha}(y) +\bar{c}_{i\alpha}(x) G(x-y) G(y-x) G(x-y) c_{i\alpha}(y) \Bigg) \\ &\equiv - 2 N_v N_f g_1^2 I_1^{(3)} + 2g_1^2 I_2^{(3)} \, ,
\end{split}
\end{align}

\noindent where the following two integrals have been defined

\begin{align}
\begin{split}
I_1^{(3)} &= \mu^{-2\epsilon} \int_{x,y} \Tr[G(x-y) G(y-x)] \bar{c}_{i\alpha}(x) G(x-y) c_{i\alpha}(y) \, , \\
I_2^{(3)} &= \mu^{-2\epsilon} \int_{x,y}\bar{c}_{i\alpha}(x) G(x-y) G(y-x) G(x-y) c_{i\alpha}(y) \, .
\end{split}
\end{align}

\noindent Diagrammatically, $T_1^{(3)}$ can be represented by Fig. \ref{Fig:Feynman_diagram_3} (b). The goal is now to compute these two integrals to extract the $1/\epsilon$ UV divergence. Note that in order to regulate the IR, it will be convenient to work with the modified propagator \cite{bondi1990metric,ladovrechis2023gross}

\begin{equation}
\Tilde{G}(p) = -\frac{\ii \slashed{p}}{p^2+m^2} \, ,
\end{equation}

\noindent where $m$ will be sent to zero at the end of the calculation.

Let us begin with the evaluation of $I_1^{(3)}$. Fourier transforming yields

\begin{equation}
I_1^{(3)} = \mu^{-2\epsilon} \int_q \int_k \Bigg( \int_p \Tr[\Tilde{G}(p-k+q) \Tilde{G}(p)] \Bigg) \bar{c}_{i\alpha}(q) \Tilde{G}(k) c_{i\alpha}(q) \, .
\end{equation}

\noindent We first focus on the integral over $p$

\begin{align}
\begin{split}
\int_p \Tr[\Tilde{G}(p-k+q) \Tilde{G}(p)] = \int_p \Tr[\frac{-\I(\slashed{p}-\slashed{k}+\slashed{q})}{(p-k+q)^2 + m^2} \frac{-\ii \slashed{p}}{p^2+m^2}]= - D \int_p \frac{p^2 + (q-k)\cdot p}{[(p-k+q)^2+m^2][p^2+m^2]} \, ,
\end{split}
\end{align}

\noindent which results from using $\Tr[\gamma_{\mu} \gamma_{\nu}] = D \delta_{\mu \nu}$. We now define $\Delta_1 = x(1-x)(q-k)^2 + m^2$, $l = p+x(q-k)$ and use a Feynman parameter $(x)$, resulting in

\begin{align} \label{Eq:Int_TrGG_sunset}
\begin{split}
\int_p \Tr[\Tilde{G}(p-k+q) \Tilde{G}(p)] &= -D \int_0^1 dx \int_l \frac{l^2-x(1-x)(q-k)^2}{[l^2+\Delta_1]^2} \\ &= -\frac{D}{(4\pi)^{D/2}} \int_0^1 dx \Bigg[ \frac{D}{2} \Gamma\Big(1-\frac{D}{2}\Big) \Delta_1^{\frac{D}{2}-1} - x(1-x) (q-k)^2 \Gamma\Big(2-\frac{D}{2}\Big) \Delta_1^{\frac{D}{2}-2} \Bigg] \\ &= -\frac{D}{(4\pi)^{D/2}} \int_0^1 dx \Bigg[ \frac{D}{2} \Gamma\Big(1-\frac{D}{2}\Big) \Delta_1^{\frac{D}{2}-1} - (\Delta_1-m^2) \Gamma\Big(2-\frac{D}{2}\Big) \Delta_1^{\frac{D}{2}-2} \Bigg] \\ &= -\frac{D(D-1)}{(4\pi)^{D/2}} \Gamma\Big( 1-\frac{D}{2} \Big) \int_0^1 dx \Delta_1^{\frac{D}{2}-1} \, ,
\end{split}
\end{align}

\noindent where the integrals over $l$ have been performed using the usual general formula

\begin{equation} \label{Eq:Int_l_general}
\int \frac{d^dl}{(2\pi)^D} \frac{(l^2)^m}{(l^2 + \Delta)^n} = \frac{1}{(4\pi)^{D/2}}  \frac{\Gamma(m+D/2) \Gamma(n-m-D/2)}{\Gamma(n) \Gamma(D/2)} \Delta^{D/2+m-n} \, .
\end{equation}

\noindent The third equality follows from the definition of $\Delta_1$, while the term with $m^2$ at the numerator can be dropped. The integral $I_1^{(3)}$ thus becomes

\begin{align}
\begin{split}
I_1^{(3)} &= -\frac{D(D-1)}{(4\pi)^{D/2}} \mu^{-2\epsilon} \Gamma\Big( 1-\frac{D}{2} \Big) \int_q  \int_k \bar{c}_{i\alpha}(q) \Tilde{G}(k) c_{i\alpha}(q) \int_x \Delta_1^{\frac{D}{2}-1} \\ &= \frac{D(D-1)}{(4\pi)^{D/2}} \mu^{-2\epsilon} \Gamma\Big( 1-\frac{D}{2} \Big) \int_q \bar{c}_{i\alpha}(q) \ii \gamma_{\mu} c_{i\alpha}(q) \int_0^1 dx \int_k \Delta_1^{\frac{D}{2}-1} \frac{k_{\mu}}{k^2+m^2} \, .
\end{split}
\end{align}

\noindent Let us now compute the integral over $k$

\begin{align}\label{Eq:Int_k1}
\begin{split}
\int_k \Delta_1^{\frac{D}{2}-1} \frac{k_{\mu}}{k^2+m^2} &= [x(1-x)]^{\frac{D}{2}-1} \int_k \frac{k_{\mu}}{\big[(k-q)^2+\frac{m^2}{x(1-x)}\big]^{1-\frac{D}{2}} [k^2+m^2]} \\ &= [x(1-x)]^{\frac{D}{2}-1} \int_0^1 dy \frac{\Gamma\big( 2-\frac{D}{2} \big)}{\Gamma\big( 1-\frac{D}{2} \big)} y^{-\frac{D}{2}} \int_l \frac{y q_{\mu}}{[l^2+\Tilde{\Delta}_1]^{2-\frac{D}{2}}} \\ &= \frac{[x(1-x)]^{\frac{D}{2}-1}}{(4\pi)^{D/2}} \frac{\Gamma(2-D)}{\Gamma\big( 1-\frac{D}{2} \big)} q_{\mu} \int_0^1 dy y^{1-\frac{D}{2}} \Tilde{\Delta}_1^{D-2} \, ,
\end{split}
\end{align}

\noindent where $y$ is a second Feynman parameter and $\tilde{\Delta}_1 = y(1-y)q^2 + \big[ \frac{y}{x(1-x)}+1-y \big]m^2$. Replacing in the expression for $I_1^{(3)}$, sending $m \rightarrow 0$ and expanding at leading order in $\epsilon$ finally gives

\begin{equation}
I_1^{(3)} = \frac{D(D-1)}{(4\pi)^D} \mu^{-2\epsilon} \Gamma(2-D) \int_q \bar{c}_{i\alpha}(q) \ii \slashed{q} c_{i\alpha}(q) \int_0^1 dx dy [x(1-x)]^{\frac{D}{2}-1} y^{1-\frac{D}{2}} \Tilde{\Delta}_1^{D-2} = -\frac{2}{(4\pi)^2} \frac{1}{\epsilon} \int_x \bar{c}_{i\alpha} \slashed{\partial} c_{i\alpha} \, .
\end{equation}

Moving on to $I_2^{(3)}$, Fourier transforming yields

\begin{equation}
I_2 = \mu^{-2\epsilon} \int_q \int_k \bar{c}_{i\alpha}(q) \Bigg[ \int_p \Tilde{G}(p-k+q) \Tilde{G}(p) \Bigg] \Tilde{G}(k) c_{i\alpha}(q) \, .
\end{equation}

\noindent Consider the integral over $p$

\begin{align}
\begin{split}
\int_p \Tilde{G}(p-k+q) \Tilde{G}(p) &= \int_p \frac{-\I(\slashed{p}-\slashed{k}+\slashed{q})}{(p-k+q)^2+m^2} \frac{-\ii \slashed{p}}{p^2+m^2} \\ &= -\int_p \frac{p^2+(\slashed{q}-\slashed{k})\slashed{p}}{[(p-k+q)^2+m^2][p^2+m^2]} \\ &= - \int_p \int_0^1 dx \frac{[l-x(q-k)]^2 + (\slashed{q}-\slashed{k})[\slashed{l}-x(\slashed{q}-\slashed{k})]}{[l^2+\Delta_1]^2} \\ &= - \int_p \int_0^1 dx \frac{l^2-x(1-x)(q-k)^2}{[l^2+\Delta_1]^2} \, ,
\end{split}
\end{align}

\noindent which is the same as Eq. \ref{Eq:Int_TrGG_sunset}, up to a factor of $D=2+\mathcal{O}(\epsilon)$. From this, we conclude that

\begin{equation}
I_2^{(3)} = -\frac{1}{(4\pi)^2} \frac{1}{\epsilon} \int_x \bar{c}_{i\alpha} \slashed{\partial} c_{i\alpha} \, .
\end{equation}

\noindent Therefore, the first term contributing to the renormalization of the propagator is

\begin{equation}
T_1^{(3)} = \frac{2(2N_v N_f-1)}{(4\pi)^2} g_1^2 \frac{1}{\epsilon} \int_x \bar{c}_{i\alpha} \slashed{\partial} c_{i\alpha} \, .
\end{equation}

Following the same steps as above and using the expressions for $I_1^{(3)}$ and $I_2^{(3)}$, one can show that the three following terms are exactly the same as $T_1^{(3)}$, but with a different coupling, that is

\begin{equation}
T_i^{(3)} = \frac{2(2N_v N_f-1)}{(4\pi)^2} g_i^2 \frac{1}{\epsilon} \int_x \bar{c}_{i\alpha} \slashed{\partial} c_{i\alpha} \, , \hspace{0.3cm} \text{for} \hspace{0.1cm} i=2,3,4 \, .
\end{equation}

The next six terms have a similar structure. Let us give some details for the calculation of $T_5^{(3)}$

\begin{align}
\begin{split}
T_5^{(3)} &= \ev{S_{\text{int},1}^{(3)}[c,f] S_{\text{int},2}^{(3)}[c,f]}_f^c \\ &= \mu^{-2\epsilon} g_1 g_2 \int_{x,y} \Big[ \ev{(\bar{c}_{ai\alpha} f_{ai\alpha} \bar{f}_{bj\beta} f_{bj\beta})(x) (\bar{f}_{a'i'\alpha'} f_{a'j'\alpha'} \bar{f}_{b'j'\beta'} c_{b'i'\beta'})(y)}_f^c \\&\hspace{2cm}+ \ev{(\bar{f}_{ai\alpha} f_{ai\alpha} \bar{f}_{bj\beta} c_{bj\beta})(y) (\bar{c}_{a'i'\alpha'} f_{a'j'\alpha'} \bar{f}_{b'j'\beta'} f_{b'i'\beta'})(x)}_f^c \Big] \\ &= 2 \mu^{-2\epsilon} g_1 g_2 \int_{x,y} \Bigg( N_v \bar{c}_{i\alpha}(x) G(x-y) G(y-x) G(x-y) c_{i\alpha}(y) \\ &\hspace{2.5cm} - N_f \Tr[G(x-y) G(y-x)] \bar{c}_{i\alpha}(x) G(x-y) c_{i\alpha}(y) \Bigg) \\ &= -2 g_1 g_2 \Big( N_f I_1^{(3)} - N_v I_2^{(3)} \Big) \\ &= \frac{2(2N_f-N_v)}{(4\pi)^2} g_1 g_2 \frac{1}{\epsilon} \int_x \bar{c}_{i\alpha} \slashed{\partial} c_{i\alpha} \, .
\end{split}
\end{align}

\noindent The other five terms are computed in a very similar way, yielding

\begin{align}
\begin{split}
T_6^{(3)} &= \frac{2(2N_v-N_f)}{(4\pi)^2} g_1 g_3 \frac{1}{\epsilon} \int_x \bar{c}_{i\alpha} \slashed{\partial} c_{i\alpha} \, , \qquad T_7^{(3)} = \frac{2(2-N_v N_f)}{(4\pi)^2} g_1 g_4 \frac{1}{\epsilon} \int_x \bar{c}_{i\alpha} \slashed{\partial} c_{i\alpha} \, , \\ T_8^{(3)} &= \frac{2(2-N_v N_f)}{(4\pi)^2} g_2 g_3 \frac{1}{\epsilon} \int_x \bar{c}_{i\alpha} \slashed{\partial} c_{i\alpha} \, , \qquad T_9^{(3)} = \frac{2(2N_v- N_f)}{(4\pi)^2} g_2 g_4 \frac{1}{\epsilon} \int_x \bar{c}_{i\alpha} \slashed{\partial} c_{i\alpha} \, ,  \\ T_{10}^{(3)} &= \frac{2(2N_f- N_v)}{(4\pi)^2} g_3 g_4 \frac{1}{\epsilon} \int_x \bar{c}_{i\alpha} \slashed{\partial} c_{i\alpha} \, .
\end{split}
\end{align}

Finally, we arrive at the calculation of $T_{11}^{(3)}$, which is represented by diagram (c) of Fig. \ref{Fig:Feynman_diagram_3}

\begin{align}
\begin{split}
T_{11}^{(3)} &= 2\mu^{-2\epsilon} \frac{g^2}{4} \epsilon_{ab} \epsilon_{cd} \epsilon_{ij} \epsilon_{kl} \epsilon_{\alpha \beta \gamma \delta} \epsilon_{a'b'} \epsilon_{c'd'} \epsilon_{i'j'} \epsilon_{k'l'} \epsilon_{\alpha'\beta'\gamma'\delta'} \int_{x,y} \\ &\hspace{2cm} \times \ev{(c_{ai\alpha} f_{bj\beta} f_{ck\gamma} f_{dl\delta})(x) (\bar{c}_{a'i'\alpha'} \bar{f}_{b'j'\beta'} \bar{f}_{c'k'\gamma'} \bar{f}_{d'l'\delta'})(y)}_f^c \\ &= 2\mu^{-2\epsilon} \frac{g^2}{2} \epsilon_{ab} \epsilon_{cd} \epsilon_{ij} \epsilon_{kl} \epsilon_{\alpha \beta \gamma \delta} \epsilon_{a'b'} \epsilon_{c'd'} \epsilon_{i'j'} \epsilon_{k'l'} \epsilon_{\alpha'\beta'\gamma'\delta'} \int_{x,y} \\ &\hspace{1cm} \Bigg[ -\delta_{jj'} \delta_{\beta\beta'} \delta_{kk'} \delta_{\gamma\gamma'} \delta_{ll'} \delta_{\delta\delta'} G_{bb'}(x-y) G_{cc'}(x-y) G_{dd'}(x-y) \bar{c}_{a' i'\alpha'}(y) c_{ai\alpha}(x) \\ &\hspace{1.3cm} +2 \delta_{jk'} \delta_{\beta\gamma'} \delta_{kj'} \delta_{\gamma \beta'} \delta_{ll'} \delta_{\delta \delta'} G_{bc'}(x-y) G_{cb'}(x-y) G_{dd'}(x-y) \bar{c}_{a' i'\alpha'}(y) c_{ai\alpha}(x) \Bigg] \\ &= -12 \mu^{-2\epsilon} g^2 \epsilon_{ab} \epsilon_{cd} \epsilon_{a'b'} \epsilon_{c'd'} \int_{x,y} \Bigg[ \bar{c}_{a'i\alpha}(y) c_{ai\alpha}(x) G_{bb'}(x-y) G_{cc'}(x-y) G_{dd'}(x-y) \\ &\hspace{5cm} + \bar{c}_{a'i\alpha}(y) c_{ai\alpha}(x) G_{bc'}(x-y) G_{cb'}(x-y) G_{dd'}(x-y) \Bigg] \, .
\end{split}
\end{align}

\noindent where the last equality follows from using the Levi-Civita identities 

\begin{equation}
\epsilon_{ij} \epsilon_{ij} = 2 \, , \quad \epsilon_{ik} \epsilon_{jk} = \delta_{ij} \, , \quad \epsilon_{\alpha \beta \gamma \delta} \epsilon_{\alpha' \beta \gamma \delta} = 6 \delta_{\alpha \alpha'} \, .
\end{equation}

\noindent Note that the second term in the previous expression is the same as the first one after performing the exchange of indices $b' \leftrightarrow c'$. Hence, only the first one needs to be computed. Let us define the following integral, which is then written in momentum space

\begin{align}
\begin{split}
I_{11}^{(3)} &= \mu^{-2\epsilon} \int_{x,y} \bar{c}_{a'i\alpha}(y) c_{ai\alpha}(x) G_{bb'}(x-y) G_{cc'}(x-y) G_{dd'}(x-y) \\ &= \mu^{-2\epsilon} \int_q \bar{c}_{a'i\alpha}(q) c_{ai\alpha}(q) \int_k \Tilde{G}_{dd'}(k) \int_p \Tilde{G}_{bb'}(-p-k-q) \Tilde{G}_{cc'}(p) \, .
\end{split}
\end{align}

\noindent The integral over $p$ is performed as previously, even if the spinor structure is different

\begin{align}
\begin{split}
\int_p \Tilde{G}_{bb'}(-p-k-q) \Tilde{G}_{cc'}(p) &= \gamma_{bb'}^{\mu} \gamma_{cc'}^{\nu} \int_p \frac{(p+k+q)_{\mu} p_{\nu}}{[(p+k+q)^2+m^2][p^2+m^2]} \\ &= \gamma_{bb'}^{\mu} \gamma_{cc'}^{\nu} \int_0^1 dx \int_l \frac{l_{\mu}l_{\nu} - x(1-x)(k+q)_{\mu} (k+q)_{\nu}}{[l^2+\Delta_2]^2} \, ,
\end{split}
\end{align}

\noindent where this time $\Delta_2=x(1-x)(k+q)^2+m^2$ and $l=p+x(k+q)$. Making the replacement $l_{\mu} l_{\nu} \rightarrow \frac{l^2}{D}\delta_{\mu\nu}$ and computing the integrals over $l$ results in 

\begin{equation}
\int_p \Tilde{G}_{bb'}(-p-k-q) \Tilde{G}_{cc'}(p) = \gamma_{bb'}^{\mu} \gamma_{cc'}^{\nu} \frac{1}{(4\pi)^{D/2}} \int_0^1 dx \Bigg[ \delta_{\mu\nu} \frac{1}{2} \Gamma\Big( 1-\frac{D}{2} \Big) \Delta_2^{\frac{D}{2}-1} - x(1-x) (k+q)_{\mu} (k+q)_{\nu} \Gamma\Big( 2-\frac{D}{2} \Big) \Delta_2^{\frac{D}{2}-2} \Bigg] \, .
\end{equation}

\noindent Replacing in the expression for $I_{11}^{(3)}$ then gives

\begin{align} \label{Eq:I_11_3}
\begin{split}
I_{11}^{(3)} &= -\frac{\ii \mu^{-2\epsilon}}{(4\pi)^{D/2}} \gamma_{bb'}^{\mu} \gamma_{cc'}^{\nu} \gamma_{dd'}^{\lambda} \int_q \bar{c}_{a'i\alpha}(q) c_{ai\alpha}(q) \int_0^1 dx \Bigg[ \frac{1}{2} \delta_{\mu \nu} \Gamma\Big( 1-\frac{D}{2} \Big) \int_k \Delta_2^{\frac{D}{2}-1} \frac{k_{\lambda}}{k^2+m^2} \\ &\hspace{7cm}-x (1-x) \Gamma\Big( 2-\frac{D}{2} \Big) \int_k \Delta_2^{\frac{D}{2}-2} \frac{(k+q)_{\mu} (k+q)_{\nu} k_{\lambda}}{k^2+m^2} \Bigg] \, .
\end{split}
\end{align}

\noindent The last step is then to compute the two integrals over $k$. The first one is the same as Eq. \ref{Eq:Int_k1}, but with $q\rightarrow -q$ and thus

\begin{equation}
\int_k \Delta_2^{\frac{D}{2}-1} \frac{k_{\lambda}}{k^2+m^2} = -\frac{[x(1-x)]^{\frac{D}{2}-1}}{(4\pi)^{D/2}} \frac{\Gamma(2-D)}{\Gamma\big( 1-\frac{D}{2} \big)} q_{\mu} \int_0^1 dy y^{1-\frac{D}{2}} \Tilde{\Delta}_2^{D-2} \, ,
\end{equation}

\noindent with $\Tilde{\Delta}_2=\Tilde{\Delta}_1$. For the second integral over $k$, introducing a second Feynman parameter $(y)$, shifting the integrand and computing the integrals over $l$ using Eq. \ref{Eq:Int_l_general} results in

\begin{align}
\begin{split}
\int_k \Delta_2^{\frac{D}{2}-2} \frac{(k+q)_{\mu} (k+q)_{\nu} k_{\lambda}}{k^2+m^2} &= \frac{[x(1-x)]^{\frac{D}{2}-2}}{(4\pi)^{\frac{D}{2}} \Gamma\big( 2-\frac{D}{2} \big)} \int_0^1 dy \Bigg[ \frac{1}{2} y^{1-\frac{D}{2}}(1-y) (\delta_{\mu \lambda} q_{\nu} + \delta_{\nu \lambda} q_{\mu}) \Gamma(2-D) \Tilde{\Delta}_2^{D-2} \\ &\hspace{1cm} - \frac{1}{2} y^{2-\frac{D}{2}} \delta_{\mu \nu} q_{\lambda} \Gamma(2-D) \Tilde{\Delta}^{D-2} - y^{2-\frac{D}{2}}(1-y)^2 q_{\mu} q_{\nu} q_{\lambda} \Gamma(3-D) \Tilde{\Delta}^{D-3} \Bigg] \, .
\end{split}
\end{align}

\noindent The third term is finite and can be dropped. Using the result for the two integrals over $k$, replacing in Eq. \ref{Eq:I_11_3} and isolating the $1/\epsilon$ pole finally yields

\begin{equation}
I_{11}^{(3)} = -\frac{\I}{4(4\pi)^2} \frac{1}{\epsilon} \gamma_{bb'}^{\mu} \gamma_{cc'}^{\nu} \gamma_{dd'}^{\lambda} \int_q \bar{c}_{a'i\alpha}(q) c_{ai\alpha}(q) \Big[ \delta_{\mu \nu} q_{\lambda} + \delta_{\mu \lambda} q_{\nu} + \delta_{\nu \lambda} q_{\mu} \Big] \, .
\end{equation}

\noindent It then follows that

\begin{align}
\begin{split}
T_{11}^{(3)} &= -12 g^2 \epsilon_{ab} \epsilon_{cd} \epsilon_{a'b'} \epsilon_{c'd'} \Big( I_{11}^{(3)} + I_{11}^{(3)}(b'\leftrightarrow c') \Big) \\ &= \frac{3\I}{(4\pi)^2} g^2 \frac{1}{\epsilon} \epsilon_{ab} \epsilon_{cd} \epsilon_{a'b'} \epsilon_{c'd'} \Big( \gamma_{bb'}^{\mu} \gamma_{cc'}^{\nu} \gamma_{dd'}^{\lambda} + \gamma_{bc'}^{\mu} \gamma_{cb'}^{\nu} \gamma_{dd'}^{\lambda} \Big) \int_q \bar{c}_{a'i\alpha}(q) c_{ai\alpha}(q) \Big[ \delta_{\mu \nu} q_{\lambda} + \delta_{\mu \lambda} q_{\nu} + \delta_{\nu \lambda} q_{\mu} \Big] \\ &= \frac{3\I}{(4\pi)^2} g^2 \frac{1}{\epsilon} \epsilon_{ab} \epsilon_{cd} \epsilon_{a'b'} \epsilon_{c'd'} \int_q \bar{c}_{a'i\alpha}(q) c_{ai\alpha}(q) \Big[ \gamma_{bb'}^{\mu} \gamma_{cc'}^{\mu} (\slashed{q})_{dd'} + \gamma_{bb'}^{\mu} \gamma_{dd'}^{\mu} (\slashed{q})_{cc'} + \gamma_{cc'}^{\mu} \gamma_{dd'}^{\mu} (\slashed{q})_{bb'} \\ &\hspace{7cm} + \gamma_{bc'}^{\mu} \gamma_{cb'}^{\mu} (\slashed{q})_{dd'} + \gamma_{bc'}^{\mu} \gamma_{dd'}^{\mu} (\slashed{q})_{cb'} + \gamma_{cb'}^{\mu} \gamma_{dd'}^{\mu} (\slashed{q})_{bc'} \Big] \, .
\end{split}
\end{align}

\noindent Using Eq. \ref{Eq:Completeness_gamma_matrix} in the six terms and simplifying the resulting expression finally leads to

\begin{equation}
T_{11}^{(3)} = \frac{3}{(4\pi)^2} g^2 \frac{1}{\epsilon} 15 \int_q \bar{c}_{i\alpha} \slashed{q} c_{i\alpha}(q) = \frac{45}{(4\pi)^2} g^2 \frac{1}{\epsilon} \int_x \bar{c}_{i\alpha} \slashed{\partial} c_{i\alpha} \, .
\end{equation}

Hence, combing the 11 terms/diagrams of Eq. \ref{Eq:EV_Sint3}, we get that

\begin{align}\label{Eq:EV_Sint3_2}
\begin{split}
\ev{\Big( S_{\text{int}}^{(3)}[c,f] \Big)^2}_f^c &= \frac{1}{(4\pi)^2} \Bigg[ 2(2N_v N_f-1) (g_1^2+g_2^2+g_3^2+g_4^2) + 4(2N_f-N_v)(g_1g_2+g_3g_4) \\ &\hspace{0.5cm}+ 4(2N_v-N_f)(g_1g_3+g_2g_4) +4(2-N_vN_f)(g_1g_4+g_2g_3) + 45 g^2 \Bigg] \frac{1}{\epsilon} \int_x \bar{c}_{i\alpha} \slashed{\partial} c_{i\alpha} \\ &\equiv \frac{1}{(4\pi)^2} C_{\psi}^{(2)} \frac{1}{\epsilon} \int_x \bar{c}_{i\alpha} \slashed{\partial} c_{i\alpha} \, ,
\end{split}
\end{align}

\noindent where $C_{\psi}^{(2)}$ denotes the bracket in the previous line.

\subsection{Calculation of RG functions}

The effective action at quadratic order in the cumulant expansion is then

\begin{align}
\begin{split}
\Gamma[c] &= Z_{\psi} \int_x \bar{c}_{i\alpha} \slashed{\partial} c_{i\alpha} - Z_1 \mu^{-\epsilon} \frac{g_1}{2} \int_x \mathcal{O}_1(c) - Z_2 \mu^{-\epsilon} \frac{g_2}{2} \int_x \mathcal{O}_2(c) - Z_3 \mu^{-\epsilon} \frac{g_3}{2} \int_x \mathcal{O}_3(c) \\ &\hspace{0.5cm} - Z_4 \mu^{-\epsilon} \frac{g_4}{2} \int_x \mathcal{O}_4(c) + Z_g \mu^{-\epsilon} \frac{g}{8} \int_x \mathcal{O}_{\text{SMG}}(c) - \frac{1}{2} \ev{\Big( S_{\text{int}}^{(2)}[c,f] \Big)^2}_f^c - \frac{1}{2} \ev{\Big( S_{\text{int}}^{(3)}[c,f] \Big)^2}_f^c \, ,
\end{split}
\end{align}

\noindent where recall that the two expectation values are given respectively by Eq. \ref{Eq:EV_Sint2_2} and \ref{Eq:EV_Sint3_2}. By canceling the UV divergences, the renormalization constants are identified to be

\begin{equation}
Z_{\psi} = 1 + \frac{C_{\psi}^{(2)}}{2(4\pi)^2} \frac{1}{\epsilon} \, , \quad Z_i = 1 + \frac{C_i^{(2)}}{\pi g_i} \frac{1}{\epsilon} \, , \quad Z_g = 1 - \frac{4 C_{\text{SMG}}^{(2)}}{\pi g} \frac{1}{\epsilon} \, ,
\end{equation}

\noindent with $i=1,2,3,4$. From Eq. \ref{Eq:Anomalous_dim}, we first get the fermion anomalous dimension

\begin{align} \label{Eq:eta_psi}
\begin{split}
\gamma_{\psi} &= \frac{1}{(4\pi)^2} \Big[ 2(2N_v N_f-1) (g_1^2+g_2^2+g_3^2+g_4^2) + 4(2N_f-N_v)(g_1 g_2+g_3 g_4) \\ &\hspace{2cm} + 4 (2N_v-N_f) (g_1 g_3+g_2 g_4) + 4(2-N_v N_f) (g_1 g_4+g_2 g_3) + 45 g^2 \Big] \, .
\end{split}
\end{align}

\noindent For the $\beta$ functions, the other anomalous dimensions are first computed and we then use Eq. \ref{Eq:Beta_functions}. Since $\gamma_{\psi}$ is quadratic in the couplings, it only contributes to the $\beta$ functions at 2-loop and can thus be dropped. Hence, we have at 1-loop

\begin{align}\label{Eq:beta_g1}
\begin{split}
\beta(g_1) &= -\epsilon g_1 + \frac{1}{\pi} \Bigg[ (N_v N_f-1) g_1^2 - \frac{5}{4} g_2^2 - \frac{5}{4} g_3^2 + \frac{(-5+2N_v N_f)}{4}  g_4^2 + (2N_f - N_v)g_1 g_2 + (2N_v-N_f) g_1 g_3 \\ &\hspace{2cm} + (2-N_v N_f) g_1 g_4 + 2 g_2 g_3 + (N_v-N_f) g_2 g_4 + (N_f-N_v) g_3 g_4 - \frac{13}{2} g^2 \Bigg] \, ,
\end{split}
\end{align}

\begin{align}\label{Eq:beta_g2}
\begin{split}
\beta(g_2) &= -\epsilon g_2 + \frac{1}{\pi} \Bigg[ \frac{(N_v+4N_f)}{4} g_2^2 + \frac{N_f}{2} g_3^2 - g_1 g_2 - N_f g_2 g_3 + g_2 g_4 - \frac{3}{2} g_3 g_4 + 7 g^2 \Bigg] \, ,
\end{split}
\end{align}

\begin{align}\label{Eq:beta_g3}
\begin{split}
\beta(g_3) &= -\epsilon g_3 + \frac{1}{\pi} \Bigg[ \frac{N_v}{2} g_2^2 + \frac{(4N_v+N_f)}{4} g_3^2 - g_1 g_3 - N_v g_2 g_3 - \frac{3}{2} g_2 g_4 + g_3 g_4 + \frac{13}{2} g^2 \Bigg] \, ,
\end{split}
\end{align}

\begin{align}\label{Eq:beta_g4}
\begin{split}
\beta(g_4) &= -\epsilon g_4 + \frac{1}{\pi} \Bigg[ -\frac{1}{2} g_2^2 - \frac{1}{2} g_3^2 + \frac{(2+N_v N_f)}{4} g_4^2 - g_1 g_4 + \frac{N_v}{2} g_2 g_4 + \frac{N_f}{2} g_3 g_4 - 7 g^2 \Bigg] \, ,
\end{split}
\end{align}

\begin{align}\label{Eq:beta_g}
\begin{split}
\beta(g) = -\epsilon g - \frac{1}{2\pi} \Big[ 3 g g_1 -6 g g_2 - 3 g g_3 + 6 g g_4 \Big] \, .
\end{split}
\end{align}

The last step is to obtain the $\beta$ functions for the five initial operators (Eq. \ref{Eq:Order_parameter_operators}). To do so, one can use the Fierz identities of Eq. \ref{Eq:Fierz_identities} to express $\mathcal{O}_2$, $\mathcal{O}_3$ and $\mathcal{O}_4$ in terms of the initial operators, resulting in

\begin{align}
\begin{split}
\mathcal{O}_2 = \frac{1}{N_v} \mathcal{O}_1 + \frac{1}{2} \mathcal{O}_{\sigma} \, , \qquad \mathcal{O}_3 = \frac{1}{N_f} \mathcal{O}_1 + \frac{1}{2} \mathcal{O}_{T} \, , \qquad \mathcal{O}_4 = \frac{1}{N_v N_f} \mathcal{O}_1 + \frac{1}{2N_f} \mathcal{O}_{\sigma} + \frac{1}{2N_v} \mathcal{O}_T + \frac{1}{4} \mathcal{O}_{\sigma T} \, .
\end{split}
\end{align}

\noindent We can then replace these in the interaction Lagrangian and regroup terms with the same operator

\begin{align}
\begin{split}
\mathcal{L}_{\text{int}} = -\frac{1}{2} \Bigg( g_1 + \frac{g_2}{N_v} + \frac{g_3}{N_f} + \frac{g_4}{N_v N_f} \Bigg) \mathcal{O}_1 - \frac{1}{2} \Bigg( \frac{g_2}{2} + \frac{g_4}{2N_f} \Bigg) \mathcal{O}_{\sigma} - \frac{1}{2} \Bigg( \frac{g_3}{2} + \frac{g_4}{2N_v} \Bigg) \mathcal{O}_T - \frac{1}{2} \Bigg( \frac{g_4}{4} \Bigg) \mathcal{O}_{\sigma T} + \frac{1}{8} g \mathcal{O}_{\text{SMG}} \, ,
\end{split}
\end{align}

\noindent from which we identify the couplings in this new basis

\begin{align}
\begin{split}
G_1 = g_1 + \frac{g_2}{N_v} + \frac{g_3}{N_f} + \frac{g_4}{N_v N_f} \, , \qquad G_{\sigma} = \frac{g_2}{2} + \frac{g_4}{2N_f} \, , \qquad G_T = \frac{g_3}{2} + \frac{g_4}{2N_v} \, , \qquad G_{\sigma T} = \frac{g_4}{4} \, , \qquad G = g \, .
\end{split}
\end{align}

\noindent Knowing this, the $\beta$ functions in the original basis are thus (with $N_v=2$, $N_f=4$)

\begin{align}
\begin{split}
\beta(G_1) &= -\epsilon G_1 + \frac{1}{\pi} \Bigg[ 
7 G_1^2 + \frac{3}{4} G_{\sigma}^2 + \frac{15}{4} G_T^2 + \frac{495}{16} G_{\sigma T}^2 - 3 G_1 G_{\sigma} - \frac{15}{2} G_1 G_T \\ &\hspace{2.5cm}  - \frac{45}{2} G_1 G_{\sigma T} + \frac{45}{4} G_{\sigma} G_{\sigma T} + \frac{45}{2} G_T G_{\sigma T} - \frac{9}{4} G^2 \Bigg] \, ,
\end{split}
\end{align}

\begin{align}
\begin{split}
\beta(G_{\sigma}) = -\epsilon G_{\sigma} + \frac{1}{\pi} \Bigg[ \frac{39}{4} G_{\sigma}^2 + \frac{15}{4} G_T^2 + \frac{135}{16} G_{\sigma T}^2 - G_1 G_{\sigma} - \frac{15}{2} G_{\sigma} G_T + \frac{15}{4} G_{\sigma} G_{\sigma T} - \frac{15}{2} G_T G_{\sigma T} + \frac{21}{8} G^2 \Bigg] \, ,
\end{split}
\end{align}

\begin{align}
\begin{split}
\beta(G_{T}) = -\epsilon G_T + \frac{1}{\pi} \Bigg[ \frac{3}{2} G_{\sigma}^2 + 6 G_T^2 + \frac{63}{8} G_{\sigma T}^2 - G_1 G_T - 3 G_{\sigma} G_T - \frac{3}{2} G_{\sigma} G_{\sigma T} - \frac{3}{2} G_T G_{\sigma T} + \frac{3}{2} G^2 \Bigg] \, ,
\end{split}
\end{align}

\begin{align}
\begin{split}
\beta(G_{\sigma T}) = -\epsilon G_{\sigma T} + \frac{1}{\pi} \Bigg[ -\frac{1}{2} G_{\sigma}^2 - \frac{1}{2} G_T^2 + \frac{31}{8} G_{\sigma T}^2 - G_1 G_{\sigma T} + \frac{7}{2} G_{\sigma} G_{\sigma T} + \frac{11}{2} G_T G_{\sigma T} - \frac{7}{4} G^2 \Bigg] \, ,
\end{split}
\end{align}

\begin{align}
\begin{split}
\beta(G) = -\epsilon G - \frac{1}{\pi} \Bigg[ \frac{3}{2} G G_1 - \frac{15}{2} G G_{\sigma} - \frac{15}{4} G G_T + \frac{75}{4} G G_{\sigma T} \Bigg] \, .
\end{split}
\end{align}

\subsection{Analysis of the RG equations}

\subsubsection{RG flow and fixed points}

To analyze the RG flow, we first solve the above RG equations for fixed points. Given the complexity of the equations, we will proceed numerically, and first consider $N_v=2$ and $N_f=4$, the case our of main interest. This results in $2^5=32$ fixed points, among which 24 are complex and thus ignored. We focus our attention on the eight real solutions.

\begin{figure}[H]
\centering
\includegraphics[width=0.35\hsize]{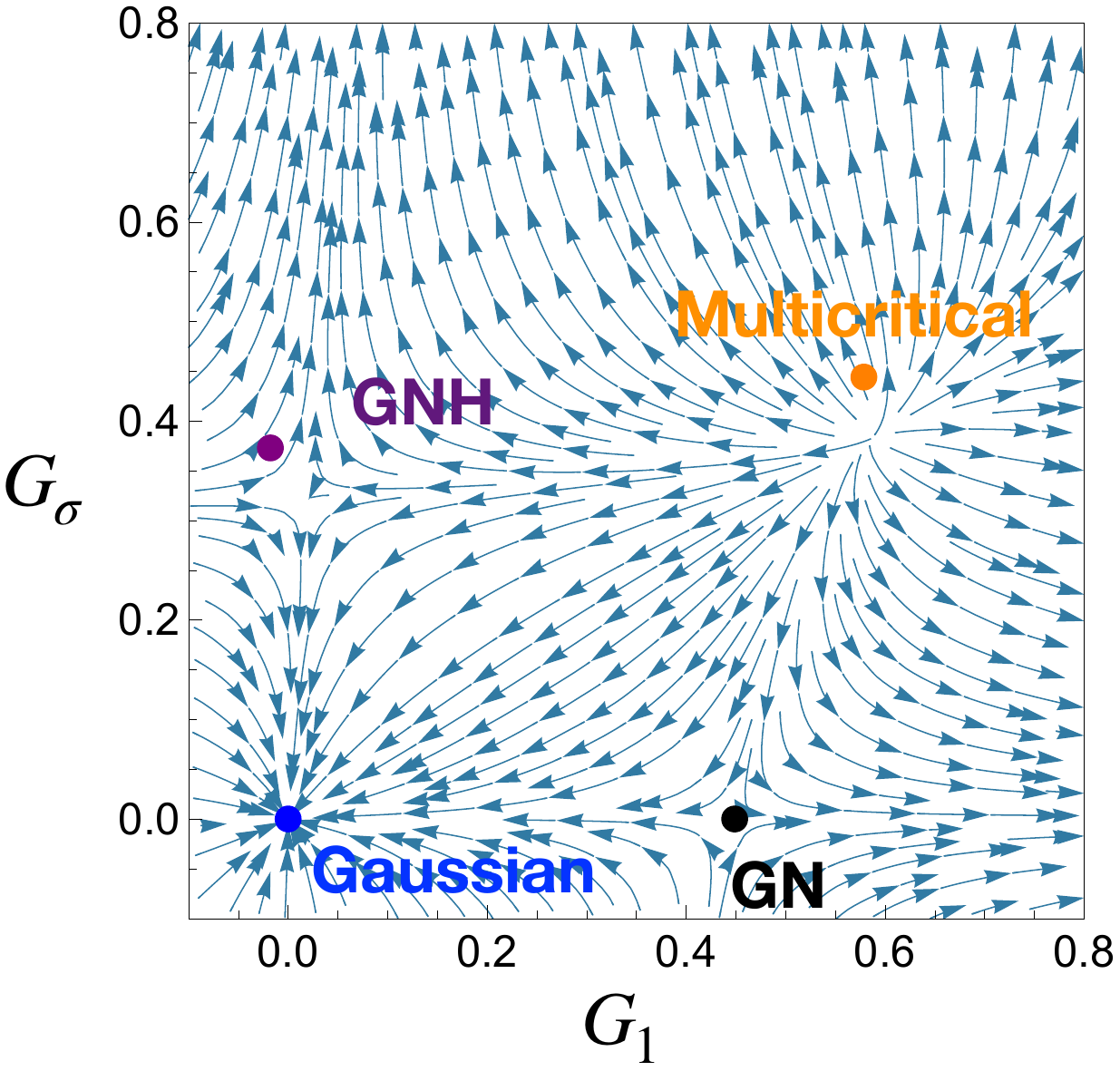}
\caption{2-d slice of the full 4-d RG flow in the $U(1)$-symmetric subspace ($G=0$), in the $(G_1,G_{\sigma})$ plane, with $G_{T},G_{\sigma T}=0$.}
\label{Fig:U(1)_RG}
\end{figure}

Let us first start by considering the four-dimensional $U(1)$-symmetric subspace, with $G=0$. In this case, there are four real fixed points. The first two are respectively the fully attractive free Dirac CFT at the origin and the Gross-Neveu (GN) fixed point, located at $(G_1,G_{\sigma},G_T,G_{\sigma T}) = (\pi \epsilon/7,0,0,0)$. The latter has a single relevant direction and describes a $\mathbb{Z}_2$ symmetry breaking phase transition between a Dirac semi-metal and an insulator. There is also a second fixed point with a single relevant direction. It is identified with the Gross-Neveu-Heisenberg (GNH) fixed point, describing a $SU(2)_f$-breaking phase transition, which will be further justified in the next section. Finally, the fourth fixed point has two relevant directions and is a multicritical point at the intersection of the three aforementioned stable phases. Fig. \ref{Fig:U(1)_RG} shows a two-dimensional slice of the RG flow in the $(G_1,G_{\sigma})$ plane, where the two other couplings $G_T$ and $G_{\sigma T}$ are set to zero. Since the GNH and the multicritical fixed points are located at nonzero values of $G_{T}$ and $G_{\sigma}$, their projection (respectively purple and orange dots) does not fully align with the apparent position of the fixed point from the RG flow lines.

We now move on to the full theory space with $G\neq 0$. There are four new fixed points, but because of the symmetry of the RG equations under $G \rightarrow -G$, two of them are redundant. We thus only focus our attention on $G>0$, since the RG flow for $G<0$ is a mirror image of the first case. Among these two new fixed points, one has a single relevant direction and is thus identified as the putative SMG fixed point, while the second has two relevant directions and is then a multicritical point.

However, the addition of a nonzero $G$ also has an additional non-trivial effect, as it destabilizes the GNH fixed point, which now acquires a second relevant direction. The $U(1)$-symmetric multicritical point also acquires an additional relevant direction and has thus three of those in the full theory. Therefore, this dictates the shape of the phase diagram presented in the main text.

\subsubsection{Calculation of critical exponents}

We are now in a position to calculate universal critical exponents. Again, we start by considering the $U(1)$-symmetric subspace. First, for both the GN and GNH fixed points, the (inverse) correlation length exponent at 1-loop is $\nu^{-1}=\epsilon$, which is a well known result in the literature\cite{gracey2018large,ladovrechis2023gross}). Next, we can also obtain the fermion anomalous dimension $\eta_{\psi}$ at 2-loop from Eq. \ref{Eq:eta_psi}. For the GN fixed point, one obtains the following analytical expression for any $N_v$ and $N_f$

\begin{equation}
\eta_{\psi}^{\text{GN}} = \frac{2N_v N_f-1}{8(N_v N_f-1)^2} \epsilon^2 \, ,
\end{equation}

\noindent which agrees with the literature\cite{gracey2018large,ladovrechis2023gross}). For $N_v=2$, $N_f=4$, this becomes $\eta_{\psi}^{\text{GN}} = \frac{15}{392} \epsilon^2 \approx 0.03827\epsilon^2$.

For the GNH fixed point, we must proceed numerically. For $N_v=2$, $N_f=4$, we get $\eta_{\psi}^{\text{GNH}} \approx 0.07366\epsilon^2$, which is remarkably close to the value obtained in \cite{ladovrechis2023gross}, despite a very different calculation scheme. Indeed, the authors of this paper work with the $D=2$ spacetime dimension spinor structure, which leads to the presence of four-fermion operators containing the Dirac matrix $\gamma^5$. On the other hand, recall that we work with Dirac spinors in $D=3$, and we thus do not have a $\gamma^5$. To push the comparison further, the anomalous dimension can be analytically computed in the large-$N_f$ limit. Solving perturbatively the RG equations \ref{Eq:beta_g1}-\ref{Eq:beta_g4} in powers of $1/N_f$, the location of the GNH fixed point is identified to be

\begin{align}
\begin{split}
g_1 &= \Big[ -\frac{1}{2N_f} - \frac{1}{2N_f^3} + \mathcal{O}(N_f^{-4}) \Big] \pi \epsilon \, , \quad g_2 = \Big[ \frac{1}{N_f} - \frac{1}{2N_f^3} + \mathcal{O}(N_f^{-4}) \Big] \pi \epsilon \, , \\ g_3 &= \Big[ \frac{1}{N_f^2} - \frac{1}{2N_f^3} + \mathcal{O}(N_f^{-4}) \Big] \pi \epsilon \, , \quad g_4 = \Big[ -\frac{1}{2N_f^2} - \frac{7}{8N_f^3} + \mathcal{O}(N_f^{-4}) \Big] \pi \epsilon \, .
\end{split}
\end{align}

\noindent The $1/N_f$ expansion of the fermion anomalous dimension is therefore

\begin{equation}
\eta_{\psi}^{\text{GNH}} = \Big[ \frac{3}{8N_f} - \frac{9}{32N_f^2} - \frac{3}{16N_f^3} + \mathcal{O}(N_f^{-4}) \Big] \epsilon^2 \, ,
\end{equation}

\noindent which agrees perfectly with \cite{ladovrechis2023gross,gracey2018large} up to $\mathcal{O}(N_f^{-4})$ (Note that Eq. 43 of \cite{ladovrechis2023gross} is missing an overall factor of $1/4$). Since our result for small $N_f$ deviates from the one in \cite{ladovrechis2023gross}, it implies that our two large-$N_f$ expansions must differ at higher order in $1/N_f$. Nevertheless, the perfect agreement up to $\mathcal{O}(1/N_f^3)$ constitutes a strong validation that the Gross-Neveu-Heisenberg fixed point has been correctly identified in our calculation.

Finally, at the SMG fixed point, the 1-loop inverse correlation length exponent is also $\nu^{-1}=\epsilon$. For the fermion anomalous dimension, using Eq. \ref{Eq:eta_psi}, one gets $\eta_{\psi}^{\text{SMG}} \approx 0.04650 \epsilon^2$. Note that most of this value comes from the $g^2$ term, as all the other terms with couplings of $U(1)$-symmetric operators almost cancel when evaluated at the SMG fixed point.

\section{Two species of Dirac fermions} \label{sec:boundary_SPT}

One can consider a simpler model of relativistic fermions in $D=2+\epsilon$ spacetime dimensions with an SMG-like term, consisting of only two flavors of Dirac spinors, given by the following Euclidean action

\begin{equation}
S = \int_x \bar{\psi}_i \slashed{\partial} \psi_i -\frac{G_1}{2} \int_x (\bar{\psi}_i \psi_i)^2 - \frac{G_{\sigma}}{2} \int_x (\bar{\psi}_i \sigma_{ij}^A \psi_j)^2 + \frac{G}{8} \epsilon_{ij}\epsilon_{kl} \int_x \Big[ \psi_{ai} \psi_{aj} \psi_{bk} \psi_{bl} + \bar{\psi}_{ai} \bar{\psi}_{aj} \bar{\psi}_{bk} \bar{\psi}_{bl} \Big] \, ,
\end{equation}

\noindent where $a=1,2$ labels the spinor component and $i=1,2$ labels the flavor, while $\sigma^A$ are the three Pauli matrices. However, this theory cannot describe the transition to a symmetric gapped phase since it does not contain enough Dirac fermions to cancel the $\mathbb{Z}_{16}$ anomaly. Nevertheless, it has a structure similar to the $SU(2)\times SU(4)$-symmetric theory previously studied and it is thus worth exploring its renormalization group structure.

The RG calculation follows closely the one in Appendix \ref{sec:app_rel}, where the hybrid dimensionality scheme is used again. The RG equations are first calculated in a basis with the operator $(\bar{\psi}_i \psi_j)(\bar{\psi}_j \psi_i)$ instead of $(\bar{\psi}_i \sigma_{ij}^A \psi_j)^2$. The two can then be related via the $SU(2)$ Fierz identity (Eq. \ref{Eq:Fierz_identities}), which thus results in

\begin{equation}
\beta(G_1) = -\epsilon G_1 + \frac{1}{\pi} \Bigg[ G_1^2 - 3 G_1 G_{\sigma} + 3 G_{\sigma}^2 - \frac{1}{2} G^2 \Bigg] \, ,
\end{equation}

\begin{equation}
\beta(G_{\sigma}) = -\epsilon G_{\sigma} + \frac{1}{\pi} \Bigg[ 3 G_{\sigma}^2 - G_1 G_{\sigma} - \frac{1}{4} G^2 \Bigg] \, ,
\end{equation}

\begin{equation}
\beta(G) = -\epsilon G - \frac{1}{2\pi} \Big[ 3 G_1 G + 9 G_{\sigma} G  \Big] \, .
\end{equation}

\begin{figure}[H]
\centering
\includegraphics[width=0.6\hsize]{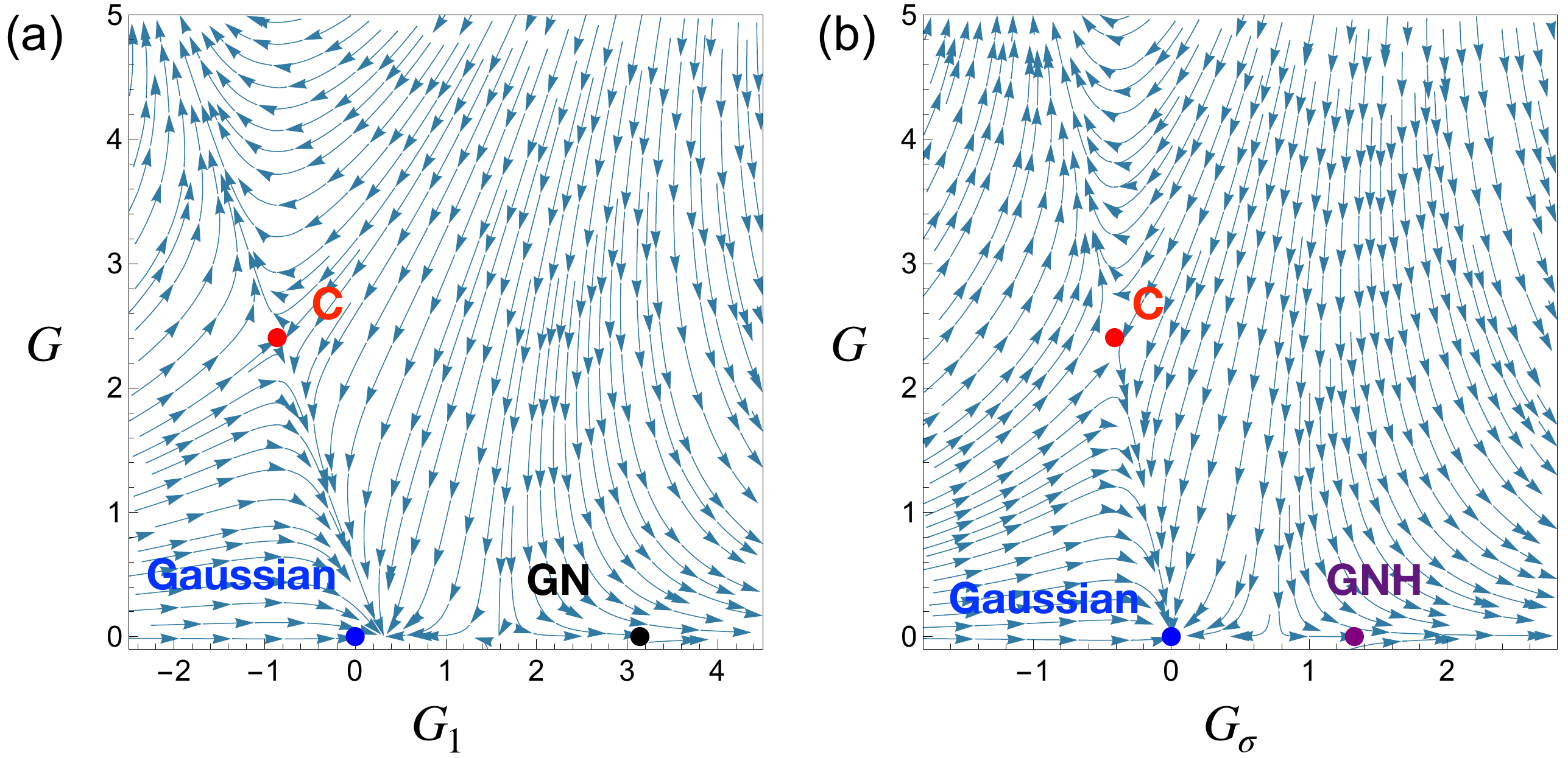}
\caption{2d slices of the full 3d RG flow. For both figures, the vertical axis represents the coupling $G$, while the horizontal axis corresponds to (a) the GN coupling $G_1$ or (b) the GNH coupling $G_{\sigma}$. In both flows, the coupling not represented is set to its value at the $G \neq 0$ fixed point with a single relevant direction, denoted by ``C'' (red). The other dots represent the projections of the Gaussian (blue), GN (black) and GNH (purple) fixed points.}
\label{Fig:Boundary_SPT}
\end{figure}

Solving for fixed points yields $2^3 = 8$ real solutions. Four of them have $G=0$ and correspond to the fully stable Gaussian fixed point $(G_1,G_{\sigma},G)=(0,0,0)$, the Gross-Neveu and Gross-Neveu-Heisenberg fixed point (both have a single relevant direction), respectively located at $(G_1,G_{\sigma},G)=(\pi \epsilon,0,0)$ and $(G_1,G_{\sigma},G)=(2-\sqrt{3},1-\sqrt{3}/3,0)\pi \epsilon$, and a multicritical point at $(G_1,G_{\sigma},G)=(2+\sqrt{3},1+\sqrt{3}/3,0)\pi \epsilon$ with two relevant directions. Note that unlike in the $SU(2)\times SU(4)$-symmetric theory, the GNH fixed point is \textit{not} destabilized by a nonzero $G$ (see Fig. \ref{Fig:Boundary_SPT} (b)).

The remaining four fixed points with $G \neq 0$ come in two pairs related by $G \rightarrow -G$, again due to the invariance of $\beta(G)$ under $G \rightarrow -G$. For $G>0$, the two fixed points are located at

\begin{equation}
(G_1,G_{\sigma},G) = \Bigg( \frac{1}{18}(17\mp \sqrt{481}), \frac{1}{54} (-29\pm \sqrt{481}), \frac{1}{9} \sqrt{\frac{2}{3}} \sqrt{1409 \mp \sqrt{481}} \Bigg) \pi \epsilon \, ,
\end{equation}

\noindent and respectively have one and two relevant directions. The former is the analog of the putative SMG fixed point identified in the $SU(2)\times SU(4)$-symmetric theory in the main text. Since this theory cannot contain an SMG phase, it is an interesting open question what phase one enters as one exits the gapless Dirac phase across the phase transition described by this fixed point.

\twocolumngrid

\end{document}